\documentclass[spanish, a4paper, 12pt, twoside, openany]{book} 

\usepackage{textcomp}                           
\usepackage[T1]{fontenc, url}                   
\usepackage[utf8]{inputenc}                     
\usepackage{titlesec}                           
\usepackage{multirow}                           
\usepackage{adjustbox}                          
\usepackage{graphicx}                           
\usepackage{amsmath,amssymb,amsfonts,dsfont,mathrsfs,amsthm,mathtools}          
\usepackage{color}                              
\usepackage{cancel}
\usepackage{subcaption}                         
\usepackage[toc,page]{appendix}                 
\usepackage{chngcntr}                           

\counterwithin{table}{section}                  
\counterwithin{figure}{section}                 
\numberwithin{equation}{section}                
\usepackage{xparse,nameref}                     
\usepackage[bottom,hang,flushmargin]{footmisc}  
\usepackage{lipsum}                             

\usepackage[spanish]{babel}                                         
\graphicspath{{Images/}{../Images/}}                                
\usepackage[left=2.5cm,top=4cm,bottom=2.5cm,right=2.5cm]{geometry}    
\usepackage{setspace}                                               
\usepackage{microtype}                                              
\usepackage{ragged2e}

\usepackage{fancyhdr}                           
\usepackage{subfiles}                           
\usepackage{array}                              
\usepackage[rightcaption]{sidecap}              
\usepackage{wrapfig}                            
\usepackage{float}                              
\usepackage[labelfont=bf]{caption}              
\usepackage[para]{threeparttable}               
\usepackage{url}                                
\usepackage[table,xcdraw,dvipsnames]{xcolor}    
\usepackage{makecell}                           
\usepackage{hhline}                             
\usepackage{textcomp}                           
\usepackage{graphicx}

\usepackage[noadjust]{cite}
\def\biblio{\clearpage\bibliographystyle{apalike}\bibliography{References}} 

\fancypagestyle{plain}{                     
\fancyhf{}                                  
\fancyhead[RO,LE]{\thepage}                 
\fancyhead[RE,LO]{\nouppercase{\leftmark}}  
\fancyfoot{}                                
}

\fancypagestyle{resumen}{                   
\fancyhf{}                                  
\fancyhead[RO,LE]{\thepage}                 
\fancyhead[RE,LO]{\nouppercase{Resumen}}    
\fancyfoot{}                                
}

\fancypagestyle{abstract}{                  
\fancyhf{}                                  
\fancyhead[RO,LE]{\thepage}                 
\fancyhead[RE,LO]{\nouppercase{Abstract}}   
\fancyfoot{}                                
}

\fancypagestyle{agradecimientos}{                   
\fancyhf{}                                          
\fancyhead[RO,LE]{\thepage}                         
\fancyhead[RE,LO]{\nouppercase{Agradecimientos}}    
\fancyfoot{}                                        
}

\fancypagestyle{frontpage}{             
	\fancyhf{}                          
	\vspace*{1\baselineskip}
	
	\   
	\fancyhead{\centering \includegraphics[width=1.8in]{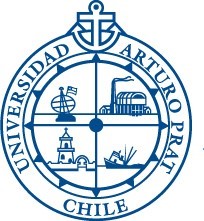}                       
	\linebreak\linebreak       Instituto de Ciencias Exactas y Naturales \\ Facultad de Ciencias} 
}

\usepackage{hyperref}   
\newcommand\myshade{85} 

\colorlet{mylinkcolor}{DarkOrchid}   
\colorlet{mycitecolor}{YellowOrange} 
\colorlet{myurlcolor}{Aquamarine}    

\hypersetup{  
  	linkcolor  = mylinkcolor!\myshade!black,    
  	citecolor  = mycitecolor!\myshade!black,    
  	urlcolor   = myurlcolor!\myshade!black,     
  	colorlinks = true,                          
}

\newcommand*{\Tr}{\operatorname{Tr}}

\newcommand{\diff}[1]{\text{d}#1}

\newcommand{\Lie}{\mathcal{L}}
\newcommand{\Lag}{\mathscr{L}}

\usepackage{hyperref}

\begin{document}
\def\biblio{}   

\begin{titlepage}
	
	\newgeometry{top=1 in, bottom=1 in, left=1 in, right= 1 in} 
	
	\thispagestyle{frontpage}
	
	\begin{center}
		
		\vspace*{8\baselineskip}

		{\Huge \textbf{Renormalización conforme de teorías escalar-tensor de la gravedad\\}}
		        \vspace*{1.5\baselineskip}

		
        \vspace*{1\baselineskip}

		\large{\textbf{Por: Mairym Eliana Busnego Barrientos}}\\ 
		
		\vspace{1,5\baselineskip}
		
		\large{Tesis presentada a la Facultad de Ciencias de la Universidad Arturo Prat y al Instituto de Ciencias Exactas y Naturales y  para optar al grado académico de Magíster en Ciencias con Mención en Física Teórica.} 
		
		\vspace{1,5\baselineskip}
		\begin{center}
            Marzo 2023 \\
            Iquique, Chile
            \end{center}
\vspace{1.5\baselineskip}

		\large{\textbf{Directores de tesis: Dr. Cristóbal Corral y  Dr. Nelson Merino}}\\ 
            \large{\textbf{Comisión: }}\\
	\end{center}
	
	\vspace*{4\baselineskip}
	
	
\end{titlepage}   
\vfill

\thispagestyle{empty}
\mbox{}                         
\vfill                          
\textcopyright\ 2023, Mairym Busnego Barrientos \\ 
Ninguna parte de esta tesis puede reproducirse o transmitirse bajo ninguna forma o por ningún medio o procedimiento, sin permiso por escrito del autor.\\\\
Se autoriza la reproducción total o parcial, con fines académicos, por cualquier medio o procedimiento, incluyendo la cita bibliográfica del documento
\vspace{1cm}    
\restoregeometry 







\thispagestyle{empty}
\begin{flushright}
\vfill
\textit{Now the world can be an unfair place at times\\
But your lows will have their compliment of highs\\
And if anyone should cheat you\\
Take advantage of or beat you raise your head\\
And wear your wounds with pride.}
\vfill
\end{flushright}
\restoregeometry        



\pagenumbering{roman}                            
\newpage
\addcontentsline{toc}{chapter}{AGRADECIMIENTOS}  
\section*{AGRADECIMIENTOS}                       

\vspace*{2\baselineskip}

Sinceramente creo que esta es la sección mas difícil de escribir, así que tratare de hacerlo de la mejor forma posible y no olvidar a nadie en el camino. En primer lugar, deseo expresar mi profundo agradecimiento a mi familia,  quienes han sido mi fuente inagotable de amor, apoyo y aliento a lo largo de esta travesía académica. A mis padres, Myriam Barrientos, mejor conocida como la Gordita y Elio Bunsego aka Gordin, quienes me han inculcado valores de perseverancia y dedicación, y me han brindado el respaldo incondicional en cada paso que he dado. A mi hermana, Nathalia por ser mi compañía y preocupación constante y por su confianza en mis capacidades. A Bastian Ayala, mi pareja, eres una inspiración constante para mi y un pilar fundamental en mi vida. Sin su constante apoyo emocional, realmente esta meta y tantas otras no habría sido posible. 

Asimismo, quisiera agradecer a mis amigos  y compañeros quienes han sido mi red de apoyo y aliento durante los momentos de estrés y desafíos. La lista es enorme, pero quisiera ser enfática también en esas personas que se tomaron el tiempo de enviarme memes y canciones, Gabriel Arenas, Camilo Nuñez, les debo el soundtrack de este logro.

No puedo dejar de reconocer a mis profesores, cuya sabiduría, conocimientos han sido invaluables en el desarrollo de mi tesis. Agradezco especialmente a Cristóbal Corral, Nelson Merino, Ignacio Araya y Giorgos Anastasiou por su dedicación y paciencia al dirigir este proyecto. Sus comentarios y sugerencias críticas han sido fundamentales para enriquecer mi trabajo y llevarlo a un nivel superior. Quisiera mencionar que siento una gratitud inmensa a la familia que se ha formado en Iquique, como primera experiencia de postgrado, Carlangas, Luchito, Luis, los profesores por ustedes fue como sentirse en casa. También me gustaría agradecer a Rodrigo Aros, quien ha sido una guía fundamental desde mi pregrado y sigue creyendo en mi como desde el primer día en que acepto ser mi tutor en la licenciatura en física. 

Además, quiero dar las gracias a las personas que me brindaron su apoyo en los cálculos e interpretaciones de este trabajo. Agradezco a Nicolás Cáceres, Gonzalo Barriga, Francisco Colipí, Rodrigo Olea, Leonardo Sanhueza y Kristiansen Lara, por su asesoramiento y su disposición para responder mis preguntas y su habilidad para simplificar conceptos complejos fueron esenciales para llevar a cabo esta investigación.

Finalmente, deseo agradecer a todas las personas que, de una forma u otra, contribuyeron a este logro. A aquellos que he conocido en escuelas, conferencias y workshops, es lindo saber que la comunidad de la física teórica se esta formando con gente tan dedicada, amable y unida, donde todos se ayudan a crecer como académicos y personas.

\vspace*{3\baselineskip}




\newpage
\addcontentsline{toc}{chapter}{Resumen} 
\section*{Resumen}                      

En esta tesis, estudiamos el método de renormalización conforme aplicado en teorías con grados de libertad más allá de los métricos. Específicamente, revisamos este método en presencia de un campo escalar. Para ello, a modo de revisión, repasamos el principio de acción de Relatividad General y las ecuaciones de Einstein, además de revisitar las condiciones para que esta teoría tenga un principio variacional bien definido cuando imponemos condiciones de borde tipo Dirichlet. Luego, examinamos diversos métodos para calcular cargas conservadas en espacios asintóticamente planos. En espacios asintóticamente anti-de Sitter, estudiamos dos esquemas de renormalización los cuales son relevantes para este trabajo. Con el fin de motivar el método aquí empleado, veremos que la teoría de Gravedad Conforme es finita para espacios que son asintóticamente anti-de Sitter, tal como se demostró en \cite{Grumiller_2014}; un principio guía que nos da una pista de cómo la simetría conforme estaría relacionada con la renormalización en espacio-tiempos con dicha asíntota. La base de esta construcción es la extensión de un tensor covariante bajo rescalamientos de Weyl compuesto de la métrica y del campo escalar propuesto en la \cite{Oliva:2011np}. Esta extensión hace que el peso conforme de dicho tensor sea igual al del tensor de Weyl. Extendemos esta realización revisando teorías tensor-escalar que gozan de simetría conforme, acopladas con la acción de Einstein-AdS escrita en la forma de MacDowell-Mansouri. A pesar de que el sector de Einstein-AdS rompe la simetría conforme, mostramos que la teoría completa aún puede ser renormalizada si el campo escalar tiene un decaimiento adecuado cuando se consideran soluciones asintóticamente anti-de Sitter. Finalmente, estudiamos las soluciones tipo agujero negro, calculando su temperatura de Hawking y la acción Euclídea on-shell, mostrando explícitamente que esta última es finita para espacios asintóticamente anti-de Sitter.

\par\vspace*{\fill} 

\newpage
\addcontentsline{toc}{chapter}{Abstract} 
\section*{Abstract}

In this thesis, we investigate the method of conformal renormalization applied to theories with degrees of freedom beyond the metric ones. Specifically, we examine this method in the presence of a scalar field. To do this, as part of a review, we revisit the action principle of General Relativity and Einstein's equations, in addition to re-examining the conditions for this theory to have a well-defined variational principle when Dirichlet boundary conditions are imposed. We then explore various methods for calculating conserved charges in asymptotically flat spaces. In asymptotically anti-de Sitter spaces, we study two renormalization schemes that are relevant to this work. To motivate the method used here, we observe that Conformal Gravity is finite for spaces that are asymptotically anti-de Sitter, as demonstrated in \cite{Grumiller_2014}. This guiding principle gives us a clue as to how conformal symmetry may be related to renormalization in spacetimes with such asymptotics. The basis of this construction is the extension of a covariant tensor under Weyl rescalings composed of the metric and the scalar field as proposed in \cite{Oliva:2011np}. This extension ensures that the conformal weight of this tensor is equal to that of the Weyl tensor. We extend this realization by considering tensor-scalar theories with conformal symmetry, coupled with the Einstein-AdS action written in the MacDowell-Mansouri form. Despite the fact that the Einstein-AdS sector breaks conformal symmetry, we show that the entire theory can still be renormalized if the scalar field has an appropriate decay when considering asymptotically anti-de Sitter solutions. Finally, we study black hole-type solutions, calculating their Hawking temperature and the Euclidean on-shell action, explicitly demonstrating that the latter is finite for asymptotically anti-de Sitter spaces.

\par\vspace*{\fill} 
\textbf{\textit{Keywords --}} General Relativity, Master Thesis, Black Holes, Renormalization 

\biblio 



\newpage
{\setstretch{1.0}   
\tableofcontents
}



\newpage
\addtocontents{toc}{\protect\setcounter{tocdepth}{4}}   
\pagenumbering{arabic}                                  
\setcounter{page}{1}                                    


\chapter{Introducción}                      

La renormalización de las teorías gravitacionales que admiten soluciones asintóticamente localmente anti-de Sitter (AlAdS) se ha convertido en un ingrediente crucial en la termodinámica de agujeros negros y en la correspondencia anti-de Sitter/teorías de campo conforme (AdS/CFT). La prescripción estándar ---denominada renormalización holográfica (HR)--- consiste en agregar el término de Gibbons-Hawking-York (GHY) para fijar el principio variacional de Dirichlet para la métrica inducida en la foliación radial y luego introducir contratérminos de borde intrínsecos para anular las divergencias que aparecen al evaluar la acción en soluciones con comportamiento AlAdS. Esto hace que la acción Euclídea on-shell y las cargas asintóticas sean finitas~\cite{Henningson:1998gx,Balasubramanian:1999re,Chamblin:1999tk,Emparan:1999pm,Chamblin:1999hg,Nojiri:1999mh,deHaro:2000vlm,Bianchi:2001kw,Skenderis:2002wp}, lo que permite definir el funcional generador para correladores de la CFT dual y obtener los datos holográficos de dicha teoría~\cite{Gubser:1998bc,Witten:1998qj}. 

Una observación interesante, que se hizo en Ref.~\cite{Papadimitriou:2005ii}, es que imponer la condición de Dirichlet para la métrica inducida en el borde de AdS está mal definido debido a que el elemento de volumen del espacio de AdS diverge. Además, fijar la condición de Dirichlet para la fuente holográfica asintóticamente del borde no requiere fijar la condición de Dirichlet para la métrica intrínseca, ya que tanto la curvatura intrínseca como la extrínseca admiten una expansión de de Fefferman-Graham (FG ) en la fuente holográfica~\cite{AST_1985__S131__95_0,Graham:1999jg}. Esto condujo al desarrollo de la prescripción de Kounterterms en Ref.~\cite{Olea:2006vd}, donde la renormalización se logra mediante la adición de un término de borde adecuado, que depende de las curvaturas tanto extrínseca como intrínseca del borde en una forma cerrada. Incluso para dimensiones pares en el volumen, el Kounterterm es la forma de Chern, que es el término de borde que aparece en el teorema de Euler. Entonces, la renormalización también se puede lograr agregando la densidad de Euler en el volumen con un acoplamiento fijo, de modo que cancela la divergencia de la configuración maximalmente simétrica (AdS global). En el caso de dimensiones de volumen impares, existe un procedimiento similar donde el término de frontera es el término de contacto de la forma de transgresión del grupo AdS~\cite{Mora:2006ka}. En ese caso, la segunda conexión de gauge describe una variedad de producto que comparte el mismo borde que la variedad dinámica.

En particular, como se muestra en Ref.~\cite{Miskovic:2009bm}, la acción de Einstein-AdS parcialmente renormalizada con Kounterterms en 4D se puede escribir en la forma de McDowell-Mansouri para el grupo de AdS~\cite{Stelle:1976gc,MacDowell:1977jt}. En el caso de los espacio-tiempos tipo Einstein-AdS, este última se puede escribir en términos del tensor de Weyl al cuadrado y la acción on-shell se convierte en la de Conformal Gravity (CG), que es la única invariante conforme local en 4D. Esta reescritura de la teoría de Einstein encajándola en CG es consistente a nivel de las ecuación de movimiento (EOM), ya que todos los espacio-tiempos tipo Einstein pertenecen al espacio de solución de la teoría, cuya EOM está dada por la condición de que el tensor de Bach es igual a cero. Dado que el conjunto de soluciones de CG contiene a los espacios tipo Einstein, la acción correspondiente se puede separar explícitamente en una parte de MacDowell-Mansouri más términos que se anulan para los espacios tipo Einstein~\cite{Anastasiou:2016jix}.

Para las variedades con asintótica de AdS tenue, se demostró que la acción de CG es finita off-shell, incluso para espacios-tiempos que no tienen un tensor de Bach que se anula~\cite{Grumiller:2013mxa}. Por lo tanto, la finitud de la acción de McDowell-Mansouri para la gravedad de Einstein-AdS se sigue de inmediato, ya que las dos acciones son equivalentes para los espacios tipo Einstein. Este fue el primer ejemplo en el que la incorporación de una teoría de gravedad en otra con invariancia conforme local en el volumen permitió obtener la forma renormalizada de la acción. Más tarde, en la Ref.~\cite{Anastasiou:2020mik}, se generalizó el mismo procedimiento para la gravedad de Einstein-AdS en 6D, siendo esta inmersa en la única acción de CG en 6D que admite espacios tipo Einstein como soluciones, construida en Ref.~\cite {Lu:2011zk}. Es importante enfatizar que, aunque la incorporación de Einstein-AdS en CG en 4D da el mismo principio de acción que los Kounterterms, esto no es cierto en 6D. Como se discutió en Ref.~\cite{Anastasiou:2020zwc}, en el caso 6D, la acción topológicamente renormalizada cancela todas las divergencias solo para espacios AlAdS con un borde conformemente plano. En el caso genérico, la última prescripción no cancela una divergencia del borde que depende del cuadrado de Weyl de la variedad del borde. Sin embargo, la inmersión en la teoría de CG en 6D reproduce correctamente todos los términos necesarios para lograr la renormalización, de modo que la acción obtenida es totalmente equivalente a la dada por HR, hasta el orden normalizable. Así, es esta prescripción la que da la renormalización correcta, generalizando los Kounterterms más allá del requisito de que la 
 variedad sea conformalmente plana en el borde.

Más allá de las teorías puramente métricas, se han considerado enfoques de renormalización para casos con grados de libertad adicionales, e.g. campos escalares. De hecho, en el caso de renormalizacion holográfica, los contratérminos para espacios AlAdS en las teorías escalar-tensor se han discutido en Refs.~\cite{Nojiri:1998dh,Padilla:2012ze,Caldarelli:2016nni,Liu:2017kml,Li:2018rgn,Agurto-Sepulveda:2022vvf}. Por lo tanto, una pregunta natural es si el uso de la simetría conforme local en el volumen se puede generalizar para estas teorías, y de esa forma determinar los términos de renormalización. Aquí, abordamos este tema y construimos acciones de gravedad renormalizadas que poseen un sector escalar-tensor acoplado conforme, cuyas soluciones han sido estudiadas en la literatura. Por lo tanto, esto constituye la primera aplicación de la idea de \emph{Renormalización Conforme} a las teorías escalar-tensor de la gravedad.

\biblio 
\clearpage                                  

\chapter{Relatividad General}
    
En el siglo XX, las teorías que explicaban el comportamiento de la naturaleza comenzaron a cambiar radicalmente. La no invarianza de las ecuaciones de Maxwell bajo las transformaciones de Galileo y  el nulo resultado del experimento de Michelson-Morley en detectar el ether---entidad en la que se propagaría la luz---, llevó a Albert Einstein a la formulación de la Relatividad Especial en 1905, pudiendo así explicar fenómenos a velocidades cercanas a las de la luz, introduciendo el concepto de espacio-tiempo. No obstante, su mayor invención llegaría en 1915 cuando introdujo la teoría de la Relatividad General. Esta nueva teoría de gravedad cumpliría con 3 principios generales: primero, sus ecuaciones son tensoriales, por lo tanto covariantes ante transformaciones de coordenadas. Además, en esta formulación, el espacio-tiempo es dinámico y no un escenario inmóvil en donde suceden las cosas. Segundo, involucran hasta segundas derivadas de la métrica, lo que las hace compatibles con las ecuaciones de movimiento que conocemos en las teorías clásicas de la física. Finalmente, como tercera propiedad, estas ecuaciones recuperan la gravedad de Newton en el llamado límite de bajas velocidades y campo débil. Siguiendo un procedimiento puramente físico, Einstein logro escribir las ecuaciones de campo de la Relatividad General, estas son
\begin{equation}\label{eomeinstein}
    R_{\mu\nu}-\frac{1}{2}g_{\mu\nu}R =\frac{8\pi G}{c^4}T_{\mu\nu}\, ,
\end{equation}
donde $R_{\mu\nu}$ es el tensor de Ricci, $g_{\mu\nu}$ es el tensor métrico, $g^{\mu\nu}R_{\mu\nu}=R$ es el escalar de Ricci y $T_{\mu\nu}$ es el tensor de energía-momentum de la materia. Estas ecuaciones han logrado describir satisfactoriamente fenómenos que hasta el momento no eran posibles de explicar con la teoría de gravitación universal de Newton, tales como la precesión del perihelio de Mercurio, deflexión de la luz por el sol, el redshift gravitacional de la luz, entre otros \cite{Will:2014kxa}.

\section{Acción de Einstein-Hilbert y ecuaciones de movimiento}
Paralelamente a Einstein, David Hilbert trabajaba en obtener las Eqs.~\eqref{eomeinstein} de un principio variacional. Para ello, propuso un funcional de la métrica sobre una variedad diferencial $\mathcal{M}$ que fuera invariante bajo transformaciones de coordenadas y que entregara las ecuaciones~\eqref{eomeinstein} al realizar variaciones arbitrarias con respecto de los campos dinámicos. A este principio de acción hoy lo conocemos como acción de Einstein-Hilbert, la cual está dada por
\begin{equation}\label{EHM}
    I[g_{\mu\nu},\psi]=\kappa \int_{\mathcal{M}} d^{4}x \sqrt{\lvert g\rvert}\, R + \int_{\mathcal{M}}d^{4}x \sqrt{\lvert g\rvert}\, \Lag_M[g_{\mu\nu},\psi] \, , 
\end{equation}
en donde $\psi$ es algún campo de materia, $\kappa=(16\pi G)^{-1}$ con $G$ la constante de Newton, $g=\det g_{\mu\nu}$ es el determinante de la métrica, y $\Lag_M$ es un Lagrangiano de materia, cuya variación con respecto de sus campos dinámicos entrega las ecuaciones de movimiento para la materia. Por otro lado, la variación de $\Lag_M$ con respecto de la métrica entrega el tensor de energía-momentum definido por
\begin{align}
T_{\mu\nu} = -\frac{2}{\sqrt{|g|}}\frac{\delta}{\delta g^{\mu\nu}}\left(\sqrt{|g|}\,\Lag_M\right)\,.
\end{align}
En esta sección, vamos a considerar la acción \eqref{EHM} sin materia, i.e. $\Lag_{M}=0$ La variación arbitraria de esta acción con respecto de la métrica entrega
\begin{align} \notag
     \delta I[g_{\mu\nu}]&=\kappa \int_{\mathcal{M}} d^{4}x \delta[\sqrt{\lvert g\rvert}\, R]\\
     &=\kappa \int_{\mathcal{M}} d^{4}x \left[\delta(\sqrt{\lvert g\rvert})R\,+ \sqrt{\lvert g\rvert}\delta R\right] \, . \label{variacion1}
\end{align}
Analicemos el primer termino de Ec.~\eqref{variacion1}. Conocemos que la variación cumple con la regla de Leibniz, por lo que se tiene
\begin{align} \notag
    \delta\sqrt{\lvert g \rvert}&=\frac{1}{2}\frac{1}{\sqrt{\lvert g \rvert}}\delta \lvert g \rvert\\ \notag
    &=\frac{1}{2}\lvert g \rvert^{-\frac{1}{2}}g^{\mu\nu}\delta g_{\mu\nu}\lvert g \rvert \\ \notag
    &=\frac{1}{2}\sqrt{\lvert g \rvert}g^{\mu\nu}\delta g_{\mu\nu} \, \\ 
    &=-\frac{1}{2}\sqrt{\lvert g \rvert}g_{\mu\nu}\delta g^{\mu\nu} \, ,\label{vd}
\end{align}
en donde para la variación del determinante de la métrica hemos usado la fórmula de Jacobi
\begin{align*}
    \delta \lvert g \rvert=\delta\left(e^{\Tr[\log(g_{\mu\nu})]}\right)&=\delta\left(\Tr[\log(g_{\mu\nu})]\right)e^{\Tr[\log(g_{\mu\nu})]}\\
    &=\Tr[\delta\left(\log(g_{\mu\nu})\right)]e^{\Tr[\log(g_{\mu\nu})]}\\
    &=\Tr\left(g^{\mu\nu}\delta g_{\mu\nu}\right)e^{\Tr[\log(g_{\mu\nu})]}\\
    &=g^{\mu\nu}\delta g_{\mu\nu}\lvert g \rvert \, ,
\end{align*}
en donde se puede ver que, en la penúltima linea, hemos tomado la traza al elemento dentro del paréntesis. Además, en la Ec.~\eqref{vd} usamos la propiedad 
\begin{align}
    \delta(g_{\mu\nu}g^{\nu\alpha})=0\, .
\end{align}
Estudiemos ahora la variación del termino $\delta R$. Comencemos por la definición del escalar de Ricci
\begin{equation} \label{varr}
    \delta R= \delta (g^{\mu\nu}R_{\mu\nu})=\delta (g^{\mu\nu}R^{\lambda}\,_{\mu\lambda\nu})=\delta g^{\mu\nu}R^{\lambda}_{\ \mu\lambda\nu}+g^{\mu\nu}\delta R^{\lambda}_{\ \mu\lambda\nu} \, .
\end{equation}
Para obtener la variación del tensor de Ricci vamos a considerar su definición 
\begin{equation}
R^{\lambda}_{\ \rho\mu\nu}=\partial_{\mu}\Gamma^{\lambda}_{\rho\nu}-\partial_{\nu}\Gamma^{\lambda}_{\rho\mu}+\Gamma^{\lambda}_{\sigma\mu}\Gamma^{\sigma}_{\rho\nu}-\Gamma^{\lambda}_{\sigma\nu}\Gamma^{\sigma}_{\rho\mu}\, ,
\end{equation}
en donde $\Gamma^{\lambda}_{\lambda\mu}$ es el símbolo de Christoffel que está dado por
\begin{equation} \label{chr}
    \Gamma^{\lambda}_{\mu\nu}=\frac{1}{2}g^{\lambda\gamma}\left(\partial_{\mu}g_{\gamma\nu}+\partial_{\nu}g_{\mu\gamma}-\partial_{\gamma}g_{\mu\nu}\right)\,.
\end{equation}
Luego, variando el tensor de Riemann obtenemos
\begin{align} \notag
\delta R^{\lambda}_{\ \rho\mu\nu}&=\delta(\partial_{\mu}\Gamma^{\lambda}_{\rho\nu})-\delta(\partial_{\nu}\Gamma^{\lambda}_{\rho\mu})+\delta(\Gamma^{\lambda}_{\sigma\mu}\Gamma^{\sigma}_{\rho\nu})-\delta(\Gamma^{\lambda}_{\sigma\nu}\Gamma^{\sigma}_{\rho\mu})\\
&=\partial_{\mu}\delta\Gamma^{\lambda}_{\rho\nu}-\partial_{\nu}\delta\Gamma^{\lambda}_{\rho\mu}+\delta(\Gamma^{\lambda}_{\sigma\mu})\Gamma^{\sigma}_{\rho\nu}+\Gamma^{\lambda}_{\sigma\mu}\delta(\Gamma^{\sigma}_{\rho\nu})-\delta(\Gamma^{\lambda}_{\sigma\nu})\Gamma^{\sigma}_{\rho\mu}-\Gamma^{\lambda}_{\sigma\nu}\delta(\Gamma^{\sigma}_{\rho\mu})\, ,
\end{align}
en donde usamos que la variación conmuta con las derivadas parciales. Ahora, se puede ver que es posible armar derivadas covariantes del siguiente modo
\begin{align}\notag
    \nabla_{\mu}(\delta\Gamma^{\lambda}_{\rho\nu})&= \partial_{\mu}(\delta\Gamma^{\lambda}_{\rho\nu})+\Gamma^{\lambda}_{\mu\sigma}(\delta\Gamma^{\sigma}_{\rho\nu})-\Gamma^{\sigma}_{\mu\rho}(\delta\Gamma^{\lambda}_{\sigma\nu})-\Gamma^{\sigma}_{\mu\nu}(\delta\Gamma^{\lambda}_{\rho\sigma})\, ,\\
    \nabla_{\nu}(\delta\Gamma^{\lambda}_{\rho\mu})&= \partial_{\nu}(\delta\Gamma^{\lambda}_{\rho\mu})+\Gamma^{\lambda}_{\nu\sigma}(\delta\Gamma^{\sigma}_{\rho\mu})-\Gamma^{\sigma}_{\nu\rho}(\delta\Gamma^{\lambda}_{\sigma\mu})-\Gamma^{\sigma}_{\nu\mu}(\delta\Gamma^{\lambda}_{\rho\sigma})\, .
\end{align}
Reemplazando las expresiones de las derivadas parciales en la variación del tensor de Riemann, encontramos que 
\begin{eqnarray*}
    \delta R^{\lambda}_{\ \rho\mu\nu}&=&\nabla_{\mu}(\delta\Gamma^{\lambda}_{\rho\nu})-\Gamma^{\lambda}_{\mu\sigma}(\delta\Gamma^{\sigma}_{\rho\nu})+\Gamma^{\sigma}_{\mu\rho}(\delta\Gamma^{\lambda}_{\sigma\nu})+\Gamma^{\sigma}_{\mu\nu}(\delta\Gamma^{\lambda}_{\rho\sigma})\\
    &-&\nabla_{\nu}(\delta\Gamma^{\lambda}_{\rho\mu})+\Gamma^{\lambda}_{\nu\sigma}(\delta\Gamma^{\sigma}_{\rho\mu})-\Gamma^{\sigma}_{\nu\rho}(\delta\Gamma^{\lambda}_{\sigma\mu})-\Gamma^{\sigma}_{\nu\mu}(\delta\Gamma^{\lambda}_{\rho\sigma})\\
     &+&\delta(\Gamma^{\lambda}_{\sigma\mu})\Gamma^{\sigma}_{\rho\nu}+\Gamma^{\lambda}_{\sigma\mu}\delta(\Gamma^{\sigma}_{\rho\nu})-\delta(\Gamma^{\lambda}_{\sigma\nu})\Gamma^{\sigma}_{\rho\mu}-\Gamma^{\lambda}_{\sigma\nu}\delta(\Gamma^{\sigma}_{\rho\mu})\,.
\end{eqnarray*}
Podemos notar que varios términos se cancelan, así nos queda un resultado importante que se conoce como Identidad de Palatini 
\begin{align}
    \delta R^{\lambda}{}_{\rho\mu\nu}=2 \nabla_{[\mu } \delta \Gamma_{\rho \mid \nu]}^\lambda \label{palatini}\, ,
\end{align}
luego si cambiamos $\mu\to\lambda$ nos queda 
\begin{equation}
    \delta R^{\lambda}\,_{\rho\lambda\nu}=\nabla_{\lambda}\delta\Gamma^{\lambda}_{\rho\nu}-\nabla_{\nu}\delta\Gamma^{\lambda}_{\rho\lambda}\, . \label{pal}
\end{equation}
Teniendo como resultado la variación del Riemann en la Ec.~\eqref{pal}, reemplazamos en la variación arbitraria~\eqref{varr} y obtenemos
\begin{align} \nonumber
    \delta R&=\delta g^{\mu\nu}R^{\lambda}\,_{\mu\lambda\nu}+g^{\mu\nu}\delta R^{\lambda}\,_{\mu\lambda\nu}\\ \notag
    &=\delta g^{\mu\nu}R^{\lambda}\,_{\mu\lambda\nu}+g^{\mu\nu}(\nabla_{\lambda}\delta\Gamma^{\lambda}_{\mu\nu}-\nabla_{\nu}\delta\Gamma^{\lambda}_{\mu\lambda})\\
    &=\delta g^{\mu\nu}R^{\lambda}\,_{\mu\lambda\nu}+2g^{\mu\nu}\nabla_{[\lambda}\delta\Gamma^{\lambda}_{\mu|\nu]}\, . \label{varRie}
\end{align}
Reemplazando los resultados de las variaciones de cada termino de la Ec.~\eqref{variacion1} y reordenando nos queda
\begin{align} \nonumber
     \delta I[g_{\mu\nu}]&=\kappa \int_{\mathcal{M}} d^{4}x \left[\left(-\frac{1}{2}\sqrt{\lvert g \rvert}g_{\mu\nu}\,\delta g^{\mu\nu}\right)R\,+ \sqrt{\lvert g\rvert}\left(\delta g^{\mu\nu}R_{\mu\nu}+2g^{\mu\nu}\nabla_{[\lambda}\delta\Gamma^{\lambda}_{\mu|\nu]}\right)\right] \\
     &=\kappa \int_{\mathcal{M}} d^{4}x \sqrt{\lvert g \rvert}\left(R_{\mu\nu}-\frac{1}{2}g_{\mu\nu}R\right)\delta g^{\mu\nu}+2\kappa \int_{\mathcal{M}} d^{4}x \sqrt{\lvert g \rvert}\,g^{\mu\nu}\nabla_{[\lambda}\delta\Gamma^{\lambda}_{\mu|\nu]}\,.
\end{align}
Analicemos el último término de la expresión que obtuvimos. Ahora, nuestro objetivo sera escribirlo como un termino de borde. Antisimetrizar los índices de la derivada covariante es equivalente escribir a la expresión anterior con deltas generalizadas, esto es, 
\begin{align}
2g^{\mu\nu}\nabla_{[\lambda}\delta\Gamma^{\lambda}_{\mu|\nu]}=2\cdot 2g^{\mu\nu}\delta^{[\rho}_{[\lambda}\delta^{\sigma]}_{\nu]}\nabla_{\rho}\delta\Gamma^{\lambda}_{\mu\sigma}\, .
\end{align}
Luego, usando la definición de la delta de Kronecker generalizada,
\begin{equation}
\delta^{\mu_{1}...\mu_{p}}_{\nu_{1}...\nu_{p}}=p!\delta^{[\mu_{1}...\mu_{p}]}_{[\nu_{1}...\nu_{p}]}=p!\delta^{[\mu_{1}...\mu_{p}]}_{\nu_{1}...\nu_{p}}=p!\delta^{\mu_{1}...\mu_{p}}_{[\nu_{1}...\nu_{p}]}\, ,
\end{equation}
nos queda lo siguiente
\begin{align}  \notag  2g^{\mu\nu}\nabla_{[\lambda}\delta\Gamma^{\lambda}_{\mu|\nu]}&=2g^{\mu\nu}\delta^{\rho\sigma}_{\lambda\nu}\nabla_{\rho}\delta\Gamma^{\lambda}_{\mu\sigma}\\ \notag
&=2\nabla_{\rho}\left(g^{\mu\nu}\delta^{\rho\sigma}_{\lambda\nu}\delta\Gamma^{\lambda}_{\mu\sigma}\right)\\
&\equiv 2\nabla_{\mu}\Theta^{\mu}\, ,
    \end{align}
con $\Theta^{\mu}=g^{\mu\nu}\delta^{\rho\sigma}_{\lambda\nu}\delta\Gamma^{\lambda}_{\mu\sigma}$ y donde hemos usado la condición de metricidad $\nabla_{\lambda}g_{\mu\nu}=0$ y que la derivada covariante de la delta generalizada es cero ya que es un tensor constante. Entonces, podemos escribir la variación de la  acción del siguiente modo
\begin{equation}
    \delta I[g_{\mu\nu}]=\kappa \int_{\mathcal{M}} d^{4}x \sqrt{\lvert g \rvert}G_{\mu\nu}\delta g^{\mu\nu}+2\kappa \int_{\mathcal{M}} d^{4}x \sqrt{\lvert g \rvert}\nabla_{\mu}\Theta^{\mu}\,, \label{varinicial}
\end{equation}
con el tensor de Einstein definido como  $G_{\mu\nu}=R_{\mu\nu}-\frac{1}{2}g_{\mu\nu}R$. Finalmente, usando el teorema de Stokes podemos escribir el último término como un término de borde. De este modo, encontramos que
\begin{equation}
    \delta I[g_{\mu\nu}]=\kappa \int_{\mathcal{M}} d^{4}x \sqrt{\lvert g \rvert}G_{\mu\nu}\delta g^{\mu\nu}+2\kappa\sigma \int_{\partial\mathcal{M}} d^{3}x \sqrt{\lvert h \rvert}\Theta^{\mu}n_{\mu}\, , \label{var3}
\end{equation} 
con $h_{\mu\nu}=g_{\mu\nu}-\sigma n_{\mu}n_{\nu}$ la métrica inducida en $\partial_\mathcal{M}$, $n_{\mu}$ el vector normal a una hipersuperficie $\partial\mathcal{M}$ cerrada que puede ser tipo tiempo o tipo espacio y $\sigma=n_{\mu}n^{\mu}=\pm 1$ donde el valor de la norma depende si el vector normal es tipo tiempo ($\sigma=-1$) o tipo espacio ($\sigma=1$). Estos elementos serán vistos con claridad en la Sección que sigue. 

Podemos reducir aún más la Ec.~\eqref{var3}, considerando la condición de metricidad
\begin{align}
\nabla_{\lambda}g_{\mu\nu}=
    \partial_{\lambda}g_{\mu\nu}-\Gamma^{\alpha}_{\mu\lambda}g_{\alpha\nu}-\Gamma^{\alpha}_{\nu\lambda}g_{\mu\alpha}=0\, , \label{met}
\end{align}
y tomando la variación a ambos lados de ésta ecuación. Así, obtenemos
\begin{eqnarray}
    \partial_{\lambda}\delta g_{\mu\nu}-\delta\Gamma^{\alpha}_{\mu\lambda}g_{\alpha\nu}-\Gamma^{\alpha}_{\mu\lambda}\delta g_{\alpha\nu}-\delta\Gamma^{\alpha}_{\nu\lambda}g_{\mu\alpha}-\Gamma^{\alpha}_{\nu\lambda}\delta g_{\mu\alpha}=0 \, .
\end{eqnarray}
Notemos que los primeros tres términos forman la derivada covariante de $\delta g_{\mu\nu}$
\begin{equation}
    \nabla_{\lambda}\delta g_{\mu\nu}=\partial_{\lambda}\delta g_{\mu\nu}-\Gamma^{\alpha}_{\mu\lambda}\delta g_{\alpha\nu}-\Gamma^{\alpha}_{\nu\lambda}\delta g_{\mu\alpha}\label{1} \, .
\end{equation}
Ahora, reemplazando la derivada covariante de la Ec.~\eqref{1} en la variación de la condición de metricidad~\eqref{met} y haciendo permutaciones cíclicas de los índices $\lambda$, $\mu$ y $\nu$, obtenemos
\begin{eqnarray} \label{2}
    \nabla_{\lambda}\delta g_{\mu\nu}&=&\delta\Gamma^{\alpha}_{\mu\lambda}g_{\alpha\nu}+\delta\Gamma^{\alpha}_{\nu\lambda}g_{\mu\alpha} \, ,\\
    \nabla_{\nu}\delta g_{\lambda\mu}&=&\delta\Gamma^{\alpha}_{\lambda\nu}g_{\alpha\mu}+\delta\Gamma^{\alpha}_{\nu\mu}g_{\alpha\lambda}\label{3}\, ,\\ 
    \nabla_{\mu}\delta g_{\nu\lambda}&=&\delta\Gamma^{\alpha}_{\nu\mu}g_{\alpha\lambda}+\delta\Gamma^{\alpha}_{\mu\lambda}g_{\nu\alpha}\label{4} \, .
\end{eqnarray}
Luego, sumamos \eqref{2}, \eqref{3} y restamos \eqref{4}, obteniendo
\begin{eqnarray*}
    \nabla_{\lambda}\delta g_{\mu\nu}+\nabla_{\nu}\delta g_{\lambda\mu}-\nabla_{\mu}\delta g_{\nu\lambda}&=&\cancel{\delta\Gamma^{\alpha}_{\mu\lambda}g_{\alpha\nu}}+\delta\Gamma^{\alpha}_{\lambda\nu}g_{\mu\alpha}+\delta\Gamma^{\alpha}_{\lambda\nu}g_{\alpha\mu}+\cancel{\delta\Gamma^{\alpha}_{\nu\mu}g_{\alpha\lambda}}-\cancel{\delta\Gamma^{\alpha}_{\nu\mu}g_{\alpha\lambda}}-\cancel{\delta\Gamma^{\alpha}_{\mu\lambda}g_{\nu\alpha}}\\
&=&2\delta\Gamma^{\alpha}_{\lambda\nu}g_{\mu\alpha} \, .
\end{eqnarray*}
Multiplicando a ambos lados por $\frac{1}{2}g^{\mu\alpha}$ para despejar la variación del símbolo de Christoffel, nos queda
\begin{equation}
    \delta\Gamma^{\alpha}_{\lambda\nu}=\frac{1}{2}g^{\mu\alpha}\left(\nabla_{\lambda}\delta g_{\mu\nu}+\nabla_{\nu}\delta g_{\lambda\mu}-\nabla_{\mu}\delta g_{\nu\lambda}\right) \, . \label{varchrr}
\end{equation}
Usando el resultado anterior multiplicado por la delta generalizada expandida podemos encontrar que
\begin{equation}
g^{\mu\nu}\delta^{\rho\sigma}_{\lambda\nu}\delta\Gamma^{\lambda}_{\mu\sigma}=\nabla_{\rho}\delta g^{\rho\mu}-g^{\sigma\rho}\nabla^{\mu}\delta g_{\rho\sigma}\,
\end{equation}
finalmente reemplazando esto en la variación de la acción de Einstein-Hilbert \eqref{var3}
\begin{equation}
    \delta I[g_{\mu\nu}] =\kappa\int_{\mathcal{M}}d^{4}x \sqrt{\lvert g\rvert}G_{\mu\nu}\delta g^{\mu\nu}+\kappa\sigma\int_{\partial\mathcal{M}}d^{3}x \sqrt{\lvert h\rvert}\left(\nabla^{\rho}\delta g_{\rho\mu}-g^{\sigma\rho}\nabla_{\mu}\delta g_{\rho\sigma}\right)n^{\mu}\, . \label{varfin}
\end{equation}
Si estudiamos con detenimiento este resultado, nos encontramos con que esta acción no tiene un principio variacional bien definido cuando imponemos condiciones de borde de Dirichlet para la métrica. Esto se debe a que hay términos que dependen tanto como de la derivada de la variación de la métrica y términos que van como la derivada normal de la variación de la métrica. En la siguiente sección veremos como solucionar el problema del principio variacional de la acción de Einstein-Hilbert.

\section{Principio variacional}

Hemos visto que la variación de la acción de Einstein-Hilbert con respecto a la métrica nos entrega la Ec.~\eqref{varfin}. El primer término nos da las ecuaciones de Einstein $G_{\mu\nu}=0$, mientras que la segunda integral es un término de borde. El primer término del paréntesis depende de la variación de la métrica $\delta g_{\mu\nu}$, por lo que al  imponer las condiciones de borde tipo Dirichlet estándar i.e., $\delta g_{\mu\nu}\big|_{\partial\mathcal{M}}=0$ se anula. Sin embargo, el segundo término del paréntesis depende de las derivadas normales de la variación de la métrica, es decir, $n^{\mu}\nabla_{\mu}\delta g_{\rho\sigma}$, por lo que no se hace cero al imponer condiciones de borde tipo Dirichlet. Entonces, podemos concluir que estas condiciones de borde no conducen a un principio variacional bien definido, ya que la variación de la acción no se anula cuando la evaluamos en soluciones a sus ecuaciones de movimiento.
La forma de resolver problema consiste en agregar un término de borde adecuado para la acción. De esta manera, las ecuaciones de movimiento no cambian y para condiciones de borde tipo Dirichlet el término de borde puede ser elegido de forma tal que su variación cancele las derivadas normales que aparecen en la acción.

\subsection{Interludio: Geometría de hipersuperficies}
Si bien se conoce que la elección de un  término de borde no es único, ya que los términos de borde no cambian las ecuaciones de movimiento, Gibbons, Hawking y York \cite{Gibbons:1976ue,York:1972sj} se dieron cuenta que agregando un término que se construía a partir de la traza de la curvatura extrínseca se cancelaban exactamente las derivadas normales problemáticas que impedían imponer condiciones de borde tipo Dirichlet.
Antes de mostrar el término de borde de Gibbons-Hawking-York, es necesario introducir la geometría de hipersuperficies.

En una variedad \textit{D}-dimensional $\mathcal{M}$, se define una hipersuperficie $\Sigma$ como una subvariedad (\textit{D}-1)-dimensional que \textit{vive} en esta variedad ambiente $\mathcal{M}$. Una hipersuperficie es, por tanto, el conjunto de soluciones de una sola ecuación
\begin{equation}
    f(x^{1},...,x^{D})=0 \, ,
\end{equation}
en donde $x^{\mu}$, con $\mu=1,..,D$, son coordenadas locales de la variedad $\mathcal{M}$. Adicionalmente, podemos definir una hipersuperficie a través del embedding $x^{\mu}=x^{\mu}(y^{i})$ siendo $y^{i}$ las coordenadas de la hipersuperficie. Este embedding induce una métrica en $\Sigma$ dada por
\begin{align}
    h_{ij}=g_{\mu\nu}\frac{\partial x^{\mu}}{\partial x^{i}}\frac{\partial x^{\nu}}{\partial x^{j}} \equiv g_{\mu\nu} e^{\mu}_{i}e^{\nu}_{j}\, .
\end{align} 
Se puede definir un vector normal a $\Sigma$ como $u_{\mu}=\nabla_{\mu}f$. Además, el vector 
\begin{align}
    n^{\mu}=\frac{u^{\mu}}{\sqrt{ \|u_{\lambda}u^{\lambda}\|}} \, , \label{norm}
\end{align}
es unitario y cumple con
\begin{align}
    n_{\mu}n^{\mu}=g_{\mu\nu}\frac{u^{\mu}}{\sqrt{ \|u_{\lambda}u^{\lambda}\|}}\frac{u^{\nu}}{\sqrt{ \|u_{\rho}u^{\rho}\|}}=\sigma \, ,
\end{align}
con $\sigma=\pm 1$ cuando $n^{\mu}$ es tipo espacio o tipo tiempo, respectivamente. Como $n^{\mu}$ no tiene componentes tangenciales, se puede ver directamente que $n_{\mu}e^{\mu}_{i}=0$. Esto implica que $ h_{ij}=e^{\mu}_{i}e^{\nu}_{j}h_{\mu\nu}$ en donde 
\begin{align}
    h_{\mu\nu}=g_{\mu\nu}-\sigma n_{\mu}n_{\nu} \, , \label{pro}
\end{align}
se conoce como la primera forma fundamental. Cuando lo escribimos como $h^{\mu}_{\nu}=\delta^{\mu}_{\nu}-\sigma n^{\mu}n_{\nu}$, actúa como proyector de tensores que viven en la variedad $\mathcal{M}$ a la hipersuperficie $\Sigma$. De hecho, si consideramos dos vectores $W^{\mu}$ y $Z^{\nu}$ que son tangentes a la hipersuperficie $\Sigma$, el tensor proyector actúa como una métrica, es decir,
\begin{align} \nonumber
h_{\mu\nu}W^{\mu}Z^{\nu}=&g_{\mu\nu}W^{\mu}Z^{\nu}-\sigma n_{\mu}n_{\nu}W^{\mu}Z^{\nu}\\
    =&g_{\mu\nu}W^{\mu}Z^{\nu}\, .
\end{align}

Es posible construir una derivada covariante en la hipersuperficie $\Sigma$. Para ello, definimos
\begin{align}
D_{\sigma}T^{\mu_{1}...\mu_{p}}\,_{\nu_{1}...\nu_{q}}=h^{\mu_{1}}_{\lambda_{1}}...h^{\mu_{p}}_{\lambda_{p}}h^{\rho_{1}}_{\nu_{1}}...h^{\rho_{q}}_{\nu_{q}}h^{\tau}_{\sigma}\nabla_{\tau}T^{\lambda_{1}...\lambda_{p}}\,_{\rho_{1}...\rho_{q}}\, .
\end{align}
Se puede demostrar que esta derivada es compatible con $h_{\mu\nu}$ de la siguiente forma
\begin{align}\notag
D_{\mu}h_{\lambda\rho}&=h^{\alpha}_{\lambda}h^{\beta}_{\rho}h^{\gamma}_{\mu}\nabla_{\gamma}h_{\alpha\beta}\\ \nonumber
&=h^{\alpha}_{\lambda}h^{\beta}_{\rho}h^{\gamma}_{\mu}\nabla_{\gamma}(g_{\alpha\beta}-\sigma n_{\alpha}n_{\beta})\\ \nonumber
&=h^{\alpha}_{\lambda}h^{\beta}_{\rho}h^{\gamma}_{\mu}\cancelto{0}{\nabla_{\gamma}g_{\alpha\beta}}-\sigma h^{\alpha}_{\lambda}\cancelto{0}{h^{\beta}_{\rho}n_{\beta}}h^{\gamma}_{\mu}\nabla_{\gamma}n_{\alpha}-\sigma \cancelto{0}{h^{\alpha}_{\lambda}n_{\alpha}}h^{\beta}_{\rho}h^{\gamma}_{\mu}\nabla_{\gamma}n_{\beta}\\ 
    &=0 \, ,
\end{align}
en donde hemos utilizado la condición de metricidad en el primer término y que el producto punto entre el vector normal y el proyector es cero en la igualdad siguiente.

Ahora, consideremos un vector tangente a $\Sigma$, digamos $v^{\mu}$, definido tal que cumple con las siguientes propiedades
\begin{align}\label{tan}
v^{\mu}&=h^{\mu}_{\nu}v^{\nu} \, , \\ 
n^{\mu}v_{\mu}&=0 \label{nor} \, .
\end{align}
Tomemos la derivada covariante $\nabla$ sobre este vector
\begin{align} \notag h^{\lambda}_{\mu}\nabla_{\lambda}v^{\nu}=&h^{\lambda}_{\mu}\delta^{\nu}_{\rho}\nabla_{\lambda}v^{\rho}\\ \notag
    =&h^{\lambda}_{\mu}(h^{\nu}_{\rho}+\sigma n^{\nu}n_{\rho})\nabla_{\lambda}v^{\rho}\\
    =&D_{\mu}v^{\nu}+\sigma h^{\lambda}_{\mu}n^{\nu}n_{\rho}\nabla_{\lambda}v^{\rho} \, . \label{dn}
\end{align}
De esta expresión vemos que aparecen componentes normales. Ahora, usando la condición~\eqref{nor}, encontramos que
\begin{align}
-v^{\rho}\nabla_{\lambda}n_{\rho}&=\nabla_{\lambda}v^{\rho}n_{\rho} \, .
\end{align}
Luego, reemplazando esta identidad en la Ec.~\eqref{dn}, obtenemos  
\begin{align} \notag
h^{\lambda}_{\mu}\nabla_{\lambda}v^{\nu}&=D_{\mu}v^{\nu}-\sigma h^{\lambda}_{\mu}n^{\nu}v^{\rho}\nabla_{\lambda}n_{\rho}\\ \notag
&=D_{\mu}v^{\nu}-\sigma h^{\lambda}_{\mu}h^{\rho}_{\sigma}\nabla_{\lambda}n_{\rho}v^{\sigma}n^{\nu}\\
&=D_{\mu}v^{\nu}-\sigma K_{\mu\sigma}v^{\sigma}n^{\nu},
\end{align}
en donde
\begin{equation}
K_{\mu\sigma}=h^{\lambda}_{\mu}h^{\rho}_{\sigma}\nabla_{\lambda}n_{\rho} \, ,
\end{equation}
es la curvatura extrínseca o también llamada segunda forma fundamental. Podemos interpretar este tensor como el cambio del proyector a lo largo de un campo vectorial normal y nos da información geométrica de cómo una hipersupeficie $\Sigma$ se encaja en la variedad ambiente $\mathcal{M}$.

Existen 2 relaciones que pueden ser útiles para trabajar con la curvatura extrínseca  y con el tensor de Riemann de la variedad $\mathcal{M}$. La primera es la ecuación de Gauss-Codazzi la cual corresponde a 
\begin{align}
\mathcal{R}^{\rho}\,_{\sigma\mu\nu}=h^{\rho}_{\alpha}h^{\beta}_{\rho}h^{\gamma}_{\mu}h^{\delta}_{\nu}R^{\alpha}\,_{\beta\gamma\delta}+\sigma(K^{\rho}_{\mu}K_{\sigma\nu}-K^{\rho}_{\nu}K_{\sigma\mu})\, ,
\end{align}
donde $\mathcal{R}^{\rho}\,_{\sigma\mu\nu}$ es la curvatura intrínseca de la hipersuperficie $\Sigma$ y es el tensor de curvatura asociado a $h_{\mu\nu}$. Al tomar la traza de esta ecuación, obtenemos el escalar de curvatura de la hipersuperficie. La segunda relación corresponde a la de Codazzi-Mainardi y corresponde a la derivada covariante de la hipersupeficie aplicada a la curvatura extrínseca 
\begin{align}
    D_{[\mu}K_{\nu]}^{\mu}=\frac{1}{2}h^{\alpha}_{\nu}R_{\rho\sigma}n_{\rho}\, .
\end{align}

\subsection{Término de borde de Gibbons-Hawking-York}

Ahora tenemos todas las herramientas para introducir el término de borde de Gibbons-Hawking-York, el cual se define como
\begin{equation}
    I_{GHY}=2\kappa\int_{\partial\mathcal{M}} d^{3}x \sqrt{\lvert h\rvert}\, K\, , \label{GHY}
\end{equation}
en donde $g^{\mu\nu}K_{\mu\nu}=K$ es la traza de la curvatura extrínseca. También podemos escribir de forma equivalente 
\begin{eqnarray}
K=g^{\mu\nu}K_{\mu\nu}=g^{\mu\nu}\nabla_{\mu}n_{\nu} \label{K} \, .
\end{eqnarray}
Variando la Ec.~\eqref{GHY} 
\begin{align} \notag
\delta I_{GHY}&=2\kappa\delta\int_{\partial\mathcal{M}} d^{3}x \sqrt{\lvert h\rvert}\, K\\ \notag
&=2\kappa\int_{\partial\mathcal{M}} d^{3}x\left(\delta\sqrt{\lvert h\rvert}\, K + \sqrt{\lvert h\rvert}\delta K\right)\\ 
&=2\kappa\int_{\partial\mathcal{M}} d^{3}x\left[\left(-\frac{1}{2}\sqrt{\lvert h \rvert}h_{\mu\nu}\delta h^{\mu\nu}\right)K+\sqrt{\lvert h\rvert}\delta K\right] \, .
\end{align}
Vamos a trabajar el último término con más detalle. Para ello, consideremos que 
\begin{align} \label{vark}
    2\kappa\int_{\partial\mathcal{M}} d^{3}x\sqrt{\lvert h\rvert}\delta K&= 2\kappa\int_{\partial\mathcal{M}} d^{3}x\sqrt{\lvert h\rvert}\delta (g^{\mu\nu}K_{\mu\nu})\, ,
\end{align}
en donde  
\begin{align}\notag
g^{\mu\nu}K_{\mu\nu}&=g^{\mu\nu}h^{\lambda}_{\mu}\nabla_{\lambda}n_{\nu}\\ \notag
&=h^{\lambda}_{\mu}\nabla_{\lambda}(g^{\mu\nu}n_{\nu})\\ \notag
&=h^{\lambda}_{\mu}\nabla_{\lambda}n^{\mu} \\ \notag
&=(\delta^{\lambda}_{\mu}-\sigma n^{\lambda}n_{\mu})\nabla_{\lambda}n^{\mu}\\ \notag
&=\delta^{\lambda}_{\mu}\nabla_{\lambda}n^{\mu}-\cancelto{0}{\sigma n^{\lambda}n_{\mu}\nabla_{\lambda}n^{\mu}}\\
&=\nabla_{\mu}n^{\mu}\, .
\end{align}
Entonces, la Ec.~\eqref{vark} quedará como
\begin{align}
    2\kappa\int_{\partial\mathcal{M}} d^{3}x\sqrt{\lvert h\rvert}\delta K&= 2\kappa\int_{\partial\mathcal{M}} d^{3}x\sqrt{\lvert h\rvert}\delta (\nabla_{\mu}n^{\mu})\, .
\end{align}
Vamos a trabajar por separado $\delta(\nabla_{\mu}n^{\mu})$
\begin{align} \notag
    \delta (\nabla_{\mu}n^{\mu})&=\delta (\partial_{\mu}n^{\mu}+\Gamma^{\mu}_{\lambda\mu}n^{\lambda})\\ \notag
&=\partial_{\mu}\delta n^{\mu}+\delta\Gamma^{\mu}_{\lambda\mu}n^{\lambda}+\Gamma^{\mu}_{\lambda\mu}\delta n^{\lambda}\\
&=\nabla_{\mu}\delta n^{\mu}+\delta\Gamma^{\mu}_{\lambda\mu}n^{\lambda} \, .\label{varnab}
\end{align}
Encontremos $\delta n^{\mu}$ en términos de $\delta g^{\mu\nu}$. Para ello, usando que el vector normal es unitario, $n^{\mu}n_{\mu}=\sigma$, obtenemos el siguiente resultado luego de variar a ambos lados
\begin{align} 
\delta n^{\mu}n_{\mu}&=-n^{\mu}\delta n_{\mu}\, . \label{normm}
\end{align}
Así, reescribiremos la variación del vector normal como
\begin{align}\label{deltann}
\delta n_{\mu}=\delta(g_{\mu\nu}n^{\nu})=\delta g_{\mu\nu}n^{\nu}+g_{\mu\nu}\delta n^{\nu}\, .
\end{align}
Para la variación de la métrica, usaremos la definición de la Ec.~\eqref{pro}, lo cual conduce a
\begin{align}
\delta g_{\mu\nu}&=\delta h_{\mu\nu}+2\sigma n_{(\mu}\delta n_{\nu)} \, .\label{vargn}
\end{align}
Reemplazando en la Ec.~\eqref{deltann} y considerando que el vector normal es ortogonal a $\delta h_{\mu\nu}$ nos queda
\begin{align} \notag
    \delta n_{\mu}&=\sigma(\delta n_{\mu}n_{\nu}+n_{\mu}\delta n_{\nu})n^{\nu}+g_{\mu\nu}\delta n^{\nu}\\ \notag
    \cancel{\delta n_{\mu}}&=\cancel{\delta n_{\mu}}+\sigma n_{\mu}n^{\nu}\delta n_{\nu}+g_{\mu\nu}\delta n^{\nu}\\ \notag
    -\sigma n_{\mu}n^{\nu}\delta n_{\nu}&=g_{\mu\nu}\delta n^{\nu}\,\, \,\,\,\,\,\, /g^{\mu\lambda}\\
    \delta n^{\lambda}&=-\sigma n^{\lambda}n^{\nu}\delta n_{\nu}\, . \label{varnorm2}
\end{align}
Luego, usando la Ec.~\eqref{normm} y renombrando índices encontramos
\begin{align}\label{deltanormal}
    \delta n^{\mu}=\sigma n^{\mu}n_{\lambda}\delta n^{\lambda}\, .
\end{align} 
Es decir, $\delta n^{
\mu}$ apunta en la dirección normal. Adicionalmente, podemos utilizar que $\delta(g^{\mu\nu}n_{\mu}n_{\nu})=0$ lo cual, utilizando la Ec.~\eqref{normm}, conduce a
\begin{align}
    n_{\lambda}\delta n^{\lambda}=\frac{1}{2}n_{\mu}n_{\nu}\delta g^{\mu\nu}\, .
\end{align}
Así, reemplazando en la Ec.~\eqref{deltanormal}, llegamos a la expresión para la variación del vector normal en términos de de la variación de la métrica, esto es, 
\begin{align}
    \delta n^{\mu}=\frac{1}{2}\sigma n^{\mu}n_{\alpha}n_{\beta}\delta g^{\alpha\beta}\, .
\end{align}\label{varnorm}
Ahora, remplazando $\sigma n^{\mu}n_{\alpha}=\delta^{\mu}_{\alpha}-h^{\mu}_{\alpha}$ en el resultado anterior, encontramos
\begin{align}\notag
\delta n^{\mu}&=\frac{1}{2}(\delta^{\mu}_{\alpha}-h^{\mu}_{\alpha})n_{\beta}\delta g^{\alpha\beta}\\ \notag
&=\frac{1}{2}\left(\delta^{\mu}_{\alpha}n_{\beta}\delta g^{\alpha\beta}-h^{\mu}_{\alpha}n_{\beta}\delta g^{\alpha\beta}\right)\\
&=\frac{1}{2}\left(n_{\lambda}\delta g^{\mu\lambda}-h^{\mu}_{\alpha}n_{\beta}\delta g^{\alpha\beta}\right)\, ,
\end{align}
en donde hemos renombrado los índices en el primer término de la última igualdad. Así, la variación del vector normal es
\begin{align}
    \delta n^{\mu}=\frac{1}{2}n_{\lambda}\delta g^{\mu\lambda}+ v^{\mu} \, ,
\end{align}
con $v^{\mu}=h^{\mu}_{\alpha}n_{\beta}\delta g^{\alpha\beta}$ siendo un vector puramente tangencial, i.e., $n_{\mu}v^{\mu}=0$.

Para el segundo término de \eqref{varnab} vamos a usar la ecuación que encontramos para la variación de la conexión~\eqref{varchrr} y queda de la forma
\begin{align}
\delta\Gamma^{\mu}_{\lambda\mu}=-\frac{1}{2}\nabla_{\lambda}(g_{\mu\alpha}\delta g^{\mu\alpha}) \, ,  
\end{align}
reemplazando estos resultados en \eqref{varnab}
\begin{align}\notag
\delta (\nabla_{\mu}n^{\mu})&=\nabla_{\mu}\left(\frac{1}{2}n_{\lambda}\delta g^{\mu\lambda}+ v^{\mu}\right)-\frac{1}{2}\nabla_{\lambda}(g_{\mu\alpha}\delta g^{\mu\alpha})n^{\lambda}\\
&=\frac{1}{2}\nabla_{\mu}n_{\lambda}\delta g^{\mu\lambda}+\frac{1}{2}n_{\lambda}\nabla_{\mu}\delta g^{\mu\lambda}+\nabla_{\mu}v^{\mu}-\frac{1}{2}\nabla_{\lambda}(g_{\mu\alpha}\delta g^{\mu\alpha})n^{\lambda}\, .
\end{align}
El tercer término es una derivada covariante del borde. Para demostrarlo, vamos a calcular $D_{\mu}v^{\mu}$ usando la definición en la Ec.~\eqref{varnorm2}, i.e.,
\begin{align}\notag
D_{\mu}v^{\mu}&=h^{\mu}_{\nu}\nabla_{\mu}v^{\nu}\\ \notag
&=(\delta^{\mu}_{\nu}-\sigma n^{\mu}n_{\nu})\nabla_{\mu}v^{\nu}\\
&=\nabla_{\nu}n^{\nu}+\sigma n^{\mu}v^{\nu}\nabla_{\mu}n_{\nu}\, ,
\end{align}
en donde en la última igualdad hemos usado que $v^{\nu}n_{\nu}=0$. Ahora, trabajemos el segundo término usando la definición de $v^{\mu}$
\begin{align}\notag
\sigma n^{\mu}v^{\nu}\nabla_{\mu}n_{\nu}&=\sigma n^{\mu}h^{\nu}_{\alpha}n_{\beta}\delta g^{\alpha\beta}\nabla_{\mu}n_{\nu}\\ \notag
&=\sigma n^{\mu}h^{\nu}_{\alpha}n_{\beta}(2\sigma n^{(\alpha}\delta n^{\beta)})\nabla_{\mu}n_{\nu}\\ \notag
&=n^{\mu}h^{\nu}_{\alpha}\nabla_{\mu}n_{\nu}n_{\beta}n^{\beta}\delta n^{\alpha}\\
&=\sigma n^{\mu}h^{\nu}_{\alpha}\nabla_{\mu}n_{\nu}\left(\frac{\sigma}{2}n^{\alpha}n_{\rho}n_{\lambda}\delta g^{\rho\lambda}\right)=0 \, .\label{zero}
\end{align}
Usando las Ecs.~\eqref{varnorm} y~\eqref{vargn} llegamos a que la expresión anterior se anula, ya que nos queda la contracción del vector normal con el proyector, por lo que se cumple lo siguiente
\begin{align}
    D_{\mu}v^{\mu}=\nabla_{\mu}v^{\mu}\, .
\end{align}
Reemplazando en la variación de la traza de la curvatura extrínseca, nos queda lo siguiente
\begin{align}
\delta (\nabla_{\mu}n^{\mu})&=\frac{1}{2}\nabla_{\mu}n_{\lambda}\delta g^{\mu\lambda}+\frac{1}{2}n_{\lambda}\nabla_{\mu}\delta g^{\mu\lambda}+D_{\mu}v^{\mu}-\frac{1}{2}\nabla_{\lambda}(g_{\mu\alpha}\delta g^{\mu\alpha})n^{\lambda}\, .
\end{align}
Luego, es conveniente sumar un cero, para ello usaremos un término como el de la Ec.~\eqref{zero} para escribir 
\begin{align}\notag
 \delta (\nabla_{\mu}n^{\mu})&=\frac{1}{2}\nabla_{\mu}n_{\lambda}\delta g^{\mu\lambda}+\frac{1}{2}n_{\lambda}\nabla_{\mu}\delta g^{\mu\lambda}+D_{\mu}v^{\mu}-\frac{1}{2}\nabla_{\lambda}(g_{\mu\alpha}\delta g^{\mu\alpha})n^{\lambda}-\frac{\sigma}{2}n_{\mu}n^{\nu}\nabla_{\nu}n_{\lambda}\delta g^{\lambda\mu}  \, \\ \notag
&=\frac{1}{2}\left(\nabla_{\mu}n_{\lambda}-\sigma n_{\mu}n^{\nu}\nabla_{\nu}n_{\lambda}\right)\delta g^{\lambda\mu}+D_{\mu}v^{\mu}-\frac{1}{2}\nabla_{\lambda}(g_{\mu\alpha}\delta g^{\mu\alpha})n^{\lambda}+\frac{1}{2}n_{\lambda}\nabla_{\mu}\delta g^{\mu\lambda}\\ \notag
&=\frac{1}{2}(\delta^{\nu}_{\mu}-\sigma n^{\nu}n_{\mu})\nabla_{\nu}n_{\lambda}\delta g^{\lambda\mu}+D_{\mu}v^{\mu}-\frac{1}{2}\nabla_{\lambda}(g_{\mu\alpha}\delta g^{\mu\alpha})n^{\lambda}+\frac{1}{2}n_{\lambda}\nabla_{\mu}\delta g^{\mu\lambda}\\
&=\frac{1}{2}K_{\lambda\mu}\delta g^{\lambda\mu}+D_{\mu}v^{\mu}+\frac{1}{2}n_{\lambda}\left(\nabla_{\mu}\delta g^{\mu\lambda}-g_{\mu\alpha}\nabla^{\lambda}\delta g^{\mu\alpha}\right)\, .
\end{align}
Finalmente, sumando todas las contribuciones la variación del término de Gibbons-Hawking-York queda
\begin{align}
    2\kappa\int_{\partial\mathcal{M}} d^{3}x\,\delta(\sqrt{\lvert h\rvert} K)&=\kappa\int_{\partial\mathcal{M}} d^{3}x\sqrt{\lvert h\rvert}\delta g^{\mu\nu}\left(-h_{\mu\nu}K+K_{\mu\nu}\right)\\
    &-\kappa\int_{\partial\mathcal{M}} d^{3}x\sqrt{\lvert h\rvert} \left(\nabla_{\mu}\delta g^{\mu\lambda}-g_{\mu\alpha}\nabla^{\lambda}\delta g^{\mu\alpha}\right)n_{\lambda}\, ,
\end{align}
en donde el término $D_{\mu}v^{\mu}$ ha sido omitido dado que $\partial{\mathcal{M}}$ es una hipersuperficie cerrada y, por lo tanto, no tiene borde. 

Notamos que esta variación reproduce exactamente el término de borde de la acción de Einstein-Hilbert, por lo que si ahora consideramos la acción 
\begin{align}\label{ehghy}
    I_{\rm EH}+I_{\rm GHY}=\kappa \int_{\mathcal{M}} d^{4}x \sqrt{\lvert g\rvert}\, R +  2\kappa\sigma\int_{\partial\mathcal{M}} d^{3}x \sqrt{\lvert h\rvert}\, K \, , 
\end{align}
tendremos un principio variacional bien definido para condiciones de frontera tipo Dirichlet, dado que su variación conduce a 
\begin{align}
    \delta I_{\rm EH+GHY}=\kappa\int_{\mathcal{M}}d^{4}x \sqrt{\lvert g\rvert}G_{\mu\nu}\delta g^{\mu\nu}+ \kappa\sigma\int_{\partial\mathcal{M}} d^{3}x\sqrt{\lvert h\rvert}(K_{\mu\nu}-h_{\mu\nu}K)\delta g^{\mu\nu}\, , \label{var4}
\end{align}
la cual se hace cero cuando $\delta g_{\mu\nu}|_{\partial{\mathcal{M}}}=0$. Se puede observar que aún existe la libertad de agregar términos de borde construidos con cantidades intrínsecas sin modificar las ecuaciones de movimiento ni las condiciones de borde tipo Dirichlet. Tomando en cuenta lo anterior y considerando que la acción debe anularse cuando se evalúa en el vacío de la teoría, agregaremos un término no dinámico a la acción del borde. Dicho lo anterior, la acción que satisface los requerimientos anteriormente mencionados es
\begin{align}
 I_{\rm Dir}=I_{\rm EH}+I_{\rm GHY}+I_{0}=\kappa \int_{\mathcal{M}} d^{4}x \sqrt{\lvert g\rvert}\, R + 2 \kappa\sigma\int_{\partial\mathcal{M}} d^{3}x \sqrt{\lvert h\rvert}\, (K-K_{0})\, , \label{acciondir}
\end{align}
en donde $K_{0}$ es la curvatura extrínseca de dicho vacío y, al ser no dinámico, $\delta K_{0}=0$. Si variamos la acción anterior, obtenemos 
\begin{align}
\delta I_{\rm Dir}=\kappa\int_{\mathcal{M}}d^{4}x \sqrt{\lvert g\rvert}G_{\mu\nu}\delta g^{\mu\nu}+\kappa\sigma\int_{\partial\mathcal{M}} d^{3}x\sqrt{\lvert h\rvert}[K_{\mu\nu}-h_{\mu\nu}(K-K_{0})]\delta h^{\mu\nu}\, .  
\end{align}
Este resultado será útil para obtener cargas conservadas. Como veremos en las siguientes secciones, existen diferentes maneras para obtener estas cantidades conservadas. En la Sec.~\ref{sec:BY}, revisaremos el formalismo de Brown-York, el cual esta basado en la conservación del tensor de energía-momentum cuasilocal que describiremos a continuación. Luego, en la Sec.~\ref{sec:NW}, veremos la aplicación del teorema de Noether en gravitación para soluciones de las ecuaciones de movimiento que poseen algún vector de Killing. Adicionalmente, en la Sec.~\ref{sec:AEO}, utilizaremos la aproximación de punto silla para relacionar la acción Euclídea on-shell con la función partición en el ensamble canónico para así poder encontrar las cantidades termodinámicas asociadas a soluciones estacionarias de las ecuaciones de movimiento.

\section{Tensor de energía-momentum cuasilocal \label{sec:BY}} 
Cuando una solución a las ecuaciones de Einstein posee una simetría dada por un vector de Killing $\xi=\xi^{\mu}\partial_{\mu}$, i.e., $\Lie_{\xi}g_{\mu\nu}=0$, podemos calcular cantidades conservadas asociadas a dicha simetría. Inspirados en el formalismo de Hamilton-Jacobi, Brown y York~\cite{Brown:1992br} encontraron una forma de definir un tensor de energía-momentum cuasilocal para calcular las cargas conservadas. Para ello, consideremos la variación dada en la Ec.~\eqref{var4} lo cual entrega
\begin{equation}
\delta I=\kappa\int_{\mathcal{M}}d^{4}x \sqrt{\lvert g\rvert} G_{\mu\nu}\delta g^{\mu\nu}-\frac{1}{2}\int_{\partial\mathcal{M}}d^{3}x \sqrt{\lvert h\rvert}\tau_{\mu\nu}\delta h^{\mu\nu} \, ,
\end{equation}
en donde hemos definido el tensor de energía-momentum cuasilocal como
\begin{eqnarray}
\tau_{\mu\nu}=-2\kappa\sigma\left[K_{\mu\nu}-h_{\mu\nu}(K-K_{0})\right] \, . \label{tau}
\end{eqnarray}
Este tensor es covariantemente conservado on-shell, ya que
\begin{align}
    D^{\nu}[K_{\mu\nu}-h_{\mu\nu}(K-K_{0})]=-R_{\rho\kappa}n^{\rho}h^{\kappa}_{\mu}\, ,
\end{align}
en donde, si se satisfacen las ecuaciones de campo, tenemos que $R_{\mu\nu}=0$. Esta ley de conservación se conoce como el constraint de momentum en el formalismo Hamiltoniano de la Relatividad General~\cite{Brown:1992br}. Utilizando este resultado,  podemos construir una corriente conservada contrayendo el tensor de energía-momentum cuasilocal con el vector de Killing $\xi$, esto es, 
\begin{align}
    J^{\mu}=\tau^{\mu\nu}\xi_{\nu} \;\;\; \to \;\;\; D_{\mu}J^{\mu}=0\, .
\end{align}
Integrando sobre una hipersuperficie de Cauchy $\Sigma$ y utilizando el teorema de Stokes, Brown y York formularon su definición de cargas conservadas como
\begin{equation}
  Q[\xi]=\int_{\partial\Sigma} d^{2}x \sqrt{\lvert \gamma \rvert}\, \tau_{\mu\nu}\xi^{\nu}u^{\mu} \, , \label{qby}
\end{equation}
en donde $\gamma=\det\gamma_{\mu\nu}$ es el determinante de la métrica inducida sobre la hipersuperficie de codimension-2, $\partial\Sigma$, y $u^{\mu}$ es su vector normal unitario tipo-tiempo. De esta forma, podemos obtener las cargas conservadas de una solución asintóticamente plana asociadas a un vector de Killing $\xi$.

Con el objetivo de ilustrar la utilidad de la resta de la curvatura extrínseca del background, consideremos primero el tensor de energía-momentum cuasilocal obtenido a partir de la Ec.~\eqref{var4}, donde no hay substracción de background. En particular, analicemos una solución a las ecuaciones de campo de Einstein con asíntota plana, e.g. la métrica de Schwarzschild en 4 dimensiones,
\begin{align}
    \diff s^2=-f(r)\diff t^2+\frac{\diff r^2}{f(r)}+r^2\left(\diff \theta^2+\sin^2\theta\diff \phi^2\right)\,\,\,\,\,\,\mbox{con}\,\,\,\,\,\, f(r)=1-\frac{2mG}{r}\, ,
\end{align} \label{schw}
donde $m$ es una constante de integración, la cual demostraremos mas adelante que está asociada con la masa del agujero negro. Notemos que esta solución tiene un vector de Killing temporal $\xi=\partial_{t}$. De este modo, en coordenadas de Schwarzschild, la hipersuperficie de Cauchy $\Sigma$ se define a $t=$ constante, lo cual induce un vector normal unitario tipo tiempo $u=f^{1/2}(r)\partial_{t}$, tal que $u^{\mu}u_{\mu}=-1$. Luego, usando la definición de carga conservada de la Ec.~\eqref{qby}, tenemos
\begin{align}\notag
Q[\xi]&=\int^{2\pi}_{0}\diff\phi\int^{\pi}_{0}\sin\theta\diff\theta\; r^2\tau_{tt}\xi^{t}u^{t} \\ \notag
&=-2\kappa\int^{2\pi}_{0}\diff\phi\int^{\pi}_{0}\sin\theta\diff\theta \; r^2\left[\frac{mG}{r^2}\sqrt{f(r)}+\frac{f(r)[f'(r)r+4f(r)]}{2r\sqrt{f(r)}}\right]\frac{1}{\sqrt{f(r)}}\\
&=2\kappa\int^{2\pi}_{0}\diff\phi\int^{\pi}_{0}\sin\theta\diff\theta \;\left[mG+\frac{f'(r)r^2}{2}+2f(r)r\right] \, .
\end{align}
Remplazando el valor de $f(r)$, su derivada e integrando obtenemos
\begin{align}
    Q[\xi]=-\frac{1}{2G}(2r-4mG)=2m-\frac{2r}{G} \, .
\end{align}
De la expresión anterior se puede ver que, en el límite $r\to\infty$, la carga es divergente, razón por la cual es necesaria la substracción de background. Para ello, consideraremos que el espacio-tiempo de referencia es Minkowski, ya que la métrica de Schwarzschild está continuamente conectada a ella en el límite $m\to 0$. En coordenadas esféricas, consideraremos una foliación radial de la métrica de Minkowski, lo cual nos entrega una curvatura extrínseca de $K_{0}=2/r$. Así, la integral de la carga quedará como
\begin{align}\notag
Q[\xi]&=-2\kappa\int^{2\pi}_{0}\diff\phi\int^{\pi}_{0}\sin\theta\diff\theta \; r^2\left[\frac{mG}{r^2}\sqrt{f(r)}+\frac{f(r)[f('r)r+4f(r)]}{2\sqrt{f(r)}r}-\frac{2}{r}\right]\frac{1}{\sqrt{f(r)}}  \\ \notag
&=-2\kappa\int^{2\pi}_{0}\diff\phi\int^{\pi}_{0}\sin\theta\diff\theta \;\left[\cancel{mG}-\cancel{mG}+2\left(1-\frac{2mG}{r}\right)r-2r\left(1-\frac{2mG}{r}\right)^{\frac{1}{2}}\right]\\
&=-2\kappa\int^{2\pi}_{0}\diff\phi\int^{\pi}_{0}\sin\theta\diff\theta \;\left[2r-4mG-2r\left(1-\frac{2mG}{r}\right)^{\frac{1}{2}}\right]\, .
\end{align}
Integrando y expandiendo en series cuando $r\to\infty$, encontramos que
\begin{align} \notag
    Q[\xi]&=-8\pi\kappa\left[2r-4mG-2r\left\{1-\frac{1}{2}\left(\frac{2mG}{r}\right)-\mathcal{O}(r^{-2})\right\}\right]\\ \notag
    &=-8\pi\kappa[\cancel{2r}-4mG-\cancel{2r}+2mG+\mathcal{O}(r^{-1})]\\ \notag
    &=-8\pi\kappa[-2mG+\mathcal{O}(r^{-1})]\\
    &=-\frac{1}{2G}[-2mG+\mathcal{O}(r^{-1})]\, .
\end{align}
Luego, tomando el límite de $r\to\infty$ nos queda 
\begin{align}
    Q[\xi]=m\, ,
\end{align}
lo que nos da el valor correcto para la carga asociada a la masa del agujero negro. A este método para renormalizar el valor de la carga se le conoce como substracción de background y, tal como vimos, consiste en restar el término de la curvatura extrínseca del vacío de la teoría. Para espacios que son asintóticamente planos, esta prescripción entrega valores finitos. Sin embargo, existen familias de soluciones que no están continuamente conectadas con el vacío. Para soluciones con distinta asíntota, este método falla. Veremos en detalle esta situación mas adelante. 

\section{Formalismo de Noether-Wald \label{sec:NW}}

Uno de los resultados más notables de la teoría de la Relatividad General clásica fue el descubrimiento de análogos matemáticos entre las leyes de la termodinámica que conocemos y las leyes de la mecánica de los agujeros negros. Un ejemplo particular fue el de Robert Wald en 1993, donde encontró una relación entre la entropía  de un agujero negro y su carga de Noether; explícitamente, el producto de la temperatura y la entropía del agujero negro es la carga de Noether asociada al vector de Killing tipo tiempo evaluada en el horizonte de bifurcación~\cite{Wald:1993nt}. Siguiendo el tratamiento de Wald y considerando teorías invariantes bajo difeomorfismos, en esta sección usaremos el teorema de Noether para encontrar cantidades conservadas para la solución asintóticamente plana dada por la Ec.~\eqref{schw}.

Consideremos una teoría gravitacional construida a partir de la métrica y sus derivadas, que es invariante de difeomorfismos, y que puede ser descrita por el principio de acción 
\begin{align}
I\left[g_{\mu\nu}\right]=\int d^D x \sqrt{|g|}\, \Lag\left(R_{\lambda \rho}^{\mu \nu}\right)\label{accgeneral} \, ,
\end{align}
donde la dependencia de la métrica está codificada en $R_{\lambda \rho}^{\mu \nu}=g^{\gamma\nu}R^{\mu}_{\ \gamma\lambda\rho}$. Variaciones arbitrarias de la acción entregan
\begin{align}\notag
\delta I &= \int d^D x\left(\delta\sqrt{|g|}\Lag+\sqrt{|g|}\delta\Lag\right)\\ \notag
&=\int d^D x \sqrt{|g|}\left[-\frac{1}{2} \delta g^{ \mu\nu }g_{\mu \nu} \Lag+\frac{\partial \Lag}{\partial R_{\lambda \rho}^{\mu \nu}} \delta R_{\lambda \rho}^{\mu \nu}\right]\\ 
&\equiv\int d^D x \sqrt{|g|}\left[-\frac{1}{2} \delta g^{ \mu\nu }g_{\mu \nu} \Lag+E_{\mu \nu}^{\lambda \rho} \delta R_{\lambda \rho}^{\mu \nu}\right]\, . \label{vargen}
\end{align} 
Luego, utilizando $\delta R_{\lambda \rho}^{\mu \nu}=\delta g^{\gamma\nu}R^{\mu}_{\ \gamma\lambda\rho}+g^{\gamma\nu}\delta R^{\mu}_{\ \gamma\lambda\rho}$ y remplazando en la ecuación anterior obtenemos
\begin{align}\notag
 \delta I&=\int d^D x \sqrt{|g|}\left[-\frac{1}{2} \delta g^{ \mu\nu }g_{\mu \nu} \Lag+E_{\mu \nu}^{\lambda \rho} (\delta g^{\nu \sigma} R^\mu{}_{\sigma \lambda \rho}+2 g^{\nu\sigma} \nabla_{[\lambda} \delta \Gamma_{\sigma |\rho]}^\mu)\right]\\
 &=\int d^D x \sqrt{|g|}\left[-\frac{1}{2} \delta g^{ \mu\nu }g_{\mu \nu} \Lag+E_{\mu \nu}^{\lambda \rho} (\delta g^{\nu \sigma} R^\mu{}_{\sigma \lambda \rho}+2 g^{\nu\sigma} \nabla_{\lambda} \delta \Gamma_{\sigma\rho}^\mu)\right]\, ,
\end{align}
en donde, en el último término hemos eliminado la antisimetría de los índices considerando que el tensor $E_{\mu \nu}^{\lambda \rho}$ tiene las mismas simetrías que el Riemann. Luego, integrando por partes, el tercer factor quedaría
\begin{align}\notag
-2g^{\nu\sigma}E_{\mu \nu}^{\lambda \rho}\nabla_{\lambda} \delta \Gamma_{\sigma\rho}^\mu&=2\nabla_{\lambda}E_{\mu \nu}^{\lambda \rho}g^{\nu\sigma}\delta\Gamma^{\mu}_{\sigma\rho}-\nabla_{\lambda}(2g^{\nu\sigma}E_{\mu \nu}^{\lambda \rho}\delta\Gamma^{\mu}_{\sigma\rho})\, .
\end{align}
Así, sustituyendo lo anterior en la variación de $I$, encontramos 
\begin{align}\notag
    \delta I&=\int d^D x \sqrt{|g|}\left[-\frac{1}{2} \delta g^{\mu\nu}g_{\mu\nu} \Lag+E_{\mu\nu}^{\lambda\rho} \delta g^{\nu\sigma} R^{\mu}_{\sigma\lambda\rho}-2\nabla_{\lambda}E_{\mu\nu}^{\lambda\rho}g^{\nu\sigma}\delta\Gamma^{\mu}_{\sigma\rho}\right.\\
&-\left.\nabla_{\lambda}(2g^{\nu\sigma}E_{\mu \nu}^{\lambda \rho}\delta\Gamma^{\mu}_{\sigma\rho})\right]\, .\label{vargen2}
\end{align}
Podemos expandir el tercer término usando la Ec~\eqref{varchrr}, quedando lo siguiente
\begin{align}\notag
-2\nabla_{\lambda}E_{\mu\nu}^{\lambda\rho}g^{\nu\sigma}\delta\Gamma^{\mu}_{\sigma\rho}&=-\cancel{2}\nabla_{\lambda}E_{\mu\nu}^{\lambda\rho}g^{\nu\sigma}\frac{1}{\cancel{2}}g^{\mu\tau}\left(\nabla_{\sigma}\delta g_{\tau\rho}+\cancelto{0}{\nabla_{\rho}\delta g_{\sigma\tau}}-\nabla_{\tau}\delta g_{\sigma\rho}\right) \\ \notag
&=-2\nabla_{\lambda}E_{\mu\nu}^{\lambda\rho}g^{\nu\sigma}g^{\mu\tau}\nabla_{[\sigma}\delta g_{\tau]\rho} \\ \notag
&=-2\nabla_{\lambda}E^{\lambda\rho\tau\sigma}\nabla_{\sigma}\delta g_{\tau\rho}\\
&=2\nabla_{\lambda}E^{\lambda\rho\sigma\tau}\nabla_{\sigma}\delta g_{\tau\rho}\, .
\end{align}
Además, usando nuevamente el resultado~\eqref{varchrr} para el último término de la Ec.~\eqref{vargen2}, obtenemos
\begin{align}\notag
\delta I&=\int d^D x \sqrt{|g|}\left[-\frac{1}{2} \delta g^{\mu\nu}g_{\mu\nu}\Lag+\delta g^{\mu\nu} E_{\mu}{}^{\lambda\rho\sigma}R_{\nu\lambda\rho\sigma}+2 \nabla_{\lambda} E^{\lambda\rho\sigma\tau}\nabla_{\sigma} \delta g_{\tau\rho}\right]  \\
&+\int d^D x \sqrt{|g|} \nabla_{\lambda}\left(2 E^{\lambda\rho\tau\sigma}\nabla_{\sigma}\delta g_{\tau \rho}\right)\, . \label{vararbitraria}
\end{align}
Luego, si nuevamente integramos por partes el último término de la ecuación anterior y utilizamos que el tensor $E^{\mu\nu}_{\lambda\rho}$ cumple con la identidad de Bianchi $E_{\mu[\nu\lambda\rho]}=0$ nos queda
\begin{align}
\delta I=\int d^D x \sqrt{|g|} \delta g^{\mu\nu} \mathcal{E}_{\mu\nu}+\int d^D x \sqrt{|g|} \nabla_{\mu} \Theta^{\mu} \, ,
\end{align}
en donde
\begin{align}
\mathcal{E}_{\mu\nu} &\equiv E_{\mu}{}^{\lambda\rho\sigma} R_{\nu\lambda\rho\sigma}-\frac{1}{2} g_{\mu\nu}\Lag-2 \nabla^{\lambda}\nabla^{\rho}E_{\mu\lambda\rho\nu}\, ,\\
\Theta{^\mu} &\equiv 2 \delta g_{\nu\sigma}\nabla_\rho E^{\rho\sigma\mu\nu}-2\nabla_{\rho}\delta g_{\nu \sigma} E^{\rho\sigma\mu\nu}\, .
\end{align}
Notemos que este tratamiento para la acción~\eqref{accgeneral} corresponde a variaciones arbitrarias. Por lo tanto, cuando evaluamos esta expresión en configuraciones que resuelven las ecuaciones de campo $\mathcal{E}_{\mu\nu}=0$, esta variación se vuelve un término de borde. 

Ahora, podríamos tomar la variación de la acción con respecto de algún grupo de simetría. En general, para teorías en la que los campos geométricos son dinámicos, usamos la variación difeomórfica. De éste modo, la variación de la acción~\eqref{accgeneral} bajo un difeomorfismo infinitesimal es
\begin{align} \notag
     \delta_{\xi} I&= \int d^{D}x \delta_{\xi}[\sqrt{\lvert g\rvert}\, \Lag]\\
     &=\int d^{D}x \left[\delta_{\xi}(\sqrt{\lvert g\rvert})\Lag\,+ \sqrt{\lvert g\rvert}\delta_{\xi} \Lag\right] \, . 
\end{align}
Veamos como cambia cada término por separado. Para la variación de la raíz del determinante de la métrica, se tiene
\begin{align}
\delta_{\xi}\sqrt{\lvert g\rvert}&=\frac{1}{2}\sqrt{\lvert g\rvert}g^{\mu\nu}\delta_{\xi}g_{\mu\nu}\, .
\end{align}
Ahora, sabemos que la métrica transforma bajo un difeomorfismo como $\delta_{\xi}g_{\mu\nu}=\Lie_{\xi}g_{\mu\nu}$, con 
\begin{align}
\Lie_{\xi}g_{\mu\nu}=\xi^{\lambda}\nabla_{\lambda}g_{\mu\nu}+g_{\lambda\nu}\nabla_{\mu}\xi^{\lambda}+g_{\mu\lambda}\nabla_{\nu}\xi^{\lambda}\, ,
\end{align}
la derivada de Lie. Usando que la conexión es libre de torsión, nos queda
\begin{align}
\Lie_{\xi}g_{\mu\nu}=\nabla_{\mu}\xi_{\nu}+\nabla_{\nu}\xi_{\mu}\, .
\end{align}
Reemplazando lo anterior en la variación de la métrica, obtenemos
\begin{align}
    \delta_{\xi}\sqrt{\lvert g\rvert}&=\frac{1}{2}\sqrt{\lvert g\rvert}g^{\mu\nu}2\nabla_{(\mu}\xi_{\nu)}=\sqrt{\lvert g\rvert}\nabla_{\mu}(g^{\mu\nu}\xi_{\nu})=\sqrt{\lvert g\rvert}\nabla_{\mu}\xi^{\mu}\, . \label{varxi}
\end{align}
Por otro lado, el Lagrangiano transforma como escalar bajo difeomorfismos, esto es, 
\begin{align}
    \delta_{\xi} \Lag=\Lie_{\xi} \Lag=\xi^{\mu}\nabla_{\mu}\Lag\, .
\end{align}
Así, la variación difeomórfica de la acción de la Ec.~\eqref{varinicial} será
\begin{align}\notag
    \delta_{\xi} I
     &=\int d^{D}x \sqrt{\lvert g\rvert}\left(\nabla_{\mu}\xi^{\mu}\Lag+\xi^{\mu}\nabla_{\mu}\Lag\right)\\
     &=\int d^{D}x \sqrt{\lvert g\rvert}\nabla_{\mu}\left(\xi^{\mu}\Lag\right)\, . \label{vardifeo}
\end{align}

El teorema de Noether nos dice que cualquier simetría presente en un sistema físico conduce a una ley de conservación \cite{Noether1918}. En este nivel, tenemos dos variaciones que son distintas en naturaleza: (i) variaciones arbitrarias de la acción con campos que satisfacen las ecuaciones de movimiento, y (ii) variaciones difeomórficas con campos arbitrarios. Por otro lado, la aplicación del teorema consiste en igualar estas variaciones pidiendo que en (i) la variación sea difeomórfica y en (ii) los campos satisfagan las ecuaciones de movimiento. Entonces, igualando los resultados~\eqref{vararbitraria} y~\eqref{vardifeo}, tenemos
\begin{align} \notag
\int_{\mathcal{M}} d^{D} x \sqrt{|g|} \nabla_{\mu} \Theta^{\mu}(\bar{g},\Lie_{\xi}g)&=\int_{\mathcal{M}} d^{D}x \sqrt{\lvert g\rvert}\nabla_{\mu}\left(\xi^{\mu}\Lag[\bar{R}^{\mu\nu}_{\lambda\rho}]\right)\\
\int_{\mathcal{M}} d^{D} x \sqrt{|g|}\nabla_{\mu}\left(\Theta^{\mu}(\bar{g},\Lie_{\xi}g)-\xi^{\mu}\Lag[\bar{R}^{\mu\nu}_{\lambda\rho}]\right)&=0 \, ,
\end{align}
en donde $\bar{g}$ es la métrica que satisface las ecuaciones de movimiento y a su vez el tensor de Riemann $\bar{R}^{\mu\nu}_{\lambda\rho}$ está construido con dicha métrica.
Así, para una variedad ${\mathcal{M}}$ arbitraria, nos queda que
\begin{align}
\nabla_{\mu}\left(\Theta^{\mu}(\bar{g},\Lie_{\xi}g)-\xi^{\mu}\Lag[\bar{R}^{\mu\nu}_{\lambda\rho}]\right)=\nabla_{\mu}J^{\mu}=0\, ,\label{conscorr}
\end{align}
lo que nos entrega una corriente de Noether conservada. El Lema de Poincaré nos dice que, localmente, $\nabla_{\mu}J^{\mu}=0\to J^{\mu}=\nabla_{\nu}q^{\mu\nu}$, en donde $q^{\mu\nu}=-q^{\nu\mu}$ se conoce como prepotencial de Noether. Con los resultados anteriores podemos encontrar $q^{\mu\nu}$ a partir de 
\begin{align}\notag
    J^{\mu}&=2 \Lie_{\xi}g_{\nu\sigma}\nabla_{\rho} E^{\rho\sigma\mu\nu}-2\nabla_{\rho}\Lie_{\xi}g_{\nu \sigma} E^{\rho\sigma\mu\nu}-\xi^{\mu}\Lag\\
    &=2\left(\nabla_\nu \xi_{\sigma}+\nabla_{\sigma} \xi_{\nu}\right) \nabla_{\rho} E^{\rho \sigma \mu\nu}-2 \nabla_{\rho}\left(\nabla_{\nu} \xi_{\sigma}+\nabla_{\sigma} \xi_{\nu}\right) E^{\rho \sigma \mu\nu}-\xi^{\mu}\Lag\, , \label{corriente}
\end{align}
donde en la segunda línea hemos remplazado $\mathcal{L}_\xi g_{\mu\nu}=2\nabla_{(\mu}\xi_{\nu)}$. Notemos que en la última expresión todo depende de $\bar{g}$, por lo tanto a partir de este punto omitiremos la barra. De las ecuaciones de movimiento $\mathcal{E}_{\mu\nu}=0$, podemos encontrar el ultimo término 
\begin{align}\notag
&\mathcal{E}_{\nu}{^\mu}=E^{\mu \lambda \rho\sigma} R_{\nu \lambda \rho \sigma}-\frac{1}{2} \delta_\nu^\mu \Lag-2 \nabla^{\lambda} \nabla^{\rho} E^{\mu}_{\lambda \rho \nu}=0 \,\,\,\,\, / \cdot \xi^{\nu}\\ \notag
&\xi^{\nu} \mathcal{E}_{\nu}^{\mu}=\xi^{\nu} E^{\mu \lambda \rho \sigma} R_{\nu \lambda g \sigma}-\frac{1}{2} \xi^{\mu} \Lag-2 \xi^{\nu} \nabla^{\lambda} \nabla^{\rho} E^{\mu}_{\lambda\rho\nu}\\ \notag
& \xi^{\mu} \Lag=2 \xi^{\nu} E^{\mu \lambda \rho \sigma} R_{\nu \lambda \rho \sigma}-4 \xi^{\nu} \nabla^{\lambda} \nabla_{\rho} E^{\mu}_{\lambda\rho\nu} -\cancelto{0}{2\xi^{\nu} \mathcal{E}^{\mu}_{\nu}}\\
&\xi^{\mu}\Lag=2\xi{^\lambda} E^{\mu \nu \rho\sigma} R_{\lambda \nu \rho \sigma}-4 \xi_{\sigma} \nabla_{\nu} \nabla_{\rho} E^{\mu \nu \rho \sigma} \, ,
\end{align}
donde hemos usado que estamos evaluando en soluciones que satisfacen las ecuaciones de movimiento para cancelar el término $2\xi^{\nu} \mathcal{E}^{\mu}_{\nu}$. Luego, usando que $[\nabla_{\rho},\nabla_{\sigma}]\xi_{\nu}=-R^{\lambda}\,_{\nu\rho\sigma} \xi_{\lambda}$, obtenemos 
\begin{align}
    \xi^{\mu} \Lag=-2\left[\nabla_{\rho}, \nabla_{\sigma}\right] \xi_{\nu} E^{\mu \nu\rho \sigma} -4 \xi_{\sigma} \nabla_{\nu} \nabla_{\rho} E^{\mu \nu \rho \sigma}\, .
\end{align}
Así, remplazando en la corriente~\eqref{corriente} nos quedará
\begin{align}\notag
    J^{\mu}&=2\left(\nabla_\nu \xi_{\sigma}+\nabla_{\sigma} \xi_{\nu}\right) \nabla_{\rho} E^{\rho \sigma \mu\nu}-2 \nabla_{\rho}\left(\nabla_{\nu} \xi_{\sigma}+\nabla_{\sigma} \xi_{\nu}\right) E^{\rho \sigma \mu\nu}+2\left[\nabla_{\rho}, \nabla_{\sigma}\right] \xi_{\nu} E^{\mu \nu\rho \sigma}\\ \notag
    &+4 \xi_{\sigma} \nabla_{\nu} \nabla_{\rho} E^{\mu \nu \rho \sigma}\\ \notag
    &=2\nabla_\nu \xi_{\sigma}\nabla_{\rho} E^{\rho \sigma \mu\nu}+2\nabla_{\sigma} \xi_{\nu} \nabla_{\rho} E^{\rho \sigma \mu\nu}-2 \nabla_{\rho}\nabla_{\nu} \xi_{\sigma}E^{\rho \sigma \mu\nu}-2 \nabla_{\rho}\nabla_{\sigma} \xi_{\nu}E^{\rho \sigma \mu\nu}\\ \notag
    &+ 2\nabla_{\rho}\nabla_{\sigma}E^{\mu \nu \rho \sigma}-2\nabla_{\sigma}\nabla_{\rho}E^{\mu \nu \rho \sigma}+4 \xi_{\sigma} \nabla_{\nu} \nabla_{\rho} E^{\mu \nu \rho \sigma}\\ \notag
    &=2\nabla_\nu \xi_{\sigma}\nabla_{\rho}( E^{\rho \sigma \mu\nu}+E^{\rho\nu\mu\sigma})-2 \nabla_{\rho}\nabla_{\nu} \xi_{\sigma}(E^{\rho \sigma \mu\nu}+E^{\mu\sigma\nu\rho})-\cancel{2 \nabla_{\rho}\nabla_{\sigma} \xi_{\nu}E^{\rho \sigma \mu\nu}}\\ \notag
    &+ \cancel{2\nabla_{\rho}\nabla_{\sigma}E^{\rho \sigma\mu \nu }}+4 \xi_{\sigma} \nabla_{\nu} \nabla_{\rho} E^{\mu \nu \rho \sigma}\\ 
    &=2\nabla_\nu \xi_{\sigma}\nabla_{\rho}( E^{\rho \sigma \mu\nu}+E^{\rho\nu\mu\sigma})+4 \xi_{\sigma} \nabla_{\nu} \nabla_{\rho} E^{\mu \nu \rho \sigma}-2 \nabla_{\rho}\nabla_{\nu} \xi_{\sigma}(E^{\rho\nu\sigma\nu})\, ,
\end{align}
donde hemos usado $E^{\rho[\sigma\mu\nu]}=0$. Luego, integrando por partes el ultimo término, llegaremos a una expresión para la corriente conservada dada por
\begin{align}
    J^{\mu}=-2\nabla_{\nu}(E^{\mu\nu}_{\lambda\rho}\nabla^{\lambda}\xi^{\rho}+2\xi^{\lambda}\nabla^{\rho}E^{\mu\nu}_{\lambda\rho})\, .
\end{align}
Así, usando $J^{\mu}=\nabla_{\nu}q^{\mu\nu}$, encontramos que el prepotencial de Noether es 
\begin{align}
    q^{\mu\nu}=-2\left(E_{\lambda \rho}^{\mu \nu} \nabla^{\lambda }\xi^{\rho}+2 \xi^{\lambda} \nabla^{\rho} E_{\lambda \rho}^{\mu \nu}\right)\, .
\end{align}

Para definir una carga conservada, utilizaremos la noción noción de energía a partir de la formulación Hamiltoniana, siguiendo la construcción de las Ref.~\cite{arnold1989mathematical,Lee1990LocalSA,Corichi:2018drc,Crnkovic:1986ex}. Contrayendo la Ec.~\eqref{conscorr} con el tensor de Levi-Civita $\varepsilon_{\mu \nu_2 \ldots \nu_D}=\sqrt{|g|}\epsilon_{\mu \nu_2 \ldots \nu_D}$, en donde $\epsilon_{\mu \nu_2 \ldots \nu_D}$ es el símbolo de Levi-Civita tenemos
\begin{align}
J^\mu \varepsilon_{\mu \nu_2 \ldots \nu_D}=\Theta^\mu \varepsilon_{\mu \nu_2 \ldots \nu_D}-\xi^\mu \varepsilon_{\mu \nu_2 \ldots \nu_D} \Lag \, .
\end{align}
Tomamos variaciones de la expresión anterior. Considerando que los parámetros no transforman, i.e., $\delta \xi^\mu=0$ y que $\delta(\sqrt{|g|}\Lag)= \sqrt{|g|}\left(\delta g^{\lambda \rho} \mathcal{E}_{\lambda \rho}+\nabla_\lambda \Theta^\lambda\right)$, obtenemos
\begin{align}
\delta\left(J^\mu \varepsilon_{\mu \nu_2 \ldots \nu_D}\right)=\delta\left(\Theta^\mu \varepsilon_{\mu \nu_2 \ldots \nu_D}\right)-\xi^\mu \varepsilon_{\mu \nu_2 \ldots\nu_D}(\cancelto{0}{\delta g^{\lambda\rho}\mathcal{E}_{\lambda\rho}}+\nabla_\lambda \Theta^\lambda)\, , \label{varcorri}
\end{align}
en donde hemos considerado que estamos on-shell para cancelar el factor donde aparecen las ecuaciones de movimiento. Ahora, usaremos una identidad para reescribir el último término
\begin{align}
\mathcal{L}_{\xi}\left(\Theta^\mu \varepsilon_{\mu \nu_2 \ldots \nu_D}\right)&=\mathcal{L}_{\xi} \Theta^\mu \varepsilon_{\mu \nu_2 \ldots \nu_D}+\Theta^\mu \mathcal{L}_{\xi} \varepsilon_{\mu \nu_2 \ldots \nu_D}\, ,
\end{align}
y la Ec.~\eqref{varxi}, así, nos queda lo siguiente 
\begin{align}\notag
\mathcal{L}_{\xi}\left(\Theta^\mu \varepsilon_{\mu \nu_2 \ldots \nu_D}\right)&=\left(\xi^\lambda \nabla_\lambda \Theta^\mu-\Theta^\lambda \nabla_\lambda \xi^\mu\right) \varepsilon_{\mu \nu_2 \ldots \nu_D}+\Theta^\mu \varepsilon_{\mu \nu_2 \ldots \nu_D} \nabla_\lambda \xi^\lambda\\ \notag
&=\left(\nabla_{\lambda}(\xi^{\lambda}\Theta^{\mu})-\cancel{\nabla_{\lambda}\xi^{\lambda}\Theta^{\mu}}-\nabla_{\lambda}(\Theta^{\lambda}\xi^{\mu})+\xi^{\mu}\nabla_{\lambda}\Theta^{\lambda}\right)\varepsilon_{\mu \nu_2 \ldots \nu_D}\\ \notag
&+\cancel{\Theta^\mu \varepsilon_{\mu \nu_2 \ldots \nu_D} \nabla_\lambda \xi^\lambda}\\
&=2 \nabla_\lambda\left(\xi^{[\lambda}\Theta^{\mu]}\right) \varepsilon_{\mu \nu_2 \ldots \nu_D}+\xi^\mu \varepsilon_{\mu \nu_2 \ldots \nu_D} \nabla_\lambda \Theta^\lambda \, .
\end{align}
Luego, despejando el factor que nos interesa remplazar en la corriente, obtenemos 
\begin{align}
\delta\left(J^\mu \varepsilon_{\mu \nu_2 \ldots \nu_D}\right)=\delta\left(\Theta^\mu \varepsilon_{\mu \nu_2\ldots  \nu_D}\right)-\mathcal{L}_{\xi}\left(\Theta^\mu \varepsilon_{\mu \nu_2 \ldots \nu_D}\right)\left.+2 \nabla_\lambda\left(\xi^{[\lambda} \Theta^{\mu]}\right]\right) \varepsilon_{\mu \nu_2 \ldots \nu_D}\,.
\end{align}
Contrayendo a ambos lados con $\frac{1}{(D-1) !} d x^{\nu_2} \wedge \ldots \wedge d x^{\nu_D}$ para construir el elemento de volumen de co-dimension 1 e integrando sobre una hipersuperficie de Cauchy $\mathcal{C}$, tenemos
\begin{align}
\delta \int_{\mathcal{C}} \diff^{D-1}x \sqrt{|h|} n_\mu J^\mu=\int_{\mathcal{C}} \diff^{D-1}x \sqrt{|h|} n_\mu \Omega^\mu\left(\overline{g}, \mathcal{L}_{\xi} g, \delta g\right)+2 \int \diff^{D-1}x n_\mu \nabla_\lambda\left(\xi^{[\lambda} \Theta^{\mu]}\right)\, ,
\end{align}
en donde el primer término del lado derecho es la integral de la  corriente simpléctica correspondiente a la evolución Hamiltoniana definida por el vector $\xi$, cuya integral sobre $\mathcal{C}$ define las ecuaciones de Hamilton, dadas por
\begin{align}
\delta H=\int_{\mathcal{C}} \diff^{D-1} x \sqrt{|h|} n_\mu \Omega^\mu \left(\bar{g}, \mathcal{L}_\xi g, \delta g\right)\, .
\end{align}
Podemos notar que entonces la variación del Hamiltoniano es la suma de la variación de la corriente de Noether mas una pieza extra. Así, escribiendo de manera astuta encontramos
\begin{align}
\delta H=\delta \int_{\mathcal{C}} \diff^{D-1} x \sqrt{|h|} n_\mu \nabla_\nu q^{\mu\nu}+2 \int_{\mathcal{C}} \diff^{D-1} x \sqrt{|h|} n_\mu \nabla_\nu\left(\xi^{[\mu} \Theta^{\nu]}\right)\, .
\end{align}
Por otro lado, usando que $q^{\mu\nu}$ es antisimétrico y el teorema de la divergencia proyectando el borde hacia infinito, obtendremos
\begin{align}\notag
\delta H &=\delta \int_{\mathcal{C}} \diff^{D-1} x \sqrt{|h|} D_\nu\left(n_\mu q^{\mu\nu}\right)+2 \int_{\mathcal{C}} \diff^{D-1}x \sqrt{|h|} D_\nu\left(n_\mu \xi^{[\mu} \Theta^{\nu]}\right)\\
    &=\delta \int_{\infty} q^{\mu\nu} \diff\Sigma_{\mu\nu}+2 \int_{\infty} \xi^\mu \Theta^\nu \diff\Sigma_{\mu\nu}\, . \label{variacionH}
\end{align}
Para definir una noción precisa de energía debemos incluir un término de borde, $\mathcal{B}$, que permita definir un principio variacional, de modo que cancele las derivadas normales de orden superior. Sabemos de las secciones anteriores que para el caso de Relatividad General, este término de borde es el Gibbons-Hawking-York. Ahora, queremos encontrar $\mathcal{B}$ de forma general, para ello, pedimos que se satisfaga la siguiente condición de integrabilidad 
\begin{align}
\delta \int_{\infty} 2 \xi^\mu \mathcal{B}^\nu \diff\Sigma_{\mu \nu}=-2 \int_{\infty} \xi^\mu \Theta^\nu \diff\Sigma_{\mu\nu}\, .
\end{align}
Así, la variación del término de borde que agregaremos debe cancelar el término de borde proveniente de la variación de la acción. Si dicho término es de la forma $I_{B}=\int \diff^{D-1 }x \sqrt{|h|}\mathcal{B}$, entonces comprando con $I_{B}=\int \mathcal{B}^{\mu}d\Sigma_\mu$, concluimos que 
\begin{align}
\mathcal{B}^\mu=\sigma n^\mu \mathcal{B} \, .
\end{align}
Por tanto, remplazando lo anterior en~\eqref{variacionH}, obtenemos que 
\begin{align}
\delta H=\delta \int_{\infty} q^{\mu \nu} \diff\Sigma_{\mu\nu}-2 \sigma \delta \int_{\infty} B \xi^{[\mu} n^{\nu]} \diff \Sigma_{\mu \nu}
\end{align}
Integrando en el espacio de fase, y considerando que $\xi$ es un vector de Killing que genera una isometría temporal, tenemos que 
\begin{align}
H[\xi]=E=\int_{\infty}\left(q^{\mu\nu}-2 \sigma \xi^{[\mu} n^{\nu]} B\right) \diff\Sigma_{\mu\nu}\, .
\end{align}

\subsection{Ley cero de la termodinámica}
Ahora, queremos profundizar como se relacionan las cantidades conservadas con la termodinámica. La gravedad superficial es una cantidad fundamental asociada a la termodinámica de los agujeros negros. Stephen Hawking demostró en la Ref.~\cite{Hawking:1975vcx} que la temperatura de la radiación emitida por un agujero negro es
\begin{align}
T_{H}=\frac{\kappa_s}{2\pi}\, ,\label{temp}
\end{align}
en donde $\kappa_s$ es la gravedad superficial, definida como 
\begin{align}
\xi^{\mu} \nabla_{\mu} \xi^{\nu}=\left.\kappa_s \xi^\nu\right|_H \, \label{surfgrav}
\end{align}
con $\xi=\xi^\mu \partial_\mu$ un vector de Killing ortogonal al horizonte sobre el cual su norma se anula. Adicionalmente, es posible construir una definición de gravedad superficial usando las propiedades del horizonte del agujero negro. Recordando el teorema de Frobenius para un vector $\xi=\xi^\mu \partial_\mu$ ortogonal en el horizonte de Killing, tenemos
\begin{align}
    \xi_{[\mu} \nabla_{\nu} \xi_{\lambda]}=0 \, .
\end{align}
Usando la ecuación de Killing, $\nabla_{\mu}\xi_{\nu}=-\nabla_{\nu}\xi_{\mu}$, esto implica que en horizonte se cumple
\begin{align}
\xi_{\lambda} \nabla_{\mu}\xi_{\nu}=-2\xi_{[\mu}\nabla_{\nu]}\xi_{\lambda}\, .
\end{align}
Ahora, contrayendo con $\nabla^{\mu}\xi^{\nu}$, obtenemos
\begin{align}\notag
\xi_{\lambda}(\nabla^{\mu}\xi^{\nu}) (\nabla_{\mu}\xi_{\nu})&=-2(\xi_{\mu}\nabla^{\mu}\xi^{\nu})(\nabla_{\mu}\xi_{\nu})\\ \notag
&=-2\kappa_{s}\xi^{\nu}\nabla_{\nu}\xi_{\lambda}\\ 
&=-2\kappa_{s}^2 \xi_{\lambda}\, ,
\end{align}
así, encontramos una ecuación para el cuadrado de la gravedad superficial dada por
\begin{align}
    \kappa_{s}^2=-\frac{1}{2}(\nabla^{\mu}\xi^{\nu}) (\nabla_{\mu}\xi_{\nu})|_H\, . \label{surfgrav2}
\end{align} 
Esta cantidad se puede interpretar como el valor límite en el horizonte de la fuerza que debe ser ejercida en el infinito para mantener una partícula de prueba en ese lugar. 

Ahora, la ley cero de la termodinámica de agujeros negros nos dice que $\kappa_s$ es constante sobre el horizonte de bifurcación $\Sigma$, en donde este último se define como un horizonte de Killing, en donde el vector de Killing $\xi$ se hace cero. También, $\Sigma$ se puede interpretar como un lugar en donde se tiene un equilibrio termodinámico para los agujeros negros. Lo anterior se puede demostrar considerando 
\begin{align}
\left.\xi^{\mu} \nabla_{\mu} \kappa_{s}^2\right|_H=-\frac{1}{2} \xi^{\mu} \nabla_{\mu}\left(\nabla_{\lambda} \xi_{\rho} \nabla^{\lambda} \xi^{\rho}\right)=-\left.\xi^{\mu} \nabla_{\lambda} \xi_{\rho} \nabla_{\mu} \nabla^{\alpha} \xi^{\rho}\right|_H \, ,
\end{align}
además, usando que $\nabla_\mu \nabla_\nu \xi^\lambda=R_{\rho \mu \nu}{}^\lambda \xi^\rho$, tenemos
\begin{align}
\left.\xi{^\mu} \nabla_{\mu} \kappa_s^{2}\right|_H=-\left.\xi^{\mu} \nabla^{\lambda} \xi^{\rho} R_{\sigma \mu \lambda \rho} \xi^{\sigma}\right|_H=0 \, .
\end{align}
Luego, $\kappa_s$ es constante a lo largo de las orbitas de $\xi$. Asimismo, consideremos hora un vector $v=v^{\mu} \partial_{\mu}$ que es tangente al horizonte de bifurcación $\Sigma$. Entonces 
\begin{align}
\left.v^{\mu} \nabla_{\mu} \kappa_s{^2}\right|_{\Sigma}=-\left.v^{\mu} \nabla^{\lambda }\xi^{\rho} R_{\sigma \mu \lambda \rho} \xi^{\sigma}\right|_{\Sigma}=0  \, ,
\end{align}
ya que $\xi|_{\Sigma}=0$. Entonces, para un vector tangente a $\Sigma$ arbitrario, concluimos que $\kappa_s$ es constante sobre el horizonte de bifurcación.

\subsection{Primera ley de la termodinámica}

Ahora, es posible derivar la primera ley de la termodinámica si tomamos nuevamente en cuenta la variación de la corriente de Noether~\eqref{varcorri} sobre una hipersuperficie de Cauchy $\mathcal{C}$; esta vez la hipersuperficie tiene un borde interior $\Sigma$, que, como ya conocemos, representa a la superficie de bifurcación del agujero negro. Luego, considerando el vector de Killing $\xi$
\begin{align}
\delta\left(J^\mu \varepsilon_{\mu \nu_2 \ldots \nu_D}\right)=2 \nabla_\lambda\left(\xi^{[\lambda} \theta^{\mu]}\right) \varepsilon_{\mu \nu_2 \ldots \nu_D}\, ,
\end{align}
donde en el lado derecho solo sobrevive el termino de borde ya que el otro término depende de $\delta g$.
Ahora, si recordamos el diagrama de Penrose de la parte exterior de un agujero negro estacionario 

\begin{figure}[H]
\includegraphics[width=6cm]{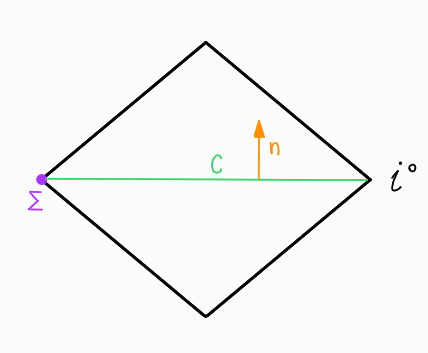}
\centering
\caption{Diagrama de Penrose de la parte exterior de un agujero negro. Donde $i^{0}$ es el infinito espacial, $\Sigma$ corresponde a la superficie de bifurcación y $\mathcal{C}$ es la hipersuperficie de Cauchy.}
\end{figure}
Podemos ver que la hipersuperficie de Cauchy tiene borde interior $\Sigma$ y borde exterior $i^{0}$. Ahora, integraremos la variación de la corriente de Noether sobre $\mathcal{C}$. Luego, contrayendo con $\frac{1}{(D-2) !} \diff x^{\nu_2} \wedge \ldots \wedge \diff x^{\nu_D} $ e integrando sobre $\mathcal{C}$, tenemos 
\begin{align}\notag
\delta \int_\mathcal{C} \diff^{D-1} x \sqrt{|h|}n_\mu {\nabla_\nu q^{\mu \nu}}&=2 \int_\mathcal{C} \diff^{D-1} x \sqrt{|h|}n_\mu \nabla_\lambda(\xi^{[\lambda} \Theta^{\mu]})\\ \notag
\delta \int_\mathcal{C}\diff^{D-1}x \sqrt{|h|} D_\nu\left(n_\mu q^{\mu \nu}\right)&=2 \int_\mathcal{C} \diff^{D-1}x\sqrt{|h|} D_\nu\left(n_\mu \xi^{[\nu} \Theta^{\mu]}\right)\\ 
\delta \oint_{\partial\mathcal{C}} \diff^{D-2}x \sqrt{|\gamma|} u_\nu n_\mu q^{\mu \nu}&=2 \oint_{\partial \mathcal{C}} \diff ^{D-2}x \sqrt{|\gamma|} u_\nu n_\mu \xi^{[\nu} \Theta^{\mu]}\, ,
\end{align}
donde podemos observar que en la primera línea usamos que $J^{\mu}$ es una corriente conservada. Luego, en la segunda línea utilizamos que un vector normal contraído con un tensor antisimétrico puede ser escrito como la derivada $D$ de la multiplicación de dichos objetos y, como último paso, el teorema de la divergencia. A continuación si consideramos $\partial\mathcal{C}=\Sigma\cup i^{0}$ tenemos 
\begin{align}\notag
&\delta \int_{\infty} \diff^{D-2}x\sqrt{|\gamma|} u_\nu n_\mu q^{\mu \nu}-\delta \int_{\Sigma} \diff^{D-2}x\sqrt{|\gamma|} u_\nu n_\mu q^{\mu\nu}\\
&=2 \int_{\infty} \diff^{D-2}x\sqrt{|\gamma|} u_\nu n_\mu \xi^{[\nu} \Theta^{\mu]}-\cancelto{0}{{2 \int_{\Sigma} \diff^{D-2}x\sqrt{|\gamma|} u_\nu n_\mu \xi^{[\nu} \Theta^{\mu]}}} \, ,
\end{align}
en donde el último término se anula ya que el vector de Killing en el horizonte de bifurcación es cero y además, el signo menos viene de la orientación del vector normal. Así, nos queda 
\begin{align}\notag
\delta \int_{\infty}  \diff^{D-2}x\sqrt{|\gamma|} u_\nu n_\mu q^{\mu \nu} +2 \int_{\infty} d^{D-2}x \sqrt{\sigma} u_\nu n_\mu \xi^{[\nu} \Theta^{\mu]}=\delta \int_{\Sigma}  \diff^{D-2}x\sqrt{|\gamma|} u_\nu n_\mu q^{\mu\nu}\, .
\end{align}
Luego, recordando de la derivación del teorema de Noether, habíamos dicho que para tener una noción de energía debíamos considerar un termino de borde, así 
\begin{align}
\delta \int_{\infty}  \diff^{D-2}x\sqrt{|\gamma|} u_\nu n_\mu q^{\mu \nu} -2 \delta\int_{\infty} \xi^{\mu}\mathcal{B}^{\nu}\diff\Sigma_{\mu\nu} =\delta \int_{\Sigma}  \diff^{D-2}x\sqrt{|\gamma|} u_\nu n_\mu q^{\mu\nu}\, ,
\end{align}
 y con ello, todo lo que esta al lado izquierdo de la ecuación anterior es la variación de la energía, por tanto, nos queda 
 \begin{align}
\delta E =\delta \int_{\Sigma} \diff^{D-2}x\sqrt{|\gamma|} u_\nu n_\mu q^{\mu\nu}=\delta \int_{\Sigma} q^{\mu\nu} d\Sigma_{\mu\nu}\, .
 \end{align}
Del mismo modo, Wald en la Ref.~\cite{Wald:1993nt} identificó el lado derecho con $T\cdot S$ comparando con la primera ley de la termodinámica. Lo anterior esta justificado solamente si la temperatura se mantiene constante sobre la hipersuperficie de bifurcación. Asimismo, Iyer y Wald en la Ref.~\cite{Iyer:1994ys} demostraron como teorema lo siguiente 
\begin{align}
\delta \int_{\Sigma} q^{\mu\nu} d\Sigma_{\mu\nu}=\frac{k_s}{2\pi}\delta S \, ,
\end{align}
por lo que la Entropía del agujero negro es la carga de Noether en el horizonte, esto es, 
\begin{align}
S=\frac{1}{T} \int_{\Sigma} q^{\mu\nu} d\Sigma_{\mu\nu}\, . \label{entropy}
\end{align}

\subsection{Ejemplo: Relatividad General}
Ahora, consideremos el caso de Relatividad General y como solución a las ecuaciones de movimiento la métrica~\eqref{schw}. Sabemos que el Lagrangiano es de la siguiente manera
\begin{align}
    \Lag=\kappa R\equiv\kappa \delta^{[\lambda}_{[\mu}\delta^{\rho]}_{\nu]}R^{\mu\nu}_{\lambda\rho}\, .
\end{align}
Luego, de la definición de $E^{\lambda\rho}_{\mu\nu}$ 
\begin{align}
E^{\lambda\rho}_{\mu\nu}=\frac{\partial\Lag}{\partial R^{\mu\nu}_{\lambda \rho}}=\kappa \delta^{[\lambda}_{[\mu}\delta^{\rho]}_{\nu]}=\frac{\kappa}{2}\delta^{\lambda\rho}_{\mu\nu}\, .
\end{align}
Así, podemos calcular el prepotencial de Noether 
\begin{align}\notag
q^{\mu \nu} &=-2\left(E_{\lambda \rho}^{\mu \nu} \nabla^\lambda \xi^\rho+2 \xi^\lambda \nabla^\rho E_{\lambda \rho}^{\mu \nu}\right) \\ \notag
&=-2\left(\frac{\kappa }{2} \delta_{\lambda \rho}^{\mu \nu} \nabla^\lambda \xi^\rho\right) \\ \notag
&=-\kappa\left(\delta_\lambda^\mu \delta_\rho^\nu-\delta_\rho^\mu \delta_{\lambda}^\nu\right) \nabla^\lambda \xi^\rho \\ \notag
&=-\kappa\left(\nabla^\mu \xi^\nu-\nabla^\nu \xi^\mu\right)\\
&=-2\kappa\nabla^{[\mu}\xi^{\nu]} \, ,
\end{align}
entonces, la definición de energía para Relatividad General considerando que el termino de borde es $\mathcal{B}=2 \sigma \kappa (K-K_0)$ quedará
\begin{align}
H[\xi]=E=\kappa\int_{\infty}\left(-2\nabla^{[\mu}\xi^{\nu]}+4  \xi^{[\mu} n^{\nu]} \left[K-K_0\right]\right) \diff\Sigma_{\mu\nu}\, .
\end{align}
Los objetos de la integral son conocidos de las secciones anteriores. Así, calculando el primer término para un vector de Killing $\xi=\partial_{t}$, tenemos 
\begin{align}\notag
-\kappa\int_{\infty}2\nabla^{[\mu}\xi^{\nu]}\diff\Sigma_{\mu\nu}&=-2\kappa\sigma\int_{\infty}\diff^{2}x\,\sqrt{|\gamma|}\,\nabla_{\mu}\xi_{\nu}n^{[\mu}u^{\nu]}\\ \notag
&=-2\kappa\sigma\int_{\infty}\diff^{2}x\,\sqrt{|\gamma|}\,\nabla_{\mu}(g_{\nu\lambda}\xi_{\lambda})n^{[\mu}u^{\nu]}\\ \notag
&=-2\kappa\sigma\int_{\infty}\diff^{2}x\,\sqrt{|\gamma|}\,g_{\nu\lambda}\nabla_{\mu}\xi^{\lambda}n^{[\mu}u^{\nu]}\\ \notag
&=-2\kappa\sigma\int_{\infty}\diff^{2}x\,\sqrt{|\gamma|}\,g_{\nu\lambda}(\cancelto{0}{\partial_{\mu}\xi^{t}}+\Gamma^{\lambda}_{\mu\alpha}\xi^{\alpha})n^{[\mu}u^{\nu]}\\ \notag
&=-2\kappa\sigma\int_{\infty}\diff^{2}x\,\sqrt{|\gamma|}\,g_{\nu\lambda}(\Gamma^{\lambda}_{\mu t}\xi^{t})n^{[\mu}u^{\nu]}\\
&=-\cancel{2}\kappa\sigma\int_{\infty}\diff^{2}x\,\sqrt{|\gamma|}\,g_{\nu\lambda}(\Gamma^{\lambda}_{\mu t}\xi^{t})\left[\frac{1}{\cancel{2}}(n^{\mu}u^{\nu}-n^{\nu}u^{\mu})\right] \, .
\end{align}
Luego, los símbolos de Christoffel no nulos de la solución de Schwarzschild son
\begin{equation}
    \begin{array}{cc}
\Gamma^t_{rr} = \frac{GM}{r^2 f(r)} &
\Gamma^r_{tr} = \frac{GM}{r^2 f(r)} \\
\Gamma^r_{tt} = -\frac{GM}{r^2 f(r)} &
\Gamma^\theta_{\theta r} = rf(r) \\
\Gamma^\theta_{\phi\phi} = -\sin\theta\cos\theta &
\Gamma^\phi_{\theta\phi} = \frac{\cos\theta}{\sin\theta} \\
\Gamma^\phi_{\phi\theta} = \frac{\cos\theta}{\sin\theta} &
\Gamma^r_{rr} = \frac{GM}{r^2}\\
\Gamma^t_{tr} = \Gamma^t_{rt} = \frac{GM}{r^2} &
\Gamma^r_{\theta\theta} = -\frac{GM}{r^2}f(r) \\
\Gamma^r_{\phi\phi} = -r\sin^2\theta f(r) &
\Gamma^t_{\theta\theta} = rf(r) \\
\Gamma^t_{\phi\phi} = -r\sin^2\theta f(r) &
\Gamma^\theta_{r\theta} = \Gamma^\theta_{\theta r} \\
\Gamma^\phi_{r\phi} = \Gamma^\phi_{\phi r} = \frac{1}{r}f(r) &
\end{array}
\end{equation}

Luego, remplanzando los valores de los Christoffel, obtenemos
\begin{align}\notag
-\kappa\int_{\infty}2\nabla^{[\mu}\xi^{\nu]}\diff\Sigma_{\mu\nu}&=-\kappa\sigma\int_{0}^{\pi}\int_{0}^{2\pi}\diff\theta \diff\phi\,\left(g_{tt}\frac{1}{2f(r)}f'(r)-g_{rr}\frac{1}{2}f(r)f'(r)\right)r^{2}\sin{\theta}\\ \notag
&=-\kappa\sigma\int_{0}^{\pi}\int_{0}^{2\pi}\diff\theta \diff\phi\,\left(-\cancel{f(r)}\frac{1}{2\cancel{f(r)}}f'(r)-\frac{1}{\cancel{f(r)}}\frac{1}{2}\cancel{f(r)}f'(r)\right)r^{2}\sin{\theta} \\ \notag
&=-\kappa\sigma\int_{0}^{\pi}\int_{0}^{2\pi}\diff\theta \diff\phi\,\left[f'(r)\right]r^{2}\sin{\theta} \notag \, .
\end{align}
Así, integrando y remplazando el valor de $\kappa$, nos queda
\begin{align}
-\kappa\int_{\infty}2\nabla^{[\mu}\xi^{\nu]}\diff\Sigma_{\mu\nu}=\frac{m}{2}\, .
\end{align}
Ahora, el segundo término es
\begin{align}\notag
\kappa\int_{\infty}4  \xi^{[\mu} n^{\nu]} \left[K-K_0\right]\diff\Sigma_{\mu\nu}&=4\kappa\sigma\int_{\infty}\sqrt{|\gamma|}\,\cancel{\sqrt{f(r)}}\left[\frac{f'(r)+4f(r)-4\sqrt{f(r)}}{2r\cancel{\sqrt{f(r)}}}\right]n_{[t}u_{r]}\diff^{2}x\\ \notag
&=4\kappa\sigma\int_{\infty}\sqrt{|\gamma|}\,\left[\frac{f'(r)+4f(r)-4\sqrt{f(r)}}{2r}\right]\frac{1}{2}(\cancel{n_{t}u_{r}}-n_{r}u_{t})\diff^{2}x\\ \notag
&=2\kappa\sigma\int_{\infty}\sqrt{|\gamma|}\,\left[\frac{f'(r)+4f(r)-4\sqrt{f(r)}}{2r}\right][-f(r)]\diff^{2}x\\ \notag
&=2\kappa\sigma\int_{0}^{\pi}\int_{0}^{2\pi}\diff\theta \diff\phi\,\left[\frac{f'(r)+4f(r)-4\sqrt{f(r)}}{2\cancel{r}}\right][-f(r)]r^{\cancel{2}}\sin{\theta}\, .
\end{align}
Entonces, sustiyuyendo el valor de $f(r)$ e integrando, obtenemos 
\begin{align}
    \kappa\int_{\infty}4  \xi^{[\mu} n^{\nu]} \left[K-K_0\right]\diff\Sigma_{\mu\nu}=\frac{m}{2}\, .
\end{align}
Finalmente, el valor de la energía será 
\begin{align}
H[\xi]=E=\kappa\int_{\infty}\left(-2\nabla^{[\mu}\xi^{\nu]}+4  \xi^{[\mu} n^{\nu]} \left[K-K_0\right]\right) \diff\Sigma_{\mu\nu}=\frac{m}{2}+\frac{m}{2}=m\, ,
\end{align}
lo cual concuerda con los métodos usados anteriormente y adicionalmente es el valor esperado al calcular una cantidad conservada asociada a una simetría temporal.

Por otro lado, para obtener la temperatura $T$ de la Ec.~\eqref{temp}, debemos calcular la gravedad superficial. Entonces usando~\eqref{surfgrav2} 
\begin{align} \notag
\kappa^{2}_{s}&=-\frac{1}{2}(\nabla_{\mu}\xi_{\nu})(\nabla^{\mu}\xi^{\nu})\\ \notag
&=-\frac{1}{2}\left[\nabla_{\mu}(g_{\nu\lambda}\xi^{\lambda})(g^{\mu\alpha}\nabla_{\alpha}\xi^{\nu})\right]\\ \notag
&=-\frac{1}{2}g_{\nu\lambda}(\cancelto{0}{\partial_{\mu}}\xi^{\lambda}+\Gamma^{\lambda}_{\beta\mu}\xi^{\beta})g^{\mu\alpha}(\cancelto{0}{\partial_{\alpha}}\xi^{\nu}+\Gamma^{\nu}_{\sigma\alpha}\xi^{\sigma})\\ \notag
&=-\frac{1}{2}(g_{\nu\lambda}\Gamma^{\lambda}_{\beta\mu}\xi^{\beta})(g^{\mu\alpha}\Gamma^{\nu}_{\sigma\alpha}\xi^{\sigma})\\ \notag
&=-\frac{1}{2}(g_{\nu\lambda}\Gamma^{\lambda}_{t\mu}\xi^{t}g^{\mu\alpha}\Gamma^{\nu}_{t\alpha}\xi^{t})\\ \notag
&=-\frac{1}{2}g^{\mu\alpha}(g_{tt}\Gamma^{t}_{\mu t}\Gamma^{t}_{\alpha t}+g_{rr}\Gamma^{r}_{\mu t}\Gamma^{r}_{\alpha t})\\ \notag
&=-\frac{1}{2}(g^{rr}g_{tt}\Gamma^{t}_{rt}\Gamma^{t}_{rt}+g^{tt}g_{rr}\Gamma^{r}_{tt}\Gamma^{r}_{tt})\\ \notag
&=-\frac{1}{2}\left(\frac{-1}{4}\cancel{f^{2}(r_{h})}f'^{2}(r_{h})\frac{1}{\cancel{f^{2}(r_{h})}}-\frac{1}{4}\frac{1}{\cancel{f^{2}(r_{h})}}f'^{2}(r_{h})\cancel{f^{2}(r_{h})}\right)\\ 
&=\frac{1}{4}f'^{2}(r_{h})\, .
\end{align}
Tomando raíz cuadrada a ambos lados nos queda 
\begin{align}
    \kappa_{s}=\frac{mG}{r^2_{h}}\, .
\end{align}
Además, dada la definición de horizonte podemos despejar $m$ en función de $r_h$ y viceversa 
\begin{align}
    f_{r_h}=0 \to r_{h}=2mG\, ,
\end{align}
entonces, tenemos
\begin{align}
    \kappa_{s}=\frac{mG}{4m^2 G^2}=\frac{1}{4mG}\, .
\end{align}
Finalmente, usando la Ec.~\eqref{temp} obtenemos el valor de la temperatura
\begin{align}
    T=\frac{\kappa_{s}}{2\pi}=\frac{1}{8\pi mG}\, . \label{temp2}
\end{align}

Ahora bien, con el prepotencial también podemos calcular la entropía para la solución~\eqref{schw}. Usando~\eqref{entropy} tenemos
\begin{align}\notag
TS&=-2\kappa\int_{\Sigma}\sqrt{|\gamma|}\nabla^{[\mu}\xi^{\nu]}n_{[\mu}u_{\nu]}\diff^{2}x\\ \notag
&=-2\kappa\int_{\Sigma}\sqrt{|\gamma|}\nabla_{\mu}\xi_{\nu}n^{[\mu}u^{\nu]}\diff^{2}x\\ \notag
&=-2\kappa\int_{\Sigma}\sqrt{|\gamma|}g_{\mu\lambda}\nabla_{\mu}\xi^{\lambda}n^{[\mu}u^{\nu]}\diff^{2}x\\ \notag
&=-2\kappa\int_{\infty}\sqrt{|\gamma|}\,g_{\nu\lambda}(\cancelto{0}{\partial_{\mu}\xi^{t}}+\Gamma^{\lambda}_{\mu\alpha}\xi^{\alpha})n^{[\mu}u^{\nu]}\,\diff^{2}x\\ \notag
&=-2\kappa\int_{\infty}\sqrt{|\gamma|}\,g_{\nu\lambda}(\Gamma^{\lambda}_{\mu t}\xi^{t})n^{[\mu}u^{\nu]}\,\diff^{2}x\\ \notag
&=-\cancel{2}\kappa\int_{\infty}\sqrt{|\gamma|}\,g_{\nu\lambda}(\Gamma^{\lambda}_{\mu t}\xi^{t})\left[\frac{1}{\cancel{2}}(n^{\mu}u^{\nu}-n^{\nu}u^{\mu})\right]\,\diff^{2}x \\ \notag
&=-\kappa\int_{0}^{\pi}\int_{0}^{2\pi}\diff\theta \diff\phi\,\left(g_{tt}\frac{1}{2f(r)}f'(r)-g_{rr}\frac{1}{2}f(r)f'(r)\right)r^{2}\sin{\theta}\\ \notag
&=-\kappa\int_{0}^{\pi}\int_{0}^{2\pi}\diff\theta \diff\phi\,\left(-\cancel{f(r)}\frac{1}{2\cancel{f(r)}}f'(r)-\frac{1}{\cancel{f(r)}}\frac{1}{2}\cancel{f(r)}f'(r)\right)r^{2}\sin{\theta} \\ 
&=-\kappa\int_{0}^{\pi}\int_{0}^{2\pi}\diff\theta \diff\phi\,\left[f'(r)\right]r^{2}\sin{\theta}=\frac{m}{2}  \, .
\end{align}
Luego, para el horizonte $r=r_{h}=2mG$ podemos escribir en función de $r_{h}$, siendo $m=r_{h}/2G$. Además, usando la Ec.~\eqref{temp2} y que $A=4\pi r^{2}_{h}$, nos queda 
\begin{align} \notag
S&=8\pi G \frac{m^{2}}{2}\\ \notag
    &=4\pi G \frac{r^{2}_{h}}{4G^2}\\
    &=\frac{r^{2}_{h}\pi}{G}\, ,
\end{align}
Lo cual es equivalente a 
\begin{align}
    S=\frac{r^{2}_{h}\pi}{G}=\frac{A}{4G}\, .
\end{align}
Es decir,la entropía satisface la ley de un cuarto del área. Además, estas cantidades satisfacen la primera ley de la termodinámica, $\delta M=T\delta S$

\section{Acción Euclídea on-shell \label{sec:AEO}}
Finalmente, revisaremos un formalismo para calcular cantidades conservadas que guarda estrecha relación con las leyes de la termodinámica. Con el fin de usar este método, es necesario tener un principio variacional bien definido. Por lo tanto, aún es necesaria la adición de un termino de borde~\cite{Gibbons:1976ue}. En el caso de Relatividad General, ya vimos que es suficiente sumar la curvatura extrínseca de la variedad para imponer condiciones de borde tipo Dirichlet. 

Consideremos nuevamente la métrica de Schwarzschild en 4 dimensiones dada por la Ec.~\eqref{schw} y apliquemos la rotación de Wick $t\to -i\tau$, lo que nos entrega
\begin{align}
 \diff s^2=\left(1-\frac{2mG}{r}\right)\diff \tau^2+\frac{\diff r^2}{\left(1-\frac{2mG}{r}\right)}+r^2\left(\diff \theta^2+\sin^2\theta\diff \phi^2\right) \, . \label{euclids}
\end{align}
Ahora, el elemento de línea tiene signatura Euclídea, por lo que $f(r)\geq 0$. Este cambio de signo podría inducir singularidades en la variedad, por lo que vamos a revisar con detalles el rango de $\tau$. 

Analicemos la geometría descrita por esta métrica cerca del horizonte $r=r_h$, en donde $f(r_h)=0$. Expandiendo $f(r)$ en torno al horizonte, obtenemos
\begin{align}
    f(r)=\cancelto{0}{f(r_h)}+f'(r)(r-r_h)+\cdots
\end{align}
donde el primer término es cero dado que es la definición de horizonte. Así, el elemento de linea se verá como 
\begin{align}
    \diff s^2\approx f'(r_h)(r-r_h)\diff \tau^2+\frac{\diff r^2}{f'(r)(r-r_h)}+r_h^2\left(\diff \theta^2+\sin^2\theta\diff \phi^2\right)\, .
\end{align}
A continuación, haremos el siguiente cambio de coordenadas
\begin{align}
    \diff\rho^2=\frac{\diff r^2}{f'(r_h)(r-r_h)}\;\;\;\;\;\mbox{o bien}\;\;\;\;\; r-r_h=\frac{\rho^2 f'^2(r_h)}{4}\, ,
\end{align}
lo que nos entrega
\begin{align}
    \diff s^2\approx\frac{\rho^2 f'^2(r_h)}{4}\diff\tau^2+\diff\rho^2+r_h^2\left(\diff \theta^2+\sin\theta\diff \phi^2\right)\, .
\end{align}
Luego, es posible definir una nueva variable angular, en particular 
\begin{align}
    \psi\equiv\frac{f'(r_h)}{2}\tau \;\;\;\to\;\;\; \diff\psi^2=\frac{f'^2(r_h)}{4}\diff\tau^2 \, .
\end{align}
De este modo, la métrica de Schwarzschild cerca del horizonte se ve 
\begin{align}
    \diff s^2\approx\rho^2\diff\psi^2+\diff\rho^2+r_h^2\left(\diff \theta^2+\sin^2\theta\diff \phi^2\right)\, . \label{euclideo}
\end{align}
Como habíamos mencionado anteriormente, queremos evitar singularidades. En este caso, para evitar singularidades cónicas, vamos a pedir que el periodo de la nueva variable sea $\psi\sim\psi +2\pi$, lo que implica que $\tau\sim\tau+\beta_{\tau}$, en donde 
\begin{align}
    \beta_{\tau}=\frac{4\pi}{f'(r_h)}\, , \label{btau}
\end{align}
es el período del tiempo Euclídeo. De esta forma, el rango de las coordenadas de la métrica~\eqref{euclideo} es $0\leqslant\tau\leqslant\beta_{\tau}$, $r_{h}\leqslant r<\infty$, $0\leqslant\theta\leqslant\pi$ y $0\leqslant\phi\leqslant 2\pi$, lo cual garantiza que es completamente regular. De hecho, podemos notar que las hipersuperficies de $\rho$ constante son topológicamente $\mathbb{S}^1\times\mathbb{S}^2$. 

Luego de que nos hemos asegurado que la métrica esta libre de singularidades, la mecánica estadística nos permite relacionar el período del tiempo Euclídeo con el inverso de la temperatura, esto es, $\beta_{\tau}=T_{H}^{-1}$, en donde $T_H$ es la temperatura de Hawking del agujero negro. De hecho, esta cantidad coincide con la obtenida a través de la gravedad superficial, la cual se define como~\eqref{surfgrav2}.
En el caso de la métrica de Schwarzschild encontramos
\begin{align}
    T_{H}=\frac{1}{8\pi mG} \, .
\end{align}
Éste resultado nos será de utilidad para el cálculo de la termodinámica del agujero negro a partir de la relación entre la función partición y la acción Euclídea on-shell. Esta relación la obtendremos mediante el procedimiento que es mostrado a continuación.

\subsection{Aproximación de punto silla}
Consideremos la siguiente integral
\begin{align}
    I=\int^{\infty}_{-\infty}\diff x \, \,e^{-f(x)}\, ,
\end{align}
en donde $f(x):\mathbb{R}\to\mathbb{R}$ y sea $x=x_{0}$ un extremo de la función, i.e., 
\begin{align}
    f'(x)\big|_{x=x_{0}}=0\, .
\end{align}
Expandiendo $f(x)$ en torno a este punto, tenemos
\begin{align}
    f(x)=f(x_{0})+\cancelto{0}{f'(x_{0})(x-x_{0})}+\frac{1}{2}f''(x_{0})(x-x_{0})^{2}+\cdots \, .
\end{align}
De este modo, a segundo orden en la expansión, la integral queda
\begin{align}\notag
I&\approx\int^{\infty}_{-\infty}\diff x \, \,e^{-f(x_{0})-\frac{1}{2}f''(x_{0})(x-x_{0})^{2}}\\
&\approx e^{-f(x_{0})}\int^{\infty}_{-\infty}\diff x \, \,e^{-\frac{1}{2}f''(x_{0})(x-x_{0})^{2}} \label{gauss}\, ,
\end{align}
en donde hemos utilizado que la primera integral es una constante. Luego, para que la integral no sea divergente, debemos pedir que $f''(x_{0})>0$. Esto implica que $x_0$ es un mínimo de $f(x)$. De este modo, la Ec.~\eqref{gauss} es una Gaussiana y, por lo tanto, tiene un resultado conocido que es
\begin{align}
    I\approx e^{-f(x_{0})}\int^{\infty}_{-\infty}\diff x \, \,e^{-\frac{1}{2}f''(x_{0})(x-x_{0})^{2}}= e^{-f(x_{0})}\sqrt{\frac{2\pi}{f''(x_{0})}}\, .
\end{align}
A este procedimiento se le conoce como aproximación de punto silla. 

\subsection{Función partición para una solución asintóticamente plana} \label{funcionparticion}

Tomando en cuenta el resultado de la subseccion anterior, consideremos la función partición de un sistema escrito en el formalismo de integral de camino de la mecánica cuántica
\begin{align}
     \mathcal{Z}=\int \mathcal{D}\phi \,\, e^{i\int \diff^4 x \sqrt{\lvert g \lvert} \;\Lag[\phi]}\, ,\label{funcionp}
\end{align}
en donde $\int \mathcal{D}\phi$ es la integral sobre todas las configuraciones de campo $\phi$. Realicemos una rotación de Wick $t\to -i\tau$ de modo que el elemento de volumen transforma como $\diff^4 x \sqrt{\lvert g \lvert}\to -i\diff^4 x_{E}\sqrt{\lvert g_{E} \lvert}$, en donde $E$ denota cantidades definidas en el espacio Euclídeo. Así, la integral~\eqref{funcionp} se verá como
\begin{align}
    \mathcal{Z}=\int \mathcal{D}\phi \,\, e^{\int \diff^4 x_{E} \sqrt{\lvert g_{E} \lvert} \;\Lag[\phi_{E}]}\equiv\int \mathcal{D}\phi \,\, e^{-I^{\rm E}_{\rm cl}[\phi]}\, ,
\end{align}
con $I^{\rm E}_{\rm cl}[\phi]$ siendo la acción Euclídea clásica. Usemos la aproximación de punto silla para encontrar la función partición. Consideremos una configuración que minimice la acción Euclídea, por ejemplo, una solución a las ecuaciones de movimiento que denotaremos por $\phi=\phi_{0}$. Expandiendo la acción en torno a esta configuración obtenemos  
\begin{align}
    I^{\rm E}_{\rm cl}[\phi]&=I^{\rm E}_{\rm cl}[\phi_{0}]+\cancelto{0}{\frac{\delta I^{\rm E}_{\rm cl}}{\delta\phi}\big|_{\phi=\phi_{0}}(\phi-\phi_{0})}+\frac{1}{2}\frac{\delta^{2} I^{\rm E}_{\rm cl}}{\delta\phi^{2}}\big|_{\phi=\phi_{0}}(\phi-\phi_{0})^{2}+\cdots\, ,
\end{align}
en este caso el segundo término se anula dado que la variación funcional de la acción con respecto a los campos son las ecuaciones de movimiento, por lo tanto, son cero evaluadas en una solución. Luego, la función partición queda 
\begin{align}
    \mathcal{Z}&\approx \int \mathcal{D}\phi \,\, e^{-I^{\rm E}_{\rm cl}[\phi_{0}]-\frac{1}{2}\frac{\delta^{2} I^{\rm E}_{\rm cl}}{\delta\phi^{2}}\big|_{\phi=\phi_{0}}(\phi-\phi_{0})^{2}}=C\, e^{-I^{\rm E}_{\rm cl}[\phi_{0}]}\, ,
\end{align}
en donde $C$ es una constante que sin perdida de generalidad puede ser fijada a 1. Entonces, tomando el logaritmo en ambos lados de la ecuación anterior, llegamos a la siguiente expresión 
\begin{equation}
 \ln \mathcal{Z} \approx -I^{\rm E}_{\rm cl}[\phi_{0}]\, .
\end{equation}
A partir de esta relación, es posible obtener las cantidades termodinámicas del sistema usando las relaciones de la mecánica estadística. Por ejemplo, la energía libre, la energía interna y la entropía se pueden obtener de
\begin{align}\label{free}
    F&=-T\ln\mathcal{Z}\approx\beta^{-1}I^{\rm E}_{\rm cl}\, ,\\ 
    U&=-\frac{\partial}{\partial\beta}\ln\mathcal{Z}\approx\frac{\partial}{\partial\beta}I^{\rm E}_{\rm cl} \label{internal}\, ,\\ 
    S&=\beta U+\ln\mathcal{Z}\approx\left(\beta\frac{\partial}{\partial\beta}-1\right)I^{\rm E}_{\rm cl}\label{entropy1}\, ,
\end{align}
respectivamente. Estas relaciones son de gran utilidad para derivar las leyes de la termodinámica de los agujeros negros tal como se mostró en la Ref.~\cite{Gibbons:1976ue}.

Evaluemos la acción Euclídea on-shell para la solución~\eqref{euclids}. Primero, podemos ver que dicha métrica es solución de $R_{\mu\nu}=0$, lo cual implica $R=0$. Así, observamos que el término de volumen de la acción~\eqref{acciondir} se anula, dando como resultado solamente el término de borde on-shell, esto es,
\begin{align}
    I^{\rm E}_{\rm cl}=-2\kappa\int \diff^3 x \sqrt{\lvert h_E \rvert}\, (K-K_0)\, . \label{aeosch}
\end{align}
Aquí, hemos elegido una foliación radial, cuyo  vector normal unitario a las hipersuperficies con $r$ constante es $n=n^{\mu}\partial_\mu=\sqrt{f(r)}\partial_r$. Los valores de $K$ y $K_0$ ya los obtuvimos cuando calculamos la carga con el método de Brown-York, siendo estos
\begin{align}
    K&=\frac{f'(r)r+4f(r)}{2r\sqrt{f(r)}}\, \label{curvextri},\\
    K_0&=\frac{2}{r}\, ,
\end{align}
para Schwarzschild (con $f(r)=1-2mG/r$) y  Minkowski, respectivamente. Antes de evaluar la acción, debemos considerar que ambas métricas deben estar en el mismo ensamble a temperatura fija. Dicha condición se traduce a que ambos elementos de línea deben tener las mismas condiciones de borde. Con el fin de conseguir esto, debemos asegurarnos que las longitudes de arco de ambas métricas coincidan para toda hipersuperficie de $r$ constante. En este punto, podemos fijar $\theta$ y $\phi$ sin pérdida de generalidad, ya que $h_{\theta\theta}$ y $h_{\phi\phi}$ coinciden con $h^{(0)}_{\theta\theta}$ y $h^{(0)}_{\phi\phi}$; no así $h_{\tau\tau}$ con $h^{(0)}_{\tau\tau}$, donde $h^{(0)}_{\mu\nu}$ es la métrica inducida del background a $r$ constante. Pedir esta condición se visualiza de la siguiente manera 
\begin{align}\notag
\int \sqrt{\diff s^2}\big|_{r=R}&=\int \sqrt{\diff s^2_{(0)}}\big|_{r=R}\\ \notag
\int_{0}^{\beta_\tau}\sqrt{f(R)}\diff\tau&=\int_{0}^{\beta_0}\diff\tau\\
\beta_\tau\sqrt{f(R)}&=\beta_0 \label{periodo}\,
\end{align}
Esta condición fija $\beta_0$ en función de $\beta_\tau$. Con esto en mente, evaluemos la acción Euclídea on-shell~\eqref{aeosch}
\begin{align}\notag
I^{\rm E}_{\rm cl}&=-2\kappa\int \diff^3 x \sqrt{\lvert h_E \rvert}\, (K-K_0)\\ \notag
&=-2\kappa\int^{2\pi}_{0}\diff\phi\int^{\pi}_{0}\diff\theta\sin\theta\int^{\beta_\tau}_{0}\diff\tau\left[\sqrt{f(r)}r^2\left(\frac{f'(r)r+4f(r)}{2r\sqrt{f(r)}}-\frac{2}{r}\right)\right]\\ \notag
&=-8\pi\kappa\int^{\beta_\tau}_{0}\diff\tau\left(\frac{r^{2}f'(r)+4f(r)r}{2}-2\sqrt{f(r)}r\right)\\ \notag
&=-8\pi\kappa\int^{\beta_\tau}_{0}\diff\tau\left(mG+2r-4mG-2r\sqrt{1-\frac{2mG}{r}}\right)\\ \notag
&=-8\pi\kappa\int^{\beta_\tau}_{0}\diff\tau\left(mG+2r-4mG-2r\left\{1-\frac{1}{2}\left(\frac{2mG}{r}\right)-\mathcal{O}(r^{-2})\right\}\right)\\ \notag
&=8\pi\kappa\beta_\tau mG\\
 &=\frac{1}{2}\beta_\tau m\, ,
\end{align}
donde en el último paso reemplazamos el valor de $\kappa$. De la definición de horizonte y de la Ec.~\eqref{btau} podemos reescribir el resultado anterior en función del radio de Schwarzschild $r_h$, quedando
\begin{align}
    I^{\rm E}_{\rm cl}=\frac{1}{2}4\pi r_h\frac{r_h}{2G}=\frac{\pi r_{h}^2}{G}\, .
\end{align}
Luego, podemos calcular las cantidades termodinámicas de las Ecs.~\eqref{free},~\eqref{internal} y~\eqref{entropy1} , obteniendo así 
\begin{align}
    F&\approx\beta^{-1}I^{\rm E}_{\rm cl}=\frac{r_h}{4G}\, ,\\ 
    U&\approx\frac{\partial}{\partial\beta}I^{\rm E}_{\rm cl} =M\, ,\\ 
    S&\approx\left(\beta\frac{\partial}{\partial\beta}-1\right)I^{\rm E}_{\rm cl}=\frac{\pi r_{h}^2}{G}\, .
\end{align}
Vale la pena destacar que la entropía cumple con la ley del cuarto del área de la fórmula de Bekenstein-Hawking. Además, la energía interna coincide con el valor de la masa del agujero negro.

En síntesis, en este capítulo revisamos en la teoría de relatividad general, su principio variacional y distintos métodos para calcular cargas conservadas para una solución asintoticamente plana. Si bien, hemos obtenido resultados correctos y compatibles con la teoría veremos en los capítulos que siguen que al tratar con soluciones con distinta asíntota los métodos ya mencionados fallan y es necesaria una nueva prescripción para el cálculo de las cargas. Además, como estamos interesados en una teoría que está acoplada con materia vamos a añadir campos escalares y estudiar como se comporta dicha acción e introducir un nuevo método de renormalizacion que se basa en las simetrías de ésta.

\biblio 

\clearpage

\chapter{Gravedad en espacios anti-de Sitter}

Modificar la Relatividad General parece un camino evidente cuando buscamos explicar fenómenos en los que la teoría falla. Entender la expansión acelerada del universo, explicar la masa faltante de objetos a escalas cosmológicas y comprender como se relaciona la gravedad con la mecánica cuántica son las razones que suelen repetirse en este marco ya que son interrogantes fundamentales de la física moderna. La extensión mas sencilla de relatividad general en 4 dimensiones es agregar el termino de constante cosmológica, lo cual esta permitido por el teorema de Lovelock \cite{Lovelock:1971yv}. La acción de Einstein-Hilbert con constante cosmológica en cuatro dimensiones esta dada por
\begin{equation}
    I_{\text{AdS}}=\kappa \int_{\mathcal{M}} \diff^{4}x \sqrt{\lvert g\rvert}\, (R-2\Lambda) + 2 \kappa\sigma\int_{\partial\mathcal{M}} \diff^{3}x \sqrt{\lvert h\rvert}\, K\, \, ,\label{ehconl}
\end{equation}
cuyas ecuaciones de movimiento estan dadas por 
\begin{align}
R_{\mu\nu}-\frac{1}{2}g_{\mu\nu}R+\Lambda g_{\mu\nu} =0\, . \label{econlambda}
\end{align}

 Una solución a estas ecuaciones que es de particular interés en este trabajo es el espacio-tiempo Anti-de Sitter (AdS), el cual se define como un tipo de espacio-tiempo con curvatura constante negativa. Esta solución fue introducida por el matemático Élie Cartan a principios del siglo XX y posteriormente utilizado por físicos en la década de 1980 para modelar ciertos sistemas gravitatorios. Actualmente tiene relevancia tanto en teorías de gravitación como en teoría de cuerdas.

Una de las características más interesantes de AdS es su simetría conforme~\cite{balasubramanian1999stress}, tema que detallaremos con mas profundidad en las siguientes secciones. Esta simetría se puede expresar matemáticamente mediante la siguiente expresión 
\begin{equation}
g_{\mu\nu}(x) \rightarrow e^{2\omega}g_{\mu\nu}(x)
\end{equation}
donde $g_{\mu\nu}(x)$ es el tensor métrico, $\omega(x)$ es una función arbitraria, y $\mu$ y $\nu$ representan índices espacio-temporales.

Esta simetría tiene importantes implicancias para la holografía, un concepto teórico que sugiere que la información contenida dentro de un sistema físico puede representarse en un espacio de menor dimensión. El principio holográfico se puede formular matemáticamente utilizando la correspondencia AdS/CFT, que relaciona el comportamiento de los sistemas gravitatorios en AdS con el de teorías de campos conforme en dimensiones inferiores~\cite{witten1998anti}. Esta correspondencia se expresa matemáticamente mediante la siguiente ecuación:
\begin{equation}
Z_{\text{AdS}}[\phi_{0}] = Z_{\text{CFT}}[\phi_{0}]
\end{equation}
donde $Z_{\text{AdS}}[\phi_{0}]$ es la función de partición de un sistema gravitatorio en AdS con una métrica de borde fija $\phi_{0}$, y $Z_{\text{CFT}}[\phi_{0}]$ es la función de partición de una teoría de campo conforme en el borde con la misma métrica fija~\cite{maldacena1999large}.

\subsection{Espacio-tiempo Anti-de Sitter}

La estructura métrica de $\rm AdS_{d+1}$ en un sistema particular de coordenadas, está representada por un hiperboloide inmerso en un espacio tiempo de Minkowski $M^{d,2}$, con coordenadas
\begin{align}
    X^{\mu}=(X^{0},\cdots,X^{d+1})\, ,
\end{align}
y el elemento de línea 
\begin{align}
    \diff s^2=\eta_{\mu\nu}\diff X^{\mu}\diff X^{\nu} \;\;\;\;\;\; \mbox{con} \;\;\;\;\; \eta_{\mu\nu}=\rm diag(-,-,+,\cdots,+)\, . \label{adsline}
\end{align}
Así, el hiperboloide esta descrito por
\begin{align}
    -(X^{0})^2-(X^{1})^2+(X^{2})^2+\cdots+(X^{d+1})^2=-\ell^2 \, . \label{ads1}
\end{align}
La ecuación anterior es llamada $\rm AdS_{d+1}$ Minkowskiano y tiene la particularidad de que pareciera tener dos direcciones temporales, siendo $\ell\neq 0$ el radio de curvaturade AdS. 

Dado que $\rm AdS_{d+1}$ está construido inmerso en $M^{d,2}$, posee las mismas isometrías del espacio ambiente, en este caso $SO(d,2)$. Dicho grupo también es el grupo de simetrías del grupo conforme en una dimensión menor. Es importante resaltar que lo anterior fue el primer indicio de una relación entre AdS como espacio-tiempo y la simetría conforme en una dimensión menor~\cite{Brown:1986nw}, lo que más tarde contribuiría al desarrollo de la conjetura AdS/CFT. El grupo $SO(d,2)$ posee exactamente el número de generadores que el grupo de isometrías de $M^{d,1}$.
\begin{align}
    SO(d,2)\to M^{d,1}\to \frac{(d+1)(d+2)}{2}\, .
\end{align}
De esta forma, es evidente que $AdS_{d+1}$ es un espacio maximalmente simétrico, es decir, posee el número máximo de vectores de Killing. Adicionalmente, dado que $AdS_{d+1}$ es maximalmente simétrico se cumple lo siguiente para sus cantidades que describen la curvatura,
\begin{align}\notag
R_{\mu\nu\lambda\rho}&=-\frac{1}{\ell^2}(g_{\mu\lambda}g_{\nu\rho}-g_{\mu\rho}g_{\nu\lambda}) \, ,\\
R_{\mu\nu}&=\frac{d}{\ell^2}g_{\mu\nu}\, ,\label{adseins}\\ \notag
R&=\frac{-d(d+1)}{\ell^2} \, . 
\end{align}
Luego, de la Ec.~\eqref{adseins} se puede ver que $AdS_{d+1}$ es un espacio tipo Einstein, puesto que estos espacios se definen como una variedad en la cual su tensor de Ricci es proporcional a la métrica de dicha variedad. Por lo tanto, tal y como habíamos mencionado, $AdS_{d+1}$ es una solución a las ecuaciones de Einstein con constante cosmológica en el vacío \eqref{econlambda}.

\subsubsection{Coordenadas globales}
Podemos parametrizar~\eqref{adsline} de la siguiente forma
\begin{align} \notag
    X^{0}&=\ell\cosh{\rho}\cos{\tau}\\ \notag
    X^{1}&=\ell\cosh{\rho}\sin{\tau}\\ \notag
    X^{i}&=\ell\Omega_{i}\sinh{\rho} \,\,\,\,\,\mbox{con}\,\,\,\,\,i=2,\cdots,d+1\,\,\,\mbox{y}\,\,\, \Omega^{2}_{i}=1 \, ,
\end{align}
con $\rho\,\in\mathbb{R}_{+}$ y $\tau\,\in\,[0,2\pi]$. Así, podemos escribir en la forma conocida como coordenadas globales. El elemento de línea toma la forma 
\begin{align}
    \diff s^2=\ell^2(-\cosh^2{\rho}\,\diff\tau^2+\diff\rho^2+\sinh^2{\rho}\,\diff\Omega^2_{d-1})\, . \label{adsglobal}
\end{align}
Como podemos observar, la métrica no depende de $\tau$, por lo que posee un vector de Killing tipo tiempo. Sin embargo, al ser periódica en la misma coordenada tenemos curvas tipo tiempo cerradas. Para evitar lo anterior, podemos extender el rango de coordenadas tal que $\tau\,\in\,[-\infty,+\infty]$ y realizar el siguiente cambio de coordenadas
\begin{align}\notag
t&=\ell\tau \, , \\ \notag 
r&=\ell\sinh{\rho}\, ,
\end{align}
de esta forma, obtenemos
\begin{align}
\diff s^{2}=-\left(1+\frac{r^2}{\ell^2}\right)\diff t^{2}+\frac{\diff r^{2}}{\left(1+\frac{r^2}{\ell^2}\right)}+r^2 \diff\Omega^{2}_{d-1} \, . \label{ads4}
\end{align}
La ecuación anterior es la representación mas conocida del elemento de línea de AdS, la cual se conoce como coordenadas estáticas. Luego, podemos tomar el límite $r\to\infty$
\begin{align}
    \lim_{r\to\infty} \diff s^{2}=-\frac{r^{2}}{\ell^{2}}\diff t^{2}+r^2 \diff\Omega^{2}_{d-1}=\frac{r^2}{\ell^{2}}\left(-\diff t^{2}+\ell^2\diff\Omega^{2}_{d-1}\right)\, , \label{ads3}
\end{align}
donde es notorio que la métrica diverge. En el límite $r\to\infty$, para cualquier espacio AdS se cumple que: (i) el volumen es infinito, (ii) posee un polo de segundo orden y (iii) se induce una estructura conforme, lo que es evidente al mirar el factor en la Ec.~\eqref{ads3}. Dicha estructura se puede considerar como un segundo indicio de una relación con la simetría conforme. Luego, conviene ver el comportamiento de la Ec.~\eqref{ads1} para valores grandes de $X^{\mu}$, entonces $\rm AdS_{d+1}$ 
\begin{align}
-(X^{0})^2-(X^{1})^2+(X^{2})^2+\cdots+(X^{d+1})^2= 0\, ,
\end{align}
esta hipersuperficie es asintóticamente tipo luz y, al igual que los conos de luz, tiene la propiedad de ser invariante conforme. Esta propiedad es importante ya que, al aplicar una transformación conforme, se preservan los ángulos entre los vectores y, por lo tanto, se mantiene la estructura causal. Entonces, al tener una hipersuperficie tipo luz como comportamiento asintótico, se induce una estructura conforme en $\rm AdS_{d+1}$. Así, tenemos nuevamente un indicio de conexión con la simetría conforme.

Ahora, si partimos de las coordenadas globales de la Ec.~\eqref{adsglobal} y aplicamos el siguiente cambio de coordenadas 
\begin{align}
    \tan{\theta}\to\sinh{\rho}\, ,
\end{align}
el parche global tomará la forma
\begin{align}
    \diff s^2=\frac{\ell^2}{\cos^2{\theta}}\left(-\diff\tau^2+\diff\theta^2+\sin^2{\theta}\,\diff\Omega^{2}_{d-1}\right)\, . \label{ads5}
\end{align}
Ahora, analicemos los rangos de coordenadas de $\theta$. En este caso, el ángulo va de $0\leqslant\theta\leqslant\pi/2$, lo cual no cubre completamente el parche. Sin embargo, se cubre el sector donde $\theta=\pi/2$. Este valor es relevante, ya que el elemento de línea $\diff s^2$ diverge y además, es donde se localiza el borde conforme. Todo espacio asintóticamente AdS (AAdS) posee un borde conforme. Este borde es tipo tiempo y nos da información sobre la geometría de la variedad. En este caso, ya no es globalmente hiperbólica. Ahora, para ir de forma suave en el borde, podemos hacer el siguiente rescalamiento 
\begin{align}
 (\diff s')^2=\frac{\cos^2{\theta}}{\ell^2}\diff s^2=-\diff\tau^2+\diff\theta^2+\sin{\theta}\,\diff\Omega^{2}_{d-1}\, ,
\end{align}
donde $\theta=\pi/2=\partial\mathcal{M}\to\diff\theta^2=0$ y $\sin{\theta}=1$, así, el elemento de línea en el borde queda 
\begin{align}
(\diff s')^2\Big|_{\partial\mathcal{M}}= \diff\tau^2+\diff\Omega^{2}_{d-1}\, ,
\end{align}
entonces la topología del borde es un cilindro, i.e., $S^{d-1}\times\mathbb{R}$. En consecuencia, el borde es conformalmente plano, lo que se puede resumir como 
\begin{align}
    W^{\alpha\beta}_{\mu\nu}({\rm bulk})=W^{ij}_{km}({\rm boundary})=0 \, ,
\end{align}
con $W^{\alpha\beta}_{\mu\nu}$ el tensor de Weyl definido como 
\begin{equation}
W_{\mu \nu}^{\alpha \beta}= R_{\mu \nu}^{\alpha \beta} -4S^{[\alpha}_{[\mu} \delta^{\beta]}_{\nu]} \;\;\;\;\; \mbox{con} \;\;\;\;\; S_{\mu\nu} = \frac{1}{2} \left(R_{\mu \nu} - \frac{1}{6} g_{\mu \nu} R\right) \,,
\label{weyltensor}
\end{equation}
con $S_{\mu\nu}$ el tensor de Schouten.

\section{Espacios asintóticamente localmente anti-de Sitter}
Ahora, queremos definir un espacio asintóticamente localmente anti-de Sitter (AlAds) generalizando las nociones anteriores. Consideremos ahora la métrica de la Ec.~\eqref{ads5}, como mencionamos anteriormente posee un polo de orden 2 en $\theta=\pi/2$ y es el lugar donde se ubica el borde. Ya sabemos cómo proceder para que el elemento de línea sea finito considerando un reescalamiento. Sin embargo, este reescalamiento no es único, ya que podemos definir también 
\begin{align}\notag
 g'_{\mu\nu}=\frac{\cos^2{\theta}}{\ell^2}g_{\mu\nu} \;\;\;\;\; \mbox{ó} \;\;\;\;\; 
 g'_{\mu\nu}=\frac{\cos^2{\theta}}{\ell^2}e^{2\sigma(x)}g_{\mu\nu}\, .
\end{align}
Así, si $\sigma(x)$ es una función sin polos en el borde y sin ceros, ambas métricas son finitas modulo una transformación conforme. Esto induce una estructura conforme en el borde de AdS, i.e., $g'=\Omega^2 g$ y $\Omega(\partial \mathcal{M})=0\to d\Omega(\partial \mathcal{M})\neq 0$. Variedades con estas propiedades se conocen como espacios asintóticamente localmente anti-de Sitter (AlAds). Para corroborar que, en efecto un espacio AlAdS se aproxima a un espacio AdS con propiedades dadas por las Ec.~\eqref{adseins}, reemplazamos~\eqref{ads5} en las ecuaciones de Einstein, obteniendo
\begin{align}\notag
    R_{\mu\nu}&=-\diff|\diff\Omega|^2 g_{\mu\nu}+\mathcal{O}^{-2}(\Omega^{-1})\\
    R_{\mu\nu\rho\sigma}&=|\diff\Omega^2|(g_{\mu\rho}g_{\nu\sigma}-g_{\mu\sigma}g_{\nu\rho})+{O}(\Omega^{-3}) \, .
\end{align}
Notemos que dado que $g$ tiene un polo de segundo orden en $\partial \mathcal{M}$, el término de mayor orden en tensor de Riemann es de orden $\Omega^{-4}$. Las ecuaciones de campo de Einstein implican entonces que 
\begin{align}
    |\diff\Omega|^2=\frac{1}{\ell^2}\;\;\;\;\;\mbox{en}\;\;\;\;\; \partial\mathcal{M}\, .
\end{align}
De ello se deduce que el tensor de Riemann de un espacio-tiempo AlAdS cerca de $\partial\mathcal{M}$ se parece al de AdS puro.

\subsubsection{Expansión de Fefferman-Graham}
El término AdS asintóticamente (localmente) sugiere que la métrica del espacio-tiempo $g_{\mu\nu}$ debería (localmente) aproximarse a~\eqref{ads4}, al menos con una adecuada elección de coordenadas. Esto está lejos de manifestarse en las definiciones anteriores. Pero resulta ser una consecuencia de las ecuaciones de Einstein. De hecho, éstas ecuaciones implican que la estructura asintótica es descrita por 
la expansión Fefferman-Graham. La métrica en coordenadas de FG se ve como 
\begin{align}
\diff s^2=\frac{\ell^2}{4\rho^2}\diff\rho^2+\frac{1}{\rho}g_{ij}(\rho,x)\diff x^{i}\diff x^{j}\, .
\end{align}

Según el teorema de Fefferman-Graham, que tenga una estructura asintótica como AdS admite una expansión de Fefferman-Graham~\cite{fefferman2008ambient} (FG). Así, AdS siendo una solución a las ecuaciones de Einstein en vacío con constante cosmológica negativa, admitirá una expansión de FG de la forma
\begin{align}
g_{ij}(\rho,x)=g_{(0)ij}(x)+\rho g_{(2)ij}(x)+\rho^{2}g_{(4)ij}(x)+\cdots+\rho^{d/2}g_{(d)ij}    \, ,
\end{align}
para dimensiones impares del borde. Para dimensiones del borde par, por otro lado, la expansión es
\begin{align}
g_{ij}(\rho,x)=g_{(0)ij}(x)+\rho g_{(2)ij}(x)+\rho^{d/2}g_{(d)ij}+\rho^{d/2}\ln{\rho}\alpha_{(d)ij}(x)+\cdots\mathcal{O}(\rho^{d/2 +1})\, .
\end{align}
El término logarítmico de la ecuación anterior expresa las diferentes características geométricas entre tener borde par o impar. Además, el coeficiente $\alpha_{d}$ está relacionado con la anomalía conforme cuando la dimensión del borde es par. Ahora, si no se incluye éste término al imponer la condición de espacios Einstein la serie se estropea.  

\section{Renormalización para espacios AlAdS}
Luego de mencionar las propiedades de los espacios AdS y AlAdS, queremos probar algunos de los métodos para calcular cargas conservadas y renormalización en espacios que solucionan las ecuaciones de Einstein con constante cosmológica negativa. 

Primeramente, calculemos la acción Euclidea on-shell para la métrica de Schwarzschild-AdS, dada por
\begin{align}
    \diff s^2=-f(r)\diff t^2+\frac{\diff r^2}{f(r)}+r^2\left(\diff \theta^2+\sin^2\theta\diff \phi^2\right)\,\,\,\,\,\,\mbox{con}\,\,\,\,\,\, f(r)=1-\frac{2mG}{r}+\frac{r^2}{\ell^2}\, .
\end{align} \label{schwads}
Recordemos que debemos tomar en cuenta un background para eliminar las divergencias. Para este caso, tomaremos como espacio tiempo de referencia AdS global. Tomando la traza la de la Ec.~\eqref{econlambda} obtenemos $R=4\Lambda$. Así, reemplazando en la acción~\eqref{ehconl}, obtenemos 
\begin{align}
I^{\text{on-shell}}_{\text{AdS}}=2\Lambda\kappa\int \diff^{4}x \sqrt{|g|}+2\sigma\kappa\int \diff^{3}x \sqrt{|h|}(K-K_{0}) \, .
\end{align}
En este caso, si calculamos el término de borde, nos queda una contribución que va como $\mathcal{O}(r^{-2})$. Luego, si tomamos el límite de $r\to\infty$ el factor correspondiente al borde no aporta al valor de la acción Euclídea on-shell. Sin embargo, el volumen como es proporcional al determinante de la métrica posee una divergencia del orden de $\mathcal{O}(r^{3})$, i.e., diverge como el volumen de AdS. 
Ahora, siguiendo el mismo espíritu del método de substracción de background, restaremos a la acción resultante el volumen de AdS global con el fin de eliminar la divergencia del término de volumen~\cite{Hawking:1982dh}. Primero, debemos fijar las longitudes de arco de la métrica, tal y como en el calculo de la acción Euclídea en la Ec.~\eqref{periodo}. Lo anterior se debe a que AdS global no tiene horizonte, así en virtud que ambas métricas estén el mismo ensamble debemos fijar el periodo del tiempo Euclídeo de la siguiente manera
\begin{align}\notag
\int \sqrt{\diff s^2}\big|_{r=R}&=\int \sqrt{\diff s^2_{(0)}}\big|_{r=R}\\ \notag
\int_{0}^{\beta_\tau}\sqrt{g^{\rm BH}_{\tau\tau}}\diff\tau&=\int_{0}^{\beta_0}\diff\tau\sqrt{g^{\rm AdS}_{\tau\tau}}\\ \notag
\beta_\tau\sqrt{\left(1-\frac{2mG}{R}+\frac{R^2}{\ell^2}\right)}&=\beta_0\sqrt{1+\frac{R^2}{\ell^2}}\\
\beta_0&=\frac{\sqrt{\left(1-\frac{2mG}{R}+\frac{R^2}{\ell^2}\right)}}{\sqrt{1+\frac{R^2}{\ell^2}}}\beta_{\tau} \, .
\end{align}
Ahora, la integral de la acción será
\begin{align}\notag
I^{\text{on-shell}}_{\text{AdS}}&=2\Lambda\kappa\int \diff^{4}x \sqrt{|g^{\rm BH}|}-2\Lambda\kappa\int \diff^{4}x \sqrt{|g^{\rm AdS}|}\\\notag
&=2\Lambda\kappa\left[\int^{\beta_{\tau}}_{0}\diff\tau\int^{2\pi}_{0}\diff\phi\int^{\pi}_{0}\sin{\theta}\diff\theta\int^{R}_{r_h}r^{2}\diff r-\int^{\beta_0}_{0}\diff\tau\int^{2\pi}_{0}\diff\phi\int^{\pi}_{0}\sin{\theta}\diff\theta\int^{R}_{0}r^{2}\diff r\right]\\\notag
&=8\pi\kappa\beta_{\tau}\Lambda\frac{1}{3}(R^3-r_{h}^3)-8\pi\kappa\beta_{0}\Lambda\frac{1}{3}R^3\\ \notag
&=\frac{8\pi\kappa\beta_{\tau}\Lambda}{3}\left(R^3-r_{h}^3-\frac{\sqrt{\left(1-\frac{2mG}{R}+\frac{R^2}{\ell^2}\right)}}{\sqrt{1+\frac{R^2}{\ell^2}}}R^3\right)\\ \notag
&=\lim_{R \to \infty}\frac{8\pi\kappa\beta_{\tau}\Lambda}{3}(R^3-r_{h}^3-R^3+mG\ell^2+\mathcal{O}(R^{-2}))\\
&=\frac{8\pi\kappa\beta_{\tau}}{\ell^2}(mG\ell^2-r_{h}^3)\, \label{AEOAdS}
\end{align}
donde en la penúltima linea expandimos en serie de Taylor el factor que acompaña a $R^3$. De este valor de la acción Euclídea on-shell usando las relaciones presentadas en la sección~\ref{sec:AEO} podemos obtener las cantidades termodinámicas del sistema. 

Ahora, veamos como se aplica el formalismo de Brown-York para el agujero negro de Schwarzschild-AdS.

Siguiendo el mismo análisis de la sección~\ref{sec:BY} vamos a calcular la masa del agujero negro a la Brown-York. El valor de la curvatura extrínseca del black hole y del background vienen dadas por la Ec.~\eqref{curvextri}, donde
\begin{align}
f(r)_{\rm BH}&=1-\frac{2mG}{r}+\frac{r^2}{\ell^2}\, ,\\
f(r)_{\rm AdS}&=1+\frac{r^2}{\ell^2}\, .
\end{align}
Luego, reemplazando los valores anteriores y el vector normal unitario definido en la sección~\ref{sec:BY} en la Ec.~\eqref{qby}, la masa va como 
\begin{align}
    Q[\xi]=\frac{r^3}{2G\ell^2}+\frac{1}{2}M+\frac{M\ell^2}{2r^2}+\mathcal{O}\left(\frac{1}{r^3}\right)\, .
\end{align}
Por ende, la carga es divergente para espacios AlAdS. De hecho, el método de substracción de background no permite cancelar las divergencias en este formalismo. Es por ello que necesitamos de una prescripción distinta.

\subsection{Renormalización holográfica}

Con el propósito de resolver el problema que de la sección anterior, Balasubramanian y Kraus \cite{balasubramanian1999stress} propusieron una manera para renormalizar el tensor de stress cuasilocal en espacios AlAdS agregando una serie de términos de borde construidos a partir de cantidades intrínsecas llamados contratérminos. Estos se añaden a la acción y cancelan las divergencias problemáticas para los cálculos de cargas. Adicionalmente, coinciden con los propuestos en las Refs.~\cite{Emparan:1999pm,deHaro:2000vlm,Skenderis:2002wp} inspirados en la renormalizacion de la acción Euclidea on-shell en el marco de la correspondencia AdS/CFT. Esta serie, para una variedad en 4 dimensiones, se ve de la forma\footnote{En general, los coeficientes de la serie de contraterminos dependen de la dimensionalidad del espacio y podría contener términos de curvatura superior~\cite{Emparan:1999pm,deHaro:2000vlm,Skenderis:2002wp}.}
\begin{equation}
    I_{ct}=\frac{1}{8\pi G}\int_{\partial\mathcal{M}}\diff^{3}x\sqrt{\lvert h\rvert}\left[\frac{2}{\ell}+\frac{\ell}{2}\mathcal{R} \right]  \, ,\label{ct}
\end{equation}
donde $\ell$ es el radio de curvatura de AdS que se relaciona con la constante cosmológica en cuatro dimensiones vía $\Lambda=-\tfrac{3}{\ell^2}$ y $\mathcal{R}$ es la curvatura intrínseca que se define a través de la relación de Gauss-Codazzi como
\begin{equation}
    \mathcal{R}^{\rho}\hspace{0.5ex}_{\sigma\mu\nu}=h^{\rho}_{\alpha}h^{\beta}_{\sigma}h^{\gamma}_{\mu}h^{\delta}_{\nu}R^{\alpha}\hspace{0.5ex}_{\beta\gamma\delta}+2\sigma K^{\rho}\hspace{0.3ex}_{[\mu}K_{\nu]\hspace{0.3ex}
    \sigma} \,.
\end{equation}
Así, la acción para AdS renormalizada con contraterminos es
\begin{align}
    I_{\rm AdS}+I_{\rm ct}=\kappa \int_{\mathcal{M}} \diff^{4}x \sqrt{\lvert g\rvert}\, (R-2\Lambda) + 2 \kappa\sigma\int_{\partial\mathcal{M}} \diff^{3}x \sqrt{\lvert h\rvert}\, K+2 \kappa\int_{\partial\mathcal{M}}\diff^{3}x\sqrt{\lvert h\rvert}\left[\frac{2}{\ell}+\frac{\ell}{2}\mathcal{R} \right]\, . 
\end{align}
Asimismo, de la Ref.~\cite{balasubramanian1999stress}, el tensor de energía-momentum cuasilocal en 4 dimensiones toma la forma 
\begin{align}
\tau_{\mu\nu}=-2\kappa\sigma\left[K_{\mu\nu}-h_{\mu\nu}K\right] -4\sigma\kappa\left[\frac{1}{\ell}h_{\mu\nu}-\frac{\ell}{2}\left(\mathcal{R}_{\mu\nu}-\frac{1}{2}h_{\mu\nu}\mathcal{R}\right)\right]\, .
\end{align}
De la definición anterior, la carga conservada asociada es 
\begin{align}
Q[\xi]=\int_{\Sigma}\diff^{2}x\sqrt{\gamma}u_{\mu}\xi_{\nu}\tau^{\mu\nu}\, ,
\end{align}
de manera que, si calculamos la integral anterior obtenemos
\begin{align}
    Q[\xi]=M\, . 
\end{align}

Es importante mencionar que con este método de renormalizacion también es posible obtener una acción Euclídea on-shell finita, obteniendo el mismo valor que la Ec.~\eqref{AEOAdS}.

\subsection{Renormalización con términos topológicos}
Ya vimos que es necesario un esquema de renormalizacion diferente a la substracción de background para calcular cantidades conservadas en la acción de Einstein-Hilbert con constante cosmológica. Existe otra prescripción para renormalizar que considera términos extrínsecos o Kounterterms, los cuales están construidos en función de la curvatura extrínseca de la variedad. En general, los métodos de renormalizacion holográfica y topológica no coinciden en cuanto al valor de las cargas. De hecho, difieren por el tensor de Weyl del borde, excepto en el caso de 4 dimensiones, ya que en este caso los bordes son conformalmente planos, i.e., $W=0$. Adicionalmente, en 4 dimensiones se puede probar que la renormalización topológica es equivalente (modulo la característica de Euler) a sumar el termino de Gauss-Bonnet \cite{Aros:1999id,Aros:1999kt,Olea_2007,Olea:2005gb}, es decir,
\begin{equation} 
    I_{EH}^{ren}=\kappa \int_{\mathcal{M}} d^{4}x \sqrt{\lvert g\rvert} (R-2\Lambda) + \kappa \alpha_{0}\int_{\mathcal{M}} d^{4}x \sqrt{\lvert g\rvert}\,\mathcal{G} \, , \label{IEGB}
\end{equation}
en donde para el valor $\alpha_{0}=\tfrac{\ell^2}{4}$  \eqref{IEGB} es finita, ya que de forma general diverge \cite{Olea:2005gb,Olea_2007,Aros:1999id,Aros:1999kt}. Hemos definido al Gauss-Bonnet como
\begin{align}
    \mathcal{G}=R^2 -4R^{\mu\nu}R_{\mu\nu}+R^{\mu\nu\lambda\rho}R_{\mu\nu\lambda\rho} \, .
\end{align}
En cuatro dimensiones, este término no cambia las ecuaciones de movimiento ya que se relaciona con la característica de Euler, $\chi(\mathcal{M})$, a través del teorema de Chern-Gauss-Bonnet
\begin{equation}
    \int d^{4}x \sqrt{\lvert g\rvert}\,\mathcal{G}=32\pi^2\chi(\mathcal{M}) + \int_{\partial{\mathcal{M}}}d^{3}x\sqrt{\lvert h\rvert}\,\mathcal{B}\,,
\end{equation}
en donde la forma de Chern está definida como
\begin{align}
    \mathcal{B}&=-4\delta^{\alpha\beta\gamma}_{\mu\nu\lambda}K^{\mu}_{\alpha}\left(\frac{1}{2}\mathcal{R}^{\nu\lambda}_{\beta\gamma}-\frac{1}{3}K^{\nu}_{\beta}K^{\lambda}_{\gamma}\right)\,.
\end{align}

Para valores arbitrarios de $\alpha_{0}$, la acción \eqref{IEGB} es divergente para espacios Einstein con comportamiento AlAdS, i.e.,
\begin{align}
    R_{\mu\nu}=-\frac{3}{\ell^2} g_{\mu\nu} \, . \label{ES}
\end{align}
Sin embargo, cuando el acoplamiento del Gauss-Bonnet toma el valor de $\alpha_{0}=\tfrac{\ell^2}{4}$, la acción puede ser escrita de la forma conocida como MacDowell-Mansouri \cite{MacDowell:1977jt}, esto es,
\begin{align}
    \label{MacdowellMansouri}
    I_{MM} &= \frac{\kappa \ell^2}{16}\int_{\mathcal{M}} d{^4x}\sqrt{\lvert g\rvert}\delta^{\mu_1\ldots\mu_4}_{\nu_1\ldots\nu_4}\left(R^{\nu_1\nu_2}_{\mu_1\mu_2} + \frac{1}{\ell^2}\delta^{\nu_1\nu_2}_{\mu_1\mu_2} \right)\left(R^{\nu_3\nu_4}_{\mu_3\mu_4} + \frac{1}{\ell^2}\delta^{\nu_3\nu_4}_{\mu_3\mu_4} \right) \, .
\end{align}
Por otro lado, la descomposición del tensor de Riemann en términos de sus piezas irreducibles nos entrega que puede ser escrito como
\begin{align}
    R^{\mu\nu}_{\lambda\rho}=W^{\mu\nu}_{\lambda\rho}+4S^{[\mu}_{[\lambda}\delta^{\nu]}_{\rho]}\, , \label{desc}
\end{align}
en donde $W^{\mu\nu}_{\lambda\rho}$ es el tensor de Weyl, mientras que  $S^{\mu}_{\nu}=g^{\mu\lambda}S_{\nu\lambda}$ es el tensor de Schouten.
Ahora, usando la condición \eqref{ES}, el tensor de Schouten es $S_{\mu\nu}=-\tfrac{1}{2\ell^2}g_{\mu\nu}$. Por lo tanto, el tensor de Weyl toma la forma 
\begin{align}
    W^{\mu\nu}_{\lambda\rho(E)} = R^{\mu\nu}_{\lambda\rho} + \frac{1}{\ell^2}\delta^{\mu\nu}_{\lambda\rho} \equiv \mathcal{F}^{\alpha \beta}_{\mu \nu}  \, .
\end{align}
El lado derecho de esta ecuación representa las piezas Riemannianas de la curvatura de gauge para el grupo de AdS. Así, la acción~\eqref{MacdowellMansouri} evaluada en un espacio Einstein~\eqref{ES} queda de la siguiente manera
\begin{equation}
    I_{MM} = \frac{\kappa\ell^2}{4}\int d{^4x}\sqrt{\lvert g\rvert}\;W^{\mu\nu}_{\lambda\rho(E)}W^{\lambda\rho}_{\mu\nu(E)} \,. \label{IMM1}
\end{equation}
 Otra característica interesante de la acción \eqref{IMM1}  es  su relación con el invariante conforme en 4 dimensiones $W^{\mu\nu}_{\lambda\rho}W^{\lambda\rho}_{\mu\nu}$.
 Esto nos invita a pensar que podría existir una relación entre la simetría conforme y la renormalizacion de acción y cargas conservadas en espacios AlAdS.

\biblio 

\clearpage
  
\chapter{Gravedad conforme}
     
Conformal Gravity fue propuesta como una teoria de gravitación modificada que podría explicar las curvas de rotación de las galaxias sin la necesidad de materia oscura~\cite{Mannheim:1988dj}. También se puede probar que en espacios AlAdS, para condiciones de borde tipo Neumann para los coeficientes de la expansión de Fefferman-Graham, la gravedad de Einstein con constante cosmológica emerge de conformal gravity en cuatro dimensiones \cite{https://doi.org/10.48550/arxiv.1105.5632}. Se ha demostrado en la Ref.~\cite{Grumiller_2014} que conformal gravity tiene acción on-shell y cargas conservadas finitas para espacios AlAdS sin necesidad de añadir contraterminos \cite{Anastasiou:2017rjf}. Adicionalmente, también se comprobó que lo anterior se cumple para Conformal Gravity en seis dimensiones \cite{Anastasiou:2021tlv}. Esto permite embeber todo el espacio de soluciones de Relatividad General en una teoría que es renormalizable para espacios AlAdS. El espacio de soluciones de CG contiene todos los espacios de Einstein. De hecho, la acción de CG se vuelve igual a la de la gravedad de Einstein-AdS renormalizada cuando se evalúa en espacios de Einstein~\cite{Anastasiou:2016jix}. En el caso de las variedades de AlAdS, la condición de Einstein se puede implementar, hasta el orden normalizable, imponiendo condiciones de borde tipo Neumann en la expansión de FG~\cite{Maldacena:2011mk}. Además, esta teoría tiene la notable propiedad de ser finita y poseer un principio variacional bien definido para espacios de AlAdS~\cite{Grumiller:2013mxa}.

\section{Principio de acción y ecuaciones de movimiento}
La acción de esta teoría representa el único funcional de cuatro dimensiones construido únicamente en términos de la métrica que permanece invariante bajo reescalamientos de Weyl locales $g_{\mu\nu}\to\tilde{g}_{\mu\nu} = e^{2\sigma(x)}g_{\mu\nu}$. Como consecuencia, la teoría involucra términos de altas derivadas lo que la hace patológica debido a la presencia de fantasmas. Sin embargo, dado que las teorías de gravedad de altas derivada tienen mejores propiedades de renormalización que la gravedad de Einstein~\cite{Capper:1975ig,Stelle:1976gc,Julve:1978xn,Fradkin:1981iu}, son considerados modelos útiles para la gravedad cuántica. La acción de conformal gravity está dada por el funcional
\begin{equation}
I_{\rm CG} = \alpha_{\rm CG} \int\diff{^4x}\sqrt{|g|} \;W^{\alpha \beta}_{\mu \nu} W^{\mu \nu}_{\alpha \beta} \,,
\label{ICGaction}
\end{equation}
donde $\alpha_{\rm CG}$ es una constante de acoplamiento adimensional, mientras que $W^{\mu\nu}_{\lambda\rho}$ es el tensor de Weyl definido en la Ec.~\eqref{weyltensor}. 
Aquí, los índices griegos nos indican el parche de coordenadas del volumen. Las ecuaciones de movimiento, que son de cuarto orden, se obtienen realizando variaciones arbitrarias de la Ec.~\eqref{ICGaction} con respecto a la métrica, lo que nos da $B_{\mu\nu}=0$, donde
\begin{align}
B_{\mu \nu} = -4 \left(\nabla^{\lambda} C_{\mu \nu \lambda} + S^{\lambda \sigma} W_{\mu \lambda \nu \sigma } \right)\;\;\;\;\; \mbox{y} \;\;\;\;\;
C_{\mu \nu \lambda} = \nabla_{\lambda} S_{\mu \nu} - \nabla_{\nu} S_{\mu \lambda} \,, 
\end{align}
son el tensor de Bach y Cotton, respectivamente. Por lo tanto, el espacio de soluciones de la teoría corresponde a espacio-tiempos Bach flat. Además, se puede observar que los espacios tipo Einstein satisfacen automáticamente esta condición. Ésto se debe a que la divergencia del tensor de Cotton se anula en virtud de que el tensor de Schouten es proporcional a la métrica, mientras que el segundo término del tensor de Bach se anula ya que el tensor de Weyl tiene traza nula en todos sus índices. Así, todos los espacios Einstein son Bach flat, aunque la afirmación inversa no es necesariamente verdadera. Por otro lado, aunque algunos espacios Bach flat  son conformemente Einstein, existen ejemplos donde esta condición no se cumple y no se pueden relacionar con soluciones en la gravedad de Einstein mediante una transformación conforme~\cite{Liu:2013fna,Dunajski:2013zta}. Existen distintas clases de soluciones a conformal gravity, entre ellas estan las conformalmente planas ($W^{\mu\nu}_{\lambda\rho}=0$), Bach-flat que no son Einstein \cite{Riegert:1984zz}, instantones gravitacionales \cite{Corral:2021xsu} y métricas de la familia Plebia\'nski-Demia\'nski \cite{Cisterna_2021}.

Una de las características más interesantes de CG es la finitud de la acción cuando se evalúa en espacio-tiempos de AlAdS. Es decir, las divergencias usuales que surgen en la acción gravitacional debido al volumen infinito de los espacios AdS están ausentes en el caso de CG y no se necesitan contratérminos adicionales. De hecho, como se muestra en la referencia~\cite{Grumiller:2013mxa} para condiciones asintóticamente débiles en el borde de AdS, tanto el tensor de energia-momentum cuasilocal como la función de respuesta parcialmente sin masa son finitos; estas son las corrientes acopladas correspondientes a los gravitones masivos y sin masa, respectivamente. 

El comportamiento asintótico de la acción de CG también se puede estudiar considerando argumentos de conteo de potencias. En particular, consideramos las condiciones genéricas de AlAdS que en el gauge de FG adquieren la forma
\begin{align}
ds^{2}&= \frac{\ell^2}{z^2} \left(dz^{2}+ \mathcal{G}_{i j} \left(z,x\right) dx^{i} dx^{j} \right)\,, \notag \\
\mathcal{G}_{i j} \left(z,x\right) &= g_{\left(0\right) ij} \left(x\right) + \frac{z}{\ell} g_{\left(1\right) ij} \left(x\right) + \frac{z^2}{\ell^2} g_{\left(2\right) ij} \left(x\right) + \frac{z^3}{\ell^3} g_{\left(3\right) ij} \left(x\right)+ \ldots \,,
\end{align}
Donde $z$ es la coordenada radial, $\ell$ es el radio de AdS, y los índices latinos indican las coordenadas en la hipersuperficie de $z$ constante de codimensión-1. Aquí, $z=0$ denota la ubicación del borde conforme. Ésta estructura define una foliación ADM radial que nos permite descomponer el cuadrado del tensor de Weyl en términos de las tres contribuciones independientes; ellas son
\begin{equation}
W^{\alpha \beta}_{\mu \nu} W^{\mu \nu}_{\alpha \beta} = W^{ij}_{km} W^{km}_{ij} + 4 W^{iz}_{jz} W^{jz}_{iz} + 4 W^{iz}_{km}W^{km}_{iz} \,.
\end{equation}
Curiosamente, para las condiciones genéricas de AlAdS descritas anteriormente, todas las componentes independientes del tensor de Weyl decaen como $\mathcal{O} \left(z^2\right)$. Por lo tanto, el Lagrangiano de CG decae como $\mathcal{O} \left(z^4\right)$, lo que lleva a que la acción se comporte como
\begin{equation}
\int\diff{^4x}\sqrt{|g|}\; W^{\alpha \beta}_{\mu \nu} W^{\mu \nu}_{\alpha \beta} \sim \int \diff{^3x} \int \diff{z}\; \frac{\sqrt{|g_{\left(0\right)}|}}{z^4}\; \mathcal{O} \left(z^4\right) \sim \mathcal{O} \left(z\right) \,.
\end{equation}
Esto último indica que la acción de CG está libre de cualquier divergencia en la región infrarroja, de acuerdo con la Ref.~\cite{Grumiller:2013mxa}. Éste comportamiento de la acción de CG pone de manifiesto la relación entre la simetría conforme del volumen y la renormalización, no solo para CG, sino también para cada subsector del espacio de soluciones de la teoría. De hecho, para los espacio-tiempos tipo Einstein-AdS, donde $S_{\mu \nu} = -\frac{1}{2\ell^2} g_{\mu \nu}$, el tensor de Weyl coincide con la curvatura del grupo AdS sin torsión, $\mathcal{F}^{\alpha \beta}_{\mu \nu}$, dado por 
\begin{equation}\label{WeylE}
W^{\alpha \beta}_{\left(E\right) \mu \nu} = R^{\alpha \beta}_{\mu \nu} + \frac{1}{\ell^2}  \delta^{\alpha \beta}_{\mu \nu} \equiv \mathcal{F}^{\alpha \beta}_{\mu \nu} \,.
\end{equation}
Esta relación indica que la acción de CG evaluada para espacio-tiempos tipo Einstein se reduce a la acción de MacDowell-Mansouri para el grupo AdS \cite{MacDowell:1977jt} como vimos anteriormente. Esta última acción corresponde a la acción de Einstein-AdS renormalizada topológicamente\footnote{De hecho, la acción resultante en la Ec. \eqref{IMM} está desplazada por un término constante que involucra la característica de Euler de la variedad, que surge naturalmente del esquema de renormalización con Kounterterms y coincide con el volumen renormalizado \cite{Anastasiou:2018mfk}.} \cite{Miskovic:2009bm}
\begin{equation}
I_{\rm CG} \left[E\right] = I^{\rm (ren)}_{\rm EAdS}=\frac{\ell^2}{256 \pi G_{N}} \int\diff{^4x}\sqrt{|g|} \delta^{\mu_1\ldots\mu_4}_{\nu_1\ldots\nu_4}\mathcal{F}^{\nu_1\nu_2}_{\mu_1\mu_2}\mathcal{F}^{\nu_3\nu_4}_{\mu_3\mu_4} \,,
\label{IMM}
\end{equation}
lo que es equivalente al tratamiento con renormalizacion holográfica. \cite{Anastasiou:2020zwc}.
Es importante destacar que la acción de la Ec. \eqref{IMM} tiene un principio variacional de Dirichlet bien definido para la métrica en el borde conforme $g_{(0)ij}$ \cite{Anastasiou:2019ldc}. Esto concuerda con la Ref. \cite{Papadimitriou:2005ii}, que mostró que la finitud y la buena formulación del principio variacional en términos de las fuentes holográficas están relacionadas.

Además, la Ec. \eqref{IMM} sugiere que los contratérminos de la gravedad de Einstein-AdS están dictados por la simetría conforme del volumen, lo que introduce el concepto de Renormalización Conforme. Por otro lado, se puede demostrar la finitud de la acción de MacDowell-Mansouri \eqref{IMM} utilizando la relación genérica off-shell entre $\mathcal{F}^{\alpha \beta}_{\mu \nu}$ y el tensor de Weyl. Esto se expresa mediante la ecuación
\begin{equation}
W^{\alpha \beta}_{\mu \nu} = \mathcal{F}^{\alpha \beta}_{\mu \nu}- X^{\alpha \beta}_{\mu \nu} \,,
\label{weylfdecomposition}
\end{equation}
\begin{equation}\label{Xtensor}
X^{\alpha \beta}_{\mu \nu} = 2 H^{[\alpha}_{[\mu} \delta^{\beta]}_{\nu]}+ \frac{1}{12} \left(R+\frac{12}{\ell^2} \right) \delta^{\alpha \beta}_{\mu \nu} \,,
\end{equation}
con $H^{\alpha}_{\mu}=R^\alpha_\mu - \tfrac{1}{4}\delta^\alpha_\mu\,R $ siendo el tensor de Ricci sin traza. Así, reemplazando en la Eq.~\eqref{IMM}, se obtiene
\begin{equation}\label{IEHren}
I^{\rm (ren)}_{\rm EAdS} = \frac{\ell^2}{256 \pi G_{N}} \int\diff{^4x}\sqrt{|g|} \left[\delta^{\mu_1\ldots\mu_4}_{\nu_1\ldots\nu_4}W^{\nu_1\nu_2}_{\mu_1\mu_2} W^{\nu_3\nu_4}_{\mu_3\mu_4}+ 8 H^{\mu}_{\nu}H^{\nu}_{\mu}+\frac{2}{3} \left(R+\frac{12}{\ell^2}\right)^{2}\right] \,.
\end{equation}
Esta acción se reduce exactamente a la de la Ec. \eqref{ICGaction} para espacios tipo Einstein, como se puede ver al notar que los dos últimos términos de la Ec. \eqref{IEHren} se anulan automáticamente bajo esta condición. Por lo tanto, dado que la acción de CG es finita en general para asintóticas AdS débiles, se concluye que la expresión obtenida está renormalizada para espacios de Einstein-AdS.

Si bien Conformal Gravity es una teoría finita, se ha demostrado que posee modos masivos de spin-2 con energía cinética negativa \cite{Lu:2011zk}. La acción \eqref{ICGaction} describe en general un campo de spin-2 sin masa asociado al gravitón, y un campo de spin-2 masivo.

\subsection{Critical gravity}\label{sec:Critical}
Consideremos la acción de gravedad cuadrática mas general en 4 dimensiones. Esta teoría está motivada desde el ángulo de teorías efectivas, en donde la Relatividad General aparece como un límite de bajas energías de una completitud ultravioleta de la gravedad. La acción viene dada por
\begin{equation}
     I_{CRG}=\kappa\int_{\mathcal{M}} d^{4}x \sqrt{\lvert g\rvert} \left(R-2\Lambda +\beta R^2 +\alpha R^{\mu\nu}R_{\mu\nu}\right) \, .\label{ICRG}
\end{equation}
Interesantemente, fue probado en la Ref.~\cite{Stelle:1976gc} que esta teoría con $\Lambda=0$ es renormalizable perturbativamente en torno al espacio de Minkowski. Sin embargo, considerar términos cuadráticos en la curvatura que aportan factores de derivadas superiores, en general inducen la existencia de ghosts en el espectro~\cite{Stelle:1976gc}. Para el caso de esta teoría, a diferencia de Conformal Gravity, tenemos un modo escalar masivo además de los ya mencionados anteriormente. Esto se puede ver escribiendo las ecuaciones de movimiento linealizadas y analizando los operadores de onda asociados a cada modo. Podemos encontrar un sector no patológico eligiendo las constantes de acoplamiento de manera que los modos masivos se desacoplen del espectro.

 Para encontrar los valores de las constante de acoplamiento en el sector no patológico, los autores de la  Ref.~\cite{Lu:2011ks} examinaron las ecuaciones de movimiento linealizadas en torno a AdS global en cuatro dimensiones. Las ecuaciones de movimiento en el régimen no lineal están dadas por
\begin{align}
    G_{\mu\nu}+\Lambda g_{\mu\nu}+E_{\mu\nu}&=0 \, ,
    \end{align}
en donde hemos definido
\begin{align}
    G_{\mu\nu}&=R_{\mu\nu}-\frac{1}{2}Rg_{\mu\nu}\,,\\
    E_{\mu\nu}&=2\alpha\left(R_{\mu\rho}R^{\rho}_{\nu}-\frac{1}{4}R^{\rho\sigma}R_{\rho\sigma}g_{\mu\nu}\right)+2\beta R\left(R_{\mu\nu}-\frac{1}{4}Rg_{\mu\nu}\right)\nonumber\\
    &+\alpha\left(\Box R_{\mu\nu}+\frac{1}{2}\Box Rg_{\mu\nu}-2\nabla_{\rho}\nabla_{(\mu}R_{\nu)}^{\rho}\right)+2\beta(g_{\mu\nu}\Box R-\nabla_{\mu}\nabla_{\nu}R)\,.
\end{align}
Escribiendo una perturbación de la métrica como $g_{\mu\nu}=\bar{g}_{\mu\nu}+h_{\mu\nu}$, en donde $\bar{g}_{\mu\nu}$ es la métrica del background e imponiendo el gauge $\nabla^{\mu}h_{\mu\nu}=\nabla_{\nu}h$, se encuentra una condición para la cual el modo escalar se desacopla, esta es
\begin{align}
    \alpha=-3\beta \, . \label{CCG}
\end{align}
En este caso, las ecuaciones de movimiento nos dicen que $h=0$. Como consecuencia de esta restricción, los términos cuadráticos  en la curvatura se pueden escribir como 
\begin{align}
     I_{CRG} = \kappa\int_{\mathcal{M}} d^{4}x \sqrt{\lvert g\rvert} \left[R-2\Lambda  -3\beta\left(R^{\mu\nu}R_{\mu\nu}-\frac{1}{3}R^2\right)\right] \, .
\end{align}
Esta acción puede ser reescrita utilizando la identidad off-shell 
\begin{align}
    W^{\mu\nu\lambda\rho}W_{\lambda\rho\mu\nu}=R^{\mu\nu\lambda\rho}R_{\lambda\rho\mu\nu}-2R^{\mu\nu}R_{\mu\nu}+\frac{1}{3}R^2 \, ,
\end{align}
lo cual implica que la acción de Critical Gravity queda
\begin{align}
       I_{CRG} = \kappa\int_{\mathcal{M}} d^{4}x \sqrt{\lvert g\rvert} \left[R-2\Lambda  -\frac{3\beta}{2}\left(W^{\mu\nu\lambda\rho}W_{\lambda\rho\mu\nu}-E_4\right)\right] \, ,
\end{align}
con $E_{4}$ el término de Gauss-Bonnet, definido como
\begin{equation*}
\mathcal{G} = \dfrac{1}{4} \delta _{\nu _{1} \ldots  \nu _{4}}^{\mu _{1} \ldots  \mu _{4}} R_{\mu _{1} \mu _{2}}^{\nu _{1} \nu _{2}} R_{\mu _{3} \mu _{4}}^{\nu _{3} \nu _{4}} = R^2 - 4R^\mu_\nu R^\nu_\mu + R^{\mu\nu}_{\lambda\rho}R^{\lambda\rho}_{\mu\nu} \,.
\end{equation*}
Luego, imponiendo \eqref{CCG} y el gauge transversal sin traza, esto es,
\begin{align}
    \nabla^{\mu}h_{\mu\nu}&=g^{\mu\nu}h_{\mu\nu}=0\,,
\end{align}
se pueden estudiar las condiciones bajo las cuales el modo masivo de spin-2 no tenga masa. En la Ref.~\cite{Lu:2011ks}, los autores encontraron que esto se logra cuando se cumple 
\begin{equation}
    \beta=-\frac{1}{2\Lambda} \label{CCG1}
\end{equation}
en donde hemos llegado a una teoría en cuatro dimensiones que solo describe gravitones sin masa. Esta teoría se conoce como Critical Gravity \cite{Lu:2011zk}. En particular, para la métrica de Schwarzschild-AdS, su masa y su entropía se anulan considerando las condiciones de criticalidad \eqref{CCG} y \eqref{CCG1}. Para el valor de $\beta$ dado por \eqref{CCG1} y reemplazando el valor de $\Lambda=-\tfrac{3}{\ell^2}$ nos queda
\begin{align}
    I_{CRG} = \kappa\int_{\mathcal{M}} d^{4}x \sqrt{\lvert g\rvert} \left[\left(R+\frac{6}{\ell^2}\right)  -\frac{\ell^2}{4}\left(W^{\mu\nu\lambda\rho}W_{\lambda\rho\mu\nu}-E_4\right)\right] \,.
\end{align}
Tal como hemos señalado para la acción \eqref{IEGB}, el valor del acoplamiento del término de Gauss-Bonnet $\alpha_0=\tfrac{\ell^2}{4}$ lleva a la acción de Einstein-AdS renormalizada. De este modo, obtenemos que la acción de Critical Gravity puede ser reescrita como~\cite{Anastasiou:2017rjf}
\begin{align}
    I_{CRG} = I_{EH}^{ren}-\frac{\kappa\ell^2}{4}\int_{\mathcal{M}} d^{4}x\sqrt{\lvert g\rvert}\,W^{\mu\nu\lambda\rho}W_{\lambda\rho\mu\nu}\,.\label{CriticalGravityAction}
\end{align}
Las ecuaciones de movimiento de esta teoría se obtienen de realizar variaciones arbitrarias de la acción~\eqref{CriticalGravityAction} con respecto de la métrica son
\begin{align}\label{eomcritical}
    G_{\mu\nu} - \frac{3}{\ell^2}g_{\mu\nu} + \frac{\ell^2}{4}B_{\mu\nu} = 0\,.
\end{align}
De aquí, es directo ver que los espacios Einstein-AdS que cumplen con~\eqref{ES} satisfacen las ecuaciones de movimiento de Critical Gravity. Además, dado que la acción~\eqref{IEGB} toma el valor de~\eqref{IMM1} para este tipo de espacios, se concluye que la acción~\eqref{CriticalGravityAction} es idénticamente cero para espacios tipo Einstein. De esta manera, se concluye que todos los espacios Einstein son el \textit{ground state} de esta teoría.


\section{Renormalización conforme}\label{sec:CR}
A continuación, presentamos una estrategia diferente para abordar la Renormalización Conforme. En lugar de encontrar una teoría invariante conforme y evaluarla en diferentes sectores del espacio de soluciones, realizamos su completitud conforme on-shell. Este será el principio guía para la derivación de la generalización a las teorías escalar-tensor.

Nuestro punto de partida es la gravedad de Einstein-AdS en cuatro dimensiones. En este caso, la densidad lagrangiana de Einstein-Hilbert con una constante cosmológica negativa $\Lambda=-3/\ell^2$ es
\begin{equation}
\Lag_{\rm EH}= \sqrt{|g|} \left(R+\frac{6}{\ell^2}\right) \, .
\end{equation}
El comportamiento del escalar de Ricci bajo rescalamientos infinitesimales locales de Weyl de la métrica, es decir, $\delta_{\sigma} g_{\mu \nu} = 2 \sigma g_{\mu \nu}$, está dado por
\begin{equation}
\delta_{\sigma} R = -2 \sigma R - 2\left(D-1\right) \Box \sigma \, .
\end{equation}
Entonces, el término de Einstein-Hilbert se transforma de acuerdo a
\begin{equation}
\delta_{\sigma} \left[\sqrt{|g|} \left(R + \frac{6}{\ell^2} \right)\right]
=2 \sqrt{|g|} \left [\sigma \left(R +\frac{12}{\ell ^{2}}\right) -3 \nabla ^{\mu } \nabla _{\mu }\sigma \right ] \, . \label{LEH4D}
\end{equation}
Para completar conformalmente el Lagrangiano de Einstein-Hilbert sin modificar las ecuaciones de campo, es necesario agregar términos de borde o topológicos. Como trabajamos en cuatro dimensiones, consideramos que la acción de Einstein-Hilbert se completa con el término de Gauss-Bonnet con una constante de acoplamiento arbitraria $c_4$, es decir,
\begin{equation}\label{LagEHGB}
\Lag_{\rm EH,GB} = \sqrt{|g|} \left(R+\frac{6}{\ell^2} + c_{4} E_{4}\right) \,,
\end{equation}
en donde $c_4$ es una constante con unidades de largo al cuadrado. En cuatro dimensiones, el término de Gauss-Bonnet no contribuye a la dinámica del volumen ya que su integral es proporcional a la suma de la característica de Euler y a la integral de la forma de Chern en un borde de codimensión uno. Sin embargo, cambia las cargas conservadas y la acción Euclídea on-shell de manera no trivial~\cite{Aros:1999id,Aros:1999kt,Miskovic:2009bm}. Considerando su variación de Weyl 
\begin{equation}
\dfrac{1}{4} \delta _{\sigma } \left(\sqrt{|g|} \;\delta _{\nu _{1} \ldots  \nu _{4}}^{\mu _{1} \ldots \mu _{4}} R_{\mu _{1} \mu _{2}}^{\nu _{1} \nu _{2}} R_{\mu _{3} \mu _{4}}^{\nu _{3} \nu _{4}}\right)
= -16 \delta _{\nu _{1} \nu _{2}}^{\mu _{1} \mu _{2}}  \nabla _{\mu _{2}}\left (\sqrt{|g|} \;S_{\mu _{1}}^{\nu _{1}}  \nabla ^{\nu _{2}}\sigma \right ) \,, \label{weylvarGB}
\end{equation}
y sumando todas las contribuciones, obtenemos
\begin{equation}\delta _{\sigma } \Lag_{\rm EH,GB} 
=2 \sqrt{|g|} \left [\sigma  \left(R +\frac{12}{\ell ^{2}}\right) -3  \Box\sigma  -8 c_{4} \delta _{\nu _{1} \nu _{2}}^{\mu _{1} \mu _{2}} \left (\dfrac{1}{2} C_{\mu _{1} \mu _{2}}^{\nu _{1}}  \nabla ^{\nu _{2}}\sigma  +S_{\mu _{1}}^{\nu _{1}}  \nabla _{\mu _{2}} \nabla ^{\nu _{2}}\sigma \right )\right ]\,.
\end{equation}
Como era de esperarse, la acción de EH con el Gauss-Bonnet no es invariante de Weyl. Sin embargo, evaluando en espacios tipo Einstein, la acción toma la siguiente forma 
\begin{equation}
\delta _{\sigma } \Lag_{\rm EH,GB}\vert _{E} =\sqrt{|g|} \left ( -6  \Box\sigma  +\frac{24}{\ell^2} c_{4}  \Box\sigma \right ) \,.
\end{equation}
Exigiendo invarianza de Weyl on-shell de la acción, la constante de acoplamiento se fija de manera única como $c_{4}=\frac{\ell^2}{4}$. Así, la acción correspondiente coincide, modulo la característica de Euler de la variedad, con la acción de Einstein-AdS renormalizada topológicamente~\cite{Miskovic:2009bm}. Recientemente, se ha demostrado que esto es equivalente a la prescripción HR~\cite{Anastasiou:2020zwc}. Este procedimiento proporciona una ruta alternativa para obtener los términos de borde que hacen que la acción sea finita.

La generalización del concepto de Renormalización Conforme en seis dimensiones ha sido presentada en la Ref. ~\cite{Anastasiou:2020mik}. En la siguiente Sección, extenderemos esta prescripción al caso en el que se incluyen campos escalares.

\biblio 

\clearpage

\chapter{Campos Escalares}

El descubrimiento del Boson de Higgs en el año 2012 por parte de las colaboraciones ATLAS y CMS en CERN, nos dio evidencia observacional de la existencia de un campo escalar fundamental como piedra angular del modelo estándar \cite{CMS:2012qbp,ATLAS:2012yve}. Incluso antes de este evento, los campos escalares tenían relevancia en ciertas áreas de la física, como por ejemplo en la cosmología de épocas tempranas, teoría de bajas energías de la cromodinámica cuántica, los pares de Bardeen–Cooper–Schrieffer en superconductividad, entre muchas otras.

En el caso de gravitación, acoplamientos no-minimales entre campos escalares y la geometría son interesantes ya que permiten describir la evolución tardía del Universo y, a su vez, evitar violaciones al teorema de no pelo; teorema que discutiremos más adelante. 

La acción para un campo escalar conformalmente acoplado a \eqref{IEGB} se ve de la siguiente manera
\begin{equation}
    I[g_{\mu\nu},\phi]=\int_{\mathcal{M}} d^{4}x \sqrt{\lvert g\rvert}\left(\kappa\,R -\frac{1}{12} R\phi^2-\frac{1}{2} \nabla^{\mu}\phi\nabla_{\mu}\phi\right)\,.
\end{equation}
Bekenstein \cite{Bekenstein:1974sf,bekenstein1973black} y Bocharova-Melnikov-Bronnikov \cite{Bocharova:1970skc} encontraron de forma independiente una solución para esta acción. Sin embargo, en esta solución el campo escalar diverge en el horizonte de eventos. Inspirados por esto, Martinez-Staforelli-Troncoso-Zanelli~\cite{Martinez:2002ru,Martinez:2005di} encontraron una generalización añadiendo un término con un potencial cuártico más la constante cosmológica. En este caso, el campo escalar es regular fuera y sobre el horizonte de eventos. Ésta solución es conocida como el MSTZ black hole y la acción está dada por
\begin{equation}
    I[g_{\mu\nu},\phi]=\int_{\mathcal{M}} d^{4}x \sqrt{\lvert g\rvert}\left[\kappa(R-2\Lambda) -\frac{1}{12} R\phi^2-\frac{1}{2} \nabla^{\mu}\phi\nabla_{\mu}\phi-\nu\phi^4\right],
\end{equation}
en donde $\nu$ es un parámetro adimensional de la teoría que caracteriza la intensidad del acoplamiento cuártico de los campos escalares. Las ecuaciones de movimiento que se obtienen de variar esta acción con respecto de la metrica y el campo escalar son
\begin{align}
 G_{\mu\nu}+\Lambda g_{\mu\nu}&=\frac{1}{2\kappa} T_{\mu\nu}\, ,\\
   \Box\phi - \frac{1}{6}\phi R - 4\nu\phi^3 &= 0\,,
\end{align}
respectivamente, en donde el tensor de energía-momentum está definido por
\begin{align}
T_{\mu\nu} &= \nabla_\mu\phi\nabla_\nu\phi - \frac{1}{2}g_{\mu\nu}\nabla_\lambda\phi\nabla^\lambda\phi + \frac{1}{6}\left(g_{\mu\nu}\Box - \nabla_\mu\nabla_\nu + G_{\mu\nu} \right)\phi^2 - \nu\phi^4 g_{\mu\nu} \,.
\label{Ttensor}
\end{align}
El valor $\nu=-\tfrac{2\Lambda G}{9}$ y la positividad de la constante cosmológica son necesarias para asegurar la existencia de una solución tipo agujero negro estático y esfericamente simétrico \cite{Martinez:2002ru}. Recientes trabajos relacionados que recuperan el agujero negro de BBMB mencionado anteriormente tomando ciertos límites en parámetros de soluciones con campos escalares pueden ser vistos aquí \cite{Barrientos:2022avi, Barrientos:2023tqb}

\section{Teorema de no Pelo}
Cuando mencionamos anteriormente el teorema de no pelo nos referimos al enunciado propuesto por Bekenstein~\cite{PhysRevLett.28.452,PhysRevD.5.1239} y desarrollado posteriormente por Israel, Wheerler, Carter y Hawking~\cite{PhysRevLett.26.331,PhysRev.164.1776} que establece que los agujeros negros en el espacio-tiempo de la Relatividad General están completamente caracterizados por solo tres propiedades físicas: masa ($\mathbf{M}$), carga eléctrica ($\mathbf{Q}$) y momento angular ($\mathbf{J}$). 

El teorema nos enuncia que, una vez que un agujero negro astrofísico se ha formado a través del colapso gravitacional de la materia, todas las demás características y propiedades del objeto original desaparecen y sólo quedan las tres cantidades mencionadas anteriormente. Actualmente, existen estudios que tratan de probar este teorema mediante ondas gravitacionales~\cite{PhysRevLett.123.111102,PhysRev.164.1776} y la observación directa de la sombra del agujero negro de M87~\cite{Tang:2022uwi}. Sin embargo, en escenarios con agujeros negros extremales~\cite{PhysRevD.103.L021502} o en mecánica cuántica~\cite{PhysRevLett.128.111301} pueden existir violaciones al teorema.

Agregar un campo escalar conformalmente acoplado puede inducir que la solución adquiera lo que se conoce como pelo primario, i.e., donde hay un parámetro nuevo no trivial, o un pelo secundario, donde este parámetro proveniente del campo escalar puede definirse en función de las constante de integración de la métrica.

A modo de ejemplo, consideremos la acción de Einstein-Klein-Gordon dada por la ecuación
\begin{align}
I_{\rm EKG}=\int\diff{^4x}\sqrt{|g|}\left(\kappa R - \frac{1}{2}\nabla^{\mu} \phi \nabla_{\mu} \phi - V(\phi)\right) \,,
\end{align}
donde $V(\phi)$ es un potencial arbitrario. Las EOM son  
\begin{align}
 G_{\mu\nu}&=\frac{1}{2\kappa} T_{\mu\nu}\, ,\\
   \Box\phi &= \frac{\diff V}{\diff\phi}\,,
\end{align}
con $T_{\mu\nu}=\nabla_{\mu}\phi\nabla_{\nu}\phi-\frac{1}{2}g_{\mu\nu}\nabla_{\lambda}\phi\nabla^{\lambda}\phi+g_{\mu\nu}V(\phi)$ el tensor de energía-momentum del campo escalar. Ahora, estudiemos la ecuación de movimiento del campo escalar. Vamos a asumir estaticidad y simetría esférica para la métrica y el campo escalar, esto es 
\begin{align}
\diff s^2=-f(r)h^2(r)\diff t^2+\frac{\diff r^2}{f(r)}+r^2\diff\Omega^2\;\;\;\;\;\mbox{y}\;\;\;\;\;\phi=\phi(r)\, .
\end{align}
Además, consideraremos la existencia de un horizonte en $r=r_h$, tal que $f(r_h)=0$. Si multiplicamos por $\phi$ e integramos ambos lados sobre una variedad $\mathcal{M}$, la cual está delimitada por el horizonte y una región asintótica $r\to\infty$, tenemos
\begin{align}
\int_{\mathcal{M}}\diff{^4x}\sqrt{|g|}\left(-\phi\Box\phi+\phi\frac{\diff V}{\diff\phi}\right)=0\, . \label{ekgm}
\end{align}
Luego, vamos a integrar por partes el término $\phi\Box\phi$, lo cual nos entrega
\begin{align}
-\phi\Box\phi=\nabla_{\mu}\phi\nabla^{\mu}\phi-\nabla_{\mu}(\phi\nabla^{\mu}\phi)\, .
\end{align}
Reemplazando en~\eqref{ekgm}, obtenemos
\begin{align}
\int_{\mathcal{M}}\diff{^4x}\sqrt{|g|}\left(\nabla_{\mu}\phi\nabla^{\mu}\phi+\phi\frac{\diff V}{\diff\phi}\right)-\int_{\mathcal{M}}\diff{^4x}\sqrt{|g|}\nabla_{\mu}(\phi\nabla^{\mu}\phi)=0\, .   
\end{align}
Como el último término es una derivada total, podemos utilizar el teorema de la divergencia, lo que nos deja con un término de borde. Así, considerando una foliación radial, el término de borde se anula tanto en el horizonte como en la región asintótica, siempre que los campos escalares decaigan lo suficientemente rápido cuando $r\to\infty$. Por lo tanto, nos queda
\begin{align}
\int_{\mathcal{M}}\diff{^4x}\sqrt{|g|}\left(\nabla_{\mu}\phi\nabla^{\mu}\phi+\phi\frac{\diff V}{\diff\phi}\right)=0\, .   
\end{align}
Luego, notemos que el primer término de la integral es semi definido positivo fuera del horizonte, ya que $g^{\mu\nu}\nabla_{\mu}\phi\nabla_{\nu}\phi=g^{rr}[\phi'(r)]^2\geq 0$ si $r\geq r_h$. Además, vamos a considerar que el potencial $V(\phi)$ cumpla con $\phi\,\diff V/\diff\phi\geq0$, como por ejemplo el termino de masa
\begin{align*}
V(\phi)=\frac{m^2}{2}\phi^2 \;\;\;\to\;\;\;\phi\frac{\diff V}{\diff\phi}=m^2\phi^2\, .
\end{align*}
Entonces, remplazando en la integral anterior, tenemos
\begin{align}
\int_{\mathcal{M}}\diff{^4x}\sqrt{|g|}\left(\nabla_{\mu}\phi\nabla^{\mu}\phi+m^2\phi^2\right)=0\, .
\end{align}
Podemos notar que tenemos la suma de dos cantidades positivas que son iguales a cero. La única forma en que esto se cumpla es si cada factor por separado es cero, lo cual implica que $\phi=0$. De esta forma, podemos ver que para que exista un horizonte y además las cantidades estén bien comportadas, el campo escalar debe ser trivial. Este es el teorema de no pelo de Bekenstein. Sin embargo, es posible relajar ciertas condiciones del teorema y encontrar soluciones tipo agujero negro con campo escalar no trivial, como veremos a continuación.

\section{Acoplamiento conforme}

El análisis realizado en el capítulo anterior nos dice que, en presencia de campos puramente métricos, la invariancia conforme en el volumen conduce a una acción finita cuando se consideran espacio-tiempos de tipo anti-de Sitter (AlAdS). En esta sección, estudiamos si esta relación se puede extender en presencia de campos escalares. Nuestro punto de partida es la teoría escalar-tensor acoplada conformalmente con un campo escalar auto-interactuante, cuya acción está dada por
\begin{equation}\label{ccscalar}
I_\phi = \int\diff{^4x}\sqrt{|g|}\left(\frac{1}{12}\phi^{2} R + \frac{1}{2}\nabla^{\mu} \phi \nabla_{\mu} \phi + \nu\phi^{4}\right) \,,
\end{equation}
donde $\nu$ es una constante de acoplamiento adimensional del campo escalar cuártico. Esta acción es cuasi-invariante conforme, es decir, transforma como un termino de borde bajo reescalamientos simultáneos de Weyl de la métrica y del campo escalar, dados por $g_{\mu\nu}\to\tilde{g}_{\mu\nu} = e^{2\sigma(x)}g_{\mu\nu}$ y $\phi\to\tilde{\phi} = e^{-\sigma(x)}\phi$.
En efecto, considerando reescalamientos infinitesimales de Weyl de los campos 
\begin{align}
\delta_{\sigma} g_{\mu \nu} = 2 \sigma g_{\mu \nu} \;\;\;\;\; \mbox{y} \;\;\;\;\; \delta_{\sigma} \phi = - \sigma \phi \,, 
\end{align}
la Ec.~\eqref{ccscalar} transforma como 
\begin{align}
\delta_{\sigma}I_{\phi} 
&=-\frac{1}{2}\int\diff{^4x}\sqrt{|g|} \nabla_{\mu} \left(\phi^{2} \nabla^{\mu} \sigma \right) \,. \label{Weylvargandphi}
\end{align}
La presencia de una derivada total indica que la acción $I_{\phi}$ debe ser complementada por un término de borde y/o una contribución topológica para restaurar la invarianza conforme local exacta de la teoría.

Para realizar la completitud conforme de la última expresión, consideramos que para un escalar $\Phi$ de dimensión de escalamiento arbitraria $\Delta$, la variación de Weyl del Laplaciano multiplicado por el elemento de volumen se expresa como
\begin{equation}
\delta_{\sigma}\left(\sqrt{|g|} \Box \Phi \right) =  \sqrt{|g|} \left[\left(D+\Delta-2\right) \sigma \Box \Phi +\Delta \Phi \Box \sigma + \left(D+2\Delta-2\right) \nabla^{\lambda} \sigma \nabla_{\lambda} \Phi \right] \,.
\end{equation}
Así, para $\Phi = \phi^2$ con $\Delta=-2$ en cuatro dimensiones, obtenemos
\begin{equation}
\delta_{\sigma}\left(\sqrt{|g|} \Box \phi^2 \right) = -2 \sqrt{|g|} \nabla_{\mu} \left(\phi^{2} \nabla^{\mu} \sigma \right) \,.
\end{equation}
Por lo tanto, la combinación 
\begin{equation}
I_{\phi,\rm cc}=I_{\phi} - \frac{1}{4}\int\diff{^4x}\sqrt{|g|} \Box \phi^2 \,,
\end{equation}
es completamente invariante bajo reescalamientos de Weyl tanto de la métrica como del campo escalar. De hecho, esta acción puede escribirse de manera equivalente como
\begin{equation}
I_{\phi,\rm cc}=\int\diff{^4}x\sqrt{|g|}\left(\frac{1}{12}\phi^2 R - \frac{1}{2}\phi\Box\phi + \nu\phi^4\right) \,.
\label{Ibulkconf}
\end{equation}
En la última expresión, queda manifiesto que la completitud conforme de la acción de campo escalar con acoplamiento no mínimo conduce a una dependencia explícita del operador Yamabe $\Delta_2$, que se expresa como sigue
\begin{equation}\label{Yamabe}
\Delta_2 = - \Box+ \frac{\left(D-2\right)}{4\left(D-1\right)} R\,.
\end{equation}
Este operador diferencial ---frecuentemente llamado Laplaciano conforme--- es covariante conforme con un peso de escala $-\frac{D+2}{2}$ cuando actúa sobre escalares de dimensiones de escala $\Delta = -\frac{D-2}{2}$~\cite{Osborn:2015rna, Gover:2002ay}. Se puede extender trivialmente este operador diferencial agregando un escalar con peso conforme $\Delta=-2$. En particular, para la teoría de interés, consideramos la extensión
\begin{equation}
\tilde{\Delta}_{2} = \Delta_{2}+ c \phi^{2}\,,
\label{Yamabeext}
\end{equation}
donde $c$ es una constante de acoplamiento adimensional arbitraria. Esto, nos permite reescribir la acción \eqref{Ibulkconf} como 
\begin{equation}
I_{\phi,\rm cc}= \frac{1}{2}\int\diff{^4}x\sqrt{|g|} \phi \tilde{\Delta}_{2} \phi \,.
\label{Ibulkconf2}
\end{equation}
Como el operador Yamabe es covariante conforme, la acción~\eqref{Ibulkconf2} es explícitamente invariante bajo reescalaminetos de Weyl. Sin embargo, existen configuraciones que rompen la simetría, como los campos escalares constantes. En ese caso, la parte cinética del campo escalar se anula y solo queda la parte de Einstein-Hilbert, perdiendo la información de la presencia del escalar. Además, se puede ver que la transformación conforme se vuelve singular para un escalar constante al empezar con una configuración escalar no trivial tal que la acción~\eqref{Ibulkconf2} sea finita y el espacio-tiempo sea AlAdS. Entonces, se puede elegir la transformación conforme para que el campo escalar sea constante. Si el valor original de la acción era finito, entonces, por la invarianza conforme, debería seguir siendo finito en la configuración de campo escalar constante. Sin embargo, en ese caso, la acción diverge con el volumen AdS. Por lo tanto, la transformación debe ser singular. Esto implica que $I_{\phi,\rm cc}$ debe ser complementado por los términos compensatorios correspondientes que son necesarios para garantizar la invariancia conforme del volumen para todas las posibles configuraciones de campo de la teoría, incluido el caso de un campo escalar constante.

Para evitar este problema, es conveniente introducir el tensor prpouesto en  Ref.~\cite{Oliva:2011np}, este es,
\begin{equation}\label{Stensor}
\mathcal{S}^{\mu\nu}_{\,\lambda\rho}=\phi^2R^{\mu\nu}_{\,\lambda\rho}-4\phi\delta^{[\mu}_{[\lambda}\nabla^{\nu]}\nabla_{\rho]}\phi+8\delta^{[\mu}_{[\lambda}\nabla^{\nu]}\phi\nabla_{\rho]}\phi-\delta^{\mu\nu}_{\lambda\rho}\nabla_\alpha\phi\nabla^\alpha\phi\,,
\end{equation}
que transforma covariantemente bajo reescalamientos de Weyl en cuatro dimensiones, i.e.,
\begin{align}
\mathcal{S}^{\mu\nu}_{\lambda\rho}\to \tilde{\mathcal{S}}^{\mu\nu}_{\lambda\rho} &=  e^{-4\sigma(x)}\mathcal{S}^{\mu\nu}_{\lambda\rho}\,. 
\end{align} 
Esto implica que el tensor~\eqref{Stensor} se convierte en un objeto conveniente para construir teorías escalar-tensor de gravedad invariante conforme. De hecho, sus trazas son
\begin{align}\label{Smunu}
\mathcal{S}^\mu_\nu &\equiv \mathcal{S}^{\mu\lambda}_{\nu\lambda} = \phi^2R^\mu_\nu - \delta^\mu_\nu \phi\Box\phi - 2\phi\nabla^\mu\nabla_\nu\phi + 4\nabla^\mu\phi\nabla_\nu\phi - \delta^\mu_\nu\nabla_\lambda\phi\nabla^\lambda\phi\,, \\
\label{S}
\mathcal{S} &\equiv \mathcal{S}^{\mu\nu}_{\mu\nu} = \phi^2 R-6\phi\square\phi \, .
\end{align}
Se puede observar que la Ec.~\eqref{Yamabe} y~ \eqref{S} coinciden, módulo un factor global, si y solo si $c=0$. Una generalización natural de la Eq.~\eqref{Stensor} incluyendo la pieza faltante del operador de Yamabe en la  Eq.~\eqref{S} se puede obtener cambiando $S^{\mu\nu}_{\lambda\rho}$ en el espacio del campo, según
\begin{equation}\label{Sigma}
\Sigma^{\mu\nu}_{\lambda\rho} = \frac{1}{\phi^2}\left(\mathcal{S}^{\mu\nu}_{\lambda\rho} + 2\nu\phi^4\,\delta^{\mu\nu}_{\lambda\rho} \right)\,,
\end{equation}
donde $\delta^{\mu_1\ldots\mu_p}_{\nu_1\ldots\nu_p}=p!\,\delta^{\mu_1}_{[\nu_1}\dots\delta^{\mu_p}_{\nu_p]}$ es la delta de Kronecker generalizada de rango $p$. Escrito en esta forma, se puede ver que $\Sigma^{\mu\nu}_{\lambda\rho}$ tiene el mismo peso conforme que el tensor de Weyl. Entonces, su traza
\begin{align}\label{Yamabe2}
\phi^2\Sigma^{\mu\nu}_{\mu\nu} &=  \phi^2 R - 6 \phi\Box\phi + 24\nu\phi^4 = 6\phi\tilde{\Delta}_2\phi
\end{align}
para $c=4\nu$. Es decir, la traza completa del tensor covariante conforme $\Sigma^{\mu\nu}_{\lambda\rho}$ es equivalente, modulo un factor global, al operador de Yamabe en 4D.

Así, basado en el análisis anterior, consideramos una teoría escalar-tensor invariante conforme cuya dinámica está descrita por el principio de acción
\begin{align}\notag
I_{\phi,\rm conf} &= \frac{\zeta}{4}\int\diff{^4}x\sqrt{|g|}\,\delta^{\mu_1\ldots\mu_4}_{\nu_1\ldots\nu_4}\Sigma^{\nu_1\nu_2}_{\mu_1\mu_2}\Sigma^{\nu_3\nu_4}_{\mu_3\mu_4} \\
&= 96\zeta\nu\int\diff{^4}x\sqrt{|g|}\,\left[\frac{1}{12}\phi^2 R - \frac{1}{2}\phi\Box\phi + \nu\phi^4 + \frac{1}{96\nu}\,\left(E_4 + \nabla_\mu J^\mu\right)\right] \label{Lagphi}  \,,
\end{align}
donde $\zeta$ es un parametro adimensional y
\begin{align}
J^{\mu} &= 8 \left[\phi^{-1} G^\mu_\lambda\nabla^\lambda\phi+ \phi^{-2} \left(\nabla^{\mu} \phi \Box \phi -\nabla^{\lambda} \phi\nabla_{\lambda}\nabla^{\mu} \phi \right) + \phi^{-3} \nabla^{\mu} \phi \nabla^{\lambda} \phi \nabla_{\lambda} \phi \right]\,,
\end{align}
con $G_{\mu\nu}=R_{\mu\nu} - \frac{1}{2}g_{\mu\nu}R$. La Ec.~\eqref{Lagphi} reproduce, modulo un término de borde, exactamente el operador de Yamabe de la acción escalar-tensor acoplada conforme en la Eq.~\eqref{Ibulkconf2}. 
Como vimos anteriormente, el término de Gauss-Bonnet no transforma covariantemente bajo reescalamientos de Weyl [ver la Eq.~\eqref{weylvarGB}]. Sin embargo, dado que el lado izquierdo de la ecuación 
\begin{align}\label{GBS}
    \frac{1}{4\phi^4}\delta^{\mu_1\ldots\mu_4}_{\nu_1\ldots\nu_4}S^{\nu_1\nu_2}_{\mu_1\mu_2}S^{\nu_3\nu_4}_{\mu_3\mu_4} = \frac{1}{4}\delta^{\mu_1\ldots\mu_4}_{\nu_1\ldots\nu_4}R^{\nu_1\nu_2}_{\mu_1\mu_2}R^{\nu_3\nu_4}_{\mu_3\mu_4} + \nabla_\mu J^\mu\,,
\end{align}
transforma de forma covariante bajo reescalamientos de Weyl por construcción, se puede concluir que la divergencia de $J^\mu$ compensa la pieza no homogénea del Gauss-Bonnet bajo transformaciones conformes. 

Las ecuaciones de movimiento para la acción inicial \eqref{ccscalar} se pueden obtener realizando variaciones estacionarias de la acción de la Eq.~\eqref{Lagphi} con respecto a la métrica y el campo escalar, así, tenemos
\begin{subequations}\label{EOMLagphi}
\begin{align}\label{eomlagphig}
    T_{\mu\nu} &\equiv \nabla_\mu\phi\nabla_\nu\phi - \frac{1}{2}g_{\mu\nu}\nabla_\lambda\phi\nabla^\lambda\phi + \frac{1}{6}\left(g_{\mu\nu}\Box - \nabla_\mu\nabla_\nu + G_{\mu\nu} \right)\phi^2 - \nu\phi^4 g_{\mu\nu} = 0 \,, \\\label{eomlagphip}
   \mathcal{E} &\equiv\Box\phi - \frac{1}{6}\phi R - 4\nu\phi^3 = 0\,, 
\end{align}
\end{subequations}
respectivamente. Tomando la traza de la Ec.~\eqref{eomlagphig} y comparandola con Ec.~\eqref{eomlagphip}, se tiene que $T= g^{\mu\nu}T_{\mu\nu}=\phi\,\mathcal{E}$. Adicionalmente, el tensor $\Sigma$ está relacionado al $T_{\mu\nu}$ de la Ec.~\eqref{eomlagphig} mediante
\begin{equation}\label{SigmaT}
\phi^{2}\Sigma^\mu_\nu=6\left(T^\mu_\nu - \frac{1}{2}T\delta^\mu_\nu\right)\,.
\end{equation}
Se puede observar que una configuración constante del campo escalar, digamos $\phi=\phi_0$, reduce la teoría~\eqref{Lagphi} a la gravedad de Einstein-AdS. Este caso corresponde al marco de Einstein de la simetría de Weyl. De hecho, para que la acción se pueda escribir en términos de la constante gravitatoria usual $G_N$ y el radio de AdS $\ell$, se realiza la siguiente elección
\begin{equation}\label{couplingsEinstein}
    \nu \phi_{0}^{2} = \frac{1}{2\ell^2} \;\;\;\;\; \mbox{y} \;\;\;\;\;  \zeta = \frac{\ell^2}{64\pi G_{N}}\,.
\end{equation}
A nivel de las ecuaciones de movimiento, se puede ver que la Ec.~\eqref{eomlagphig} se convierte en la ecuación de campo de Einstein habitual al hacer la elección de gauge del tensor de Weyl como en la Ec.~\eqref{couplingsEinstein}. Además, la Ec.~\eqref{eomlagphip} simplemente implica la restricción de que el escalar de Ricci debe estar fijo en términos del radio de AdS, como ocurre en el caso de los espacios tipo Einstein. Por lo tanto, es evidente que la teoría admite la familia completa de soluciones Einstein-AdS para un campo escalar constante.

A nivel de la acción, ahora podemos verificar cómo la elección de gauge de Weyl mencionada anteriormente implica que el Lagrangiano se reduce a la gravedad de Einstein-AdS renormalizada. De hecho, para los valores de las constantes de acoplamiento en la Ec.~\eqref{couplingsEinstein}, la acción~\eqref{Ibulkconf2} se puede expresar en la forma de MacDowell-Mansouri dada en la Ec.~\eqref{IMM}. Esta es la forma exacta de la acción de Einstein-Hilbert off-shell con constante cosmológica negativa, complementada por el término de Gauss-Bonnet con un acoplamiento fijo; este último proporciona un contratermino natural para la renormalización de la acción Euclídea on-shell y las cargas conservadas para soluciones asintóticamente localmente Einstein-AdS~\cite{Aros:1999id,Aros:1999kt,Miskovic:2009bm}. La discusión anterior se resume en la siguiente relación
\begin{equation}
I_{\phi_{0}, \rm conf} = I^{\rm (ren)}_{\rm EAdS} \, .
\end{equation}
Por lo tanto, concluimos que la acción~\eqref{Lagphi} tiene un límite Einstein bien definido, impuesto a través de la elección de la ecuación~\eqref{couplingsEinstein} para un campo escalar constante.

Para estudiar la finitud de la teoría~\eqref{Lagphi} cuando se consideran espaciotiempos AlAdS, realizamos la descomposición off-shell del tensor de Weyl \eqref{weyltensor} en términos del tensor $\Sigma$ \eqref{SigmaT} y el tensor de Schouten. Dado que el tensor de Einstein aparece explícitamente en la definición de $T_{\mu \nu}$, podemos escribir el tensor de Schouten de manera equivalente como
\begin{equation}
S^{\mu}_{\nu} = \frac{1}{2} \left(G^{\mu}_{\nu} +\frac{1}{3}R \delta^{\mu}_{\nu}\right) \, .
\end{equation}
Teniendo en cuenta que en las ecuaciones~\eqref{eomlagphig} y~\eqref{eomlagphip}, se puede reemplazar el tensor de Einstein y el escalar de Ricci en términos de $T_{\mu\nu}$ y su traza correspondiente. Cuando esta última se reemplaza en la definición del tensor de Weyl \eqref{weyltensor}, se obtiene la forma off-shell del tensor de Weyl, es decir,
\begin{equation}\label{Weyldecomp}
    W^{\mu\nu}_{\alpha \beta} = \Sigma^{\mu\nu}_{\alpha\beta} - \frac{2}{\phi^2}\left(6T^{[\mu}_{[\alpha}\delta^{\nu]}_{\beta]} - T\delta^{\mu\nu}_{\alpha\beta} \right)\,.
\end{equation}
Dado que las ecuaciones de campo implican que $T_{\mu\nu}=0$, concluimos que el tensor $\Sigma$ coincide con el tensor de Weyl y la acción on-shell se convierte en
\begin{equation}
I_{\phi,\rm conf}\big|_{\rm on-shell} = \zeta\int\diff{^4}x\sqrt{|g|}\,W^{\mu\nu}_{\lambda\rho}W^{\lambda\rho}_{\mu\nu}\,.
\end{equation}
Por lo tanto, la teoría de campo escalar-tensor completamente invariante conforme es equivalente on-shell a CG, que es finita para cualquier solución AlAdS~\cite{Grumiller:2013mxa}. Un ejemplo particular de este hecho se ha mostrado recientemente en la Ref.~\cite{Barrientos:2022yoz} para soluciones  Taub-NUT-AdS y Eguchi-Hanson cargadas en presencia de campos escalares conformemente acoplados.

Análogamente al caso de Einstein-AdS discutido en la sección~\ref{sec:CR}, pedir la invarianza conforme local exacta de la acción bajo reescalamiento de Weyl tanto de la métrica como del campo escalar, dicta los contra-términos que hacen que la acción sea finita. Es decir, la relación $I_{\phi,\rm conf}=I_{\phi}^{\rm (ren)}$ es válida y conduce a la acción escalar conforme renormalizada, que se expresa como
\begin{align}
I_{\phi}^{\rm (ren)}&=\frac{1}{384\nu}\int\diff{^4}x\sqrt{|g|}\,\delta^{\mu_1\ldots\mu_4}_{\nu_1\ldots\nu_4}\Sigma^{\nu_1\nu_2}_{\mu_1\mu_2}\Sigma^{\nu_3\nu_4}_{\mu_3\mu_4}= I_{\phi}+\frac{1}{96\nu}\int \diff{^4}x\sqrt{|g|} \left(E_4 + \nabla_\mu \tilde{J}^\mu\right) \,,
\label{Iphiren}
\end{align}
en donde
\begin{equation}
\tilde{J}^{\mu} = 8 \left[\phi^{-1} G^\mu_\lambda\nabla^\lambda\phi+ \phi^{-2} \left(\nabla^{\mu} \phi \Box \phi -\nabla^{\lambda} \phi\nabla_{\lambda}\nabla^{\mu} \phi \right) + \phi^{-3} \nabla^{\mu} \phi \nabla^{\lambda} \phi \nabla_{\lambda} \phi \right]-\tfrac{1}{2}\phi\nabla^{\mu}\phi \,
\end{equation}
y $I_{\phi}$ es definido en la Ec.~\eqref{ccscalar}.
Por eso, la última expresión es conformemente invariante para todas las configuraciones permitidas por el espacio de solución de la teoría.

Es importante destacar que en la teoría de la Ec.~\eqref{ccscalar}, o equivalente en la Ec.~\eqref{Iphiren}, el campo escalar no puede ser eliminado de forma que se recupere la gravedad de Einstein-AdS~\eqref{couplingsEinstein} sin cambiar el comportamiento asintótico. Por lo tanto, al fijar la condición de AlAdS en una configuración no trivial del campo escalar, se elige el espacio-tiempo físico como aquel en el cual dicho campo escalar está presente. Así, el escalar es físico ya que contribuirá a las cargas asintóticas de la configuración y, en el contexto de AdS/CFT, a las fuentes holográficas.

\biblio 

\clearpage

\chapter{Einstein-AdS y campos escalares conformalmente acoplados}
    
En esta sección, extendemos la aplicación de la prescripción de Renormalización Conforme, en el caso en el que la gravedad de Einstein-AdS se acopla a la teoría de campo escalar con acoplamiento conforme. Una motivación relevante para estudiar este tipo de teorías viene de que en 3 dimensiones acoplar campos escalares se puede entender como una forma de estudiar efectos cuánticos en agujeros negros, como el caso de la referencia \cite{Martinez:1996gn,Casals:2016odj} en la que se estudia una generalización el agujero negro de Bañados-Teitelboim-Zanelli con pelo. Se ha demostrado que estas soluciones están relacionadas a las C-metric en 4 dimensiones \cite{Emparan:2020znc}. Adicionalmente, lo anterior ha sido explorado tambien en 2+1 dimensiones en las refrencias \cite{Arenas-Henriquez:2022www,Arenas-Henriquez:2023hur,Cisterna:2023qhh}.

En nuestro caso, la forma genérica de la acción es la siguiente
\begin{equation}
I_{\phi \rm EAdS} = \frac{1}{16 \pi G_{N}} \int \diff{^4}x\sqrt{|g|} \left(R-2 \Lambda\right) + I_{\phi} \, .
\label{IphiEads}
\end{equation}
En las dos secciones anteriores, hemos demostrado que la cancelación de las divergencias en los espacios de AlAdS equivale al requisito de que la acción sobre la superficie sea invariante bajo reescalamientos de Weyl de los campos en el bulk. Sin embargo, la completitud conforme de la acción en la ecuación~\eqref{IphiEads} para cualquier configuración del espacio de soluciones es altamente no trivial. No obstante, hay ciertos sectores de la teoría que nos permiten hacer que la acción correspondiente sea invariante conforme. Para lograrlo, consideramos por separado la métrica y el sector escalar. Como se mostró en la última sección, este último puede complementarse con términos de borde que lo hacen invariante off-shell en cuatro dimensiones. Aunque esto es finito para todas las posibles soluciones de dicha teoría, no se espera que sea cierto cuando se incluyen otros sectores. De manera similar, el sector de la teoría puramente métrica se vuelve invariante conforme para los espacios Einstein-AdS cuando se le agrega el término de Gauss-Bonnet con una constante general fija, como se ve en la ecuación \eqref{IMM} y en las referencias \cite{Maldacena:2011mk,Anastasiou:2016jix}. Sin embargo, mostramos que, siempre y cuando el campo escalar en AdS tenga un comportamiento asintótico adecuado, la acción es finita sin necesidad de considerar contraterminos de borde intrínsecos.

\section{Renormalización}
La dinámica de la teoría en la que estamos interesados está dictada por un principio de acción que contiene un sector de Einstein-AdS escrito en una forma de MacDowell-Mansouri y los escalares acoplados conformalmente renormalizados, es decir,
\begin{equation}\label{IMMphi}
    I_{\rm MM\phi} = \frac{1}{4}\int\diff{^4}x\sqrt{|g|}\,\delta^{\mu_1\ldots\mu_4}_{\nu_1\ldots\nu_4}\left(\alpha \mathcal{F}^{\nu_1\nu_2}_{\mu_1\mu_2}\mathcal{F}^{\nu_3\nu_4}_{\mu_3\mu_4} - \zeta \Sigma^{\nu_1\nu_2}_{\mu_1\mu_2}\Sigma^{\nu_3\nu_4}_{\mu_3\mu_4} \right)\,,
\end{equation}
donde $\alpha = \frac{\ell^2}{64 \pi G}$, mientras que $\mathcal{F}^{\mu\nu}_{\lambda\rho}$ y $\Sigma^{\mu\nu}_{\lambda\rho}$ están definidos en las Ecs. \eqref{WeylE} y \eqref{Sigma}, respectivamente. Las ecuaciones de campo para el campo métrico y el campo escalar obtenidas a partir de variaciones arbitrarias de la acción \eqref{IMMphi} con respecto a esos campos son
\begin{subequations}\label{eomMMphi}
\begin{align}\label{eomgMMphi}
    \mathcal{E}_{\mu\nu} &\equiv \alpha \left(G_{\mu\nu} - \frac{3}{\ell^2} g_{\mu\nu}\right)  - 12 \ell^{2}\zeta\nu  T_{\mu\nu} =0 \,, \\
    \label{eomphiMMphi}
    \mathcal{E} &\equiv \Box\phi - \frac{1}{6}\phi R - 4\nu\phi^3 = 0\,,
\end{align}    
\end{subequations}
respectivamente. Vale la pena mencionar que esta teoría admite espacios de Einstein-AdS como soluciones cuando el campo escalar es constante. De hecho, la condición de la Ec.~\eqref{couplingsEinstein} impone que las soluciones de las ecuaciones \eqref{eomMMphi} son variedades de Einstein, para las cuales la acción de la Ec.~\eqref{IMMphi} se anula de manera idéntica. En particular, esta teoría admite el espacio global de AdS como el estado fundamental cuando el campo escalar es constante.

Habiendo completado parcialmente la simetría conforme de la teoría, estudiamos bajo qué condiciones su acción resulta. Siguiendo la prescripción introducida en la última sección, podemos reescribir la acción en términos de tensor de Weyl al cuadrado, que es finito para cualquier espacio de AlAdS. De manera similar, hemos introducido una descomposición alternativa del tensor de Weyl en términos del tensor $\Sigma$ \eqref{Weyldecomp}.
Además, dado que la traza del tensor de energía-momentum $T_{\mu\nu}$ se anula en la solución, entonces la Ec.~\eqref{eomgMMphi} restringe el espacio de soluciones a tener un escalar de Ricci constante y negativo, es decir,
\begin{equation}
    R = -\frac{12}{\ell^2}\,.
\end{equation}
Esto simplifica el tensor $X$ en la ecuación \eqref{Xtensor}, ya que ahora depende explícitamente del tensor de Ricci sin traza. Reemplazando las Ecs.~\eqref{weylfdecomposition} y \eqref{Weyldecomp} en la acción $I_{\rm MM\phi}$ y teniendo en cuenta las ecuaciones de movimiento, obtenemos que
\begin{equation}
I_{\rm MM\phi}\Big|_{\rm on-shell} = \int\diff{^4}x\sqrt{|g|} \left[\left(\alpha -\zeta\right) W^{\alpha \beta}_{\mu \nu} W^{\mu \nu}_{\alpha \beta} + 2 \alpha \left( \frac{\alpha}{4 \ell^{2} \zeta \nu^{2} \phi^{4}}-1\right) H^{\mu}_{\nu} H^{\nu}_{\mu}\right] \,. \label{IMMphios}
\end{equation}
Esta acción en la solución coincide con la acción de CG para espacios de Einstein ($H^{\mu}_{\nu} = 0$) o, equivalentemente, para soluciones stealth: un campo escalar no trivial con tensor de energía-momentum nulo~\cite{Ayon-Beato:2004nzi,Ayon-Beato:2005yoq,Hassaine:2006gz,Ayon-Beato:2013bsa}. Por lo tanto, en esos casos, la teoría es finita para espacios Einstein. De hecho, si $\alpha=\zeta$, la acción se anula de manera idéntica para soluciones de Einstein, lo que proporciona un tipo diferente de criticidad en las teorías escalar-tensor. Por otro lado, las soluciones no Einstein también pueden proporcionar una acción finita en la solución si y solo si el decaimiento de la contribución no-Weyl de la última expresión es suficientemente rápida. Un ejemplo característico de esto se presenta a continuación.

\section{Aplicaciones: El agujero negro de MSTZ}
En el caso del agujero negro de MSTZ, los autores consideraron un elemento de línea que permanece localmente invariante bajo la acción de los grupos de isometría $\mbox{SO}(3)\times\mathbb{R}$, $\mbox{SO}(1,2)\times\mathbb{R}$, y $\mbox{ISO}(2)\times\mathbb{R}$. Estas condiciones producen 
\begin{equation}\label{ansatz}
    \diff{s^2} = -f(r)\diff{t^2} + \frac{\diff{r^2}}{f(r)} + r^2\diff{\Sigma_{(k)}^2} \,.
\end{equation}
En donde $\diff{\Sigma_{(k)}^2}$ es el elemento de línea de la variedad base 2-dimensional de curvatura constante $k$, describiendo localmente secciones transversas $\mathbb{S}^2$, $\mathbb{T}^2$, y $\mathbb{H}^2$ para $k=1,0,-1$, respectivamente. Así, la solución esta dada por~\cite{Martinez:2002ru,Martinez:2005di}
\begin{equation}\label{MTZ}
    f(r) = k\left(1+\frac{\mu\,G}{r}\right)^2 + \frac{r^2}{\ell^2} \;\;\;\;\; \mbox{y} \;\;\;\;\; \phi(r) = \frac{1}{\ell}\sqrt{\frac{1}{2\nu}}\,\frac{\mu\,G}{r+\mu\,G}\,,
\end{equation}
donde $\mu$ s una constante de integración, $\nu >0$ y la siguiente condición sobre los parámetros
\begin{equation}
    \zeta = \frac{\ell^2}{64\pi G}\,,
\end{equation}
se debe cumplir. De hecho, se puede fijar $\zeta=\tfrac{1}{96\nu}$ sin perdida de generalidad con el fin de obtener la misma normalización que en las Ref.~\cite{Martinez:2002ru,Martinez:2005di}. Esta solución posee una singularidad de curvatura en $r=0$. Además, se conoce que por la hipótesis de censura cósmica se requiere la existencia de un horizonte en $r=r_{h}$, este definido por las raíces positivas del polinomio $f(r_{h})=0$. Esta condición exige que $k=-1$. Para que esta solución describa un agujero negro, la topología de la sección transversal debe ser $\mathbb{H}^2/\Gamma$, donde $\Gamma$ es un subgrupo de $SO(2,1)$, tal que las hipersuperficies de $t-r$ constante tengan un área finita. En el caso $\mu>0$, la solución tiene un solo horizonte y está dado por
\begin{equation}
    r_+ = \frac{\ell}{2}\left(1+\sqrt{1+\frac{4\mu G}{\ell}} \right)\,.
\end{equation}

Como mencionamos en la Sec.~\ref{sec:AEO}, a primer orden en la aproximación del punto de silla, podemos obtener la función de partición $\mathcal{Z}$ a través de la relación $\ln \mathcal{Z} \approx - I_E$. Para la solución~\eqref{MTZ}, encontramos
\begin{equation}
T_H = \beta^{-1} = \frac{2r_+-\ell}{2\pi\ell^2}\,.
\end{equation}
Entonces, evaluando la acción Euclídea on-shell~\eqref{IMMphi} en la solución~\eqref{MTZ} obtenemos
\begin{equation}
    I_E = - \frac{\beta\omega_{(k)}\left(\ell^2-r_+\ell+r_+^2 \right)}{8\pi G\ell}\,,
\end{equation}
con $\omega_{(k)}$ el volumen de codimension-2 de la variedad base.  Notablemente, el valor de la función de partición es finito sin ninguna referencia a contratermos en el borde, a pesar de corresponder a una configuración no stealth. Esto se debe al hecho de que el decaimiento del campo escalar y el tensor de Ricci sin traza es de orden $\mathcal{O} \left(r^{-1}\right)$ y $\mathcal{O} \left(r^{-4}\right)$, respectivamente, lo que hace que la contribución no Weyl en la ecuación \eqref{IMMphios} sea subdominante y no induzca divergencias. Lo anterior se puede ver directamente a partir del hecho de que
\begin{align}
\int\diff{^4x}\sqrt{|g|}\left( \frac{\alpha}{4 \ell^{2} \zeta \nu^{2} \phi^{4}}-1\right) H^{\mu}_{\nu} H^{\nu}_{\mu} \sim \mathcal{O}(r^{-1})\,,
\end{align}
cuando se evalúa en la solución~\eqref{MTZ}. Por lo tanto, esta prescripción proporciona una definición natural de contratérminos para teorías escalar-tensor que poseen un sector Einstein y campos escalares acoplados de forma conforme, siempre que la parte no Weyl al cuadrado del Lagrangiano on-shell~\eqref{IMMphios} tenga un decaimiento al menos tan rápido como $\mathcal{O} \left(r^{-4}\right)$.

Además, notemos que la forma on-shell de la acción $I_{\rm MM\phi}$ indica que se anula idénticamente para espacio-tiempos tipo Einstein cuando $\alpha=\zeta$ o, equivalentemente, $\nu=\frac{2\pi G}{3\ell^2}$. De hecho, este es exactamente el punto en el espacio de parámetros donde existe la solución~\eqref{MTZ}, aunque la configuración no es Einstein. Esta teoría es completamente análoga a Critical Gravity de la Sec.~\ref{sec:Critical}, que es trivial para espacio-tiempos Einstein. Esto significa que $I_{\rm MM\phi}$ para el valor específico de $\nu$ corresponde a la generalización de Critical Gravity a teorías de escalar-tensor de la gravedad.

\biblio 

\clearpage

\chapter{Gravedad conforme con campos escalares}
    
Otra clase interesante de teorías de escalar-tensor con invarianza conforme manifiesta está dada por la acción CG~\eqref{ICGaction} y la acción del campo escalar con acoplamiento no mínimo $I_{\phi}$, es decir,
\begin{equation}
I_{\rm CG\phi} = I_{\rm CG} + I_{\phi} \, .
\label{ICGphi}
\end{equation}
A diferencia de la acción de Einstein-Hilbert en la sección anterior, la acción CG es invariante conforme y la única parte que debe ser completada conforme proviene del sector escalar. Esto se logra mediante la introducción de términos de borde que nos permiten llevar $I_{\phi}$ a la forma dada en la ecuación~\eqref{Iphiren}. Por lo tanto, ambas acciones son totalmente invariantes conformes y pueden ser escritas en la forma
\begin{equation}
I_{\rm CG\phi,conf}   \equiv \frac{1}{4}\int\diff{^4x}\sqrt{|g|}\,\delta^{\mu_1\ldots\mu_4}_{\nu_1\ldots\nu_4}\left(\alpha W^{\nu_1\nu_2}_{\mu_1\mu_2}W^{\nu_3\nu_4}_{\mu_3\mu_4} - \zeta\Sigma^{\nu_1\nu_2}_{\mu_1\mu_2}\Sigma^{\nu_3\nu_4}_{\mu_3\mu_4}\right)  \,.\label{Lagconformal}
\end{equation}
De hecho, esta es la teoría escalar-tensor invariante conforme más general de gravedad, construida a partir de dos tensores cuadráticos covariantes conformes en presencia de campos escalares.
Las ecuaciones de campo se obtienen mediante variaciones estacionarias de la acción con respecto a la métrica y el campo escalar, lo que resulta en
\begin{subequations}\label{eom}
\begin{align}\label{eomg}
    \mathcal{E}_{\mu\nu} &\equiv \alpha B_{\mu\nu} - 48\zeta\nu\, T_{\mu\nu} = 0\,, \\
    \label{eomp}
    \mathcal{E} &\equiv \Box\phi - \frac{1}{6}\phi R - 4\nu\phi^3 = 0\,, 
\end{align}
\end{subequations}
respectivamente. Para un campo escalar constante, estas ecuaciones de movimiento admiten espacios Einstein como soluciones ya que estos son Bach-flat y tienen tensor de energía-momentum nulo.

\section{Renormalización}
Con la finalidad de estudiar las consecuencias de la completitud conforme de $I_{\rm CG\phi}$ en su renormalizacion, escribiremos la acción en términos de el tensor de Weyl al cuadrado tomando en cuenta la descomposicion realizada en la Ec.~\eqref{weylfdecomposition}. En este caso, $I_{\rm CG\phi,conf}$ se verá como 
\begin{equation}\label{LagSigmaT}
    I_{\rm CG\phi,conf} = \int\diff{^4x}\sqrt{|g|}\left[ \frac{(\alpha-\zeta)}{4}\delta^{\mu_1\ldots\mu_4}_{\nu_1\ldots\nu_4}W^{\nu_1\nu_2}_{\mu_1\mu_2}W^{\nu_3\nu_4}_{\mu_3\mu_4} + \frac{24\zeta}{\phi^4}\left(3 T_{\mu}^{\nu}T_{\nu}^{\mu}-T^2\right)\right]\,.
\end{equation}
Lo anterior es solo una reescritura de la  Ec.~\eqref{Lagconformal}, la cual puede simplificarse aún más si se cumplen las ecuaciones de movimiento de Ec.~\eqref{eomp} o, de forma equivalente, $T=0$, dando como resultado
\begin{equation}\label{LagSigmaTon-shell}
I_{\rm CG\phi,conf}\Big|_{\rm on-shell} = \int\diff{^4x}\sqrt{|g|}\left[ \frac{(\alpha-\zeta)}{4}\delta^{\mu_1\ldots\mu_4}_{\nu_1\ldots\nu_4}W^{\nu_1\nu_2}_{\mu_1\mu_2}W^{\nu_3\nu_4}_{\mu_3\mu_4} + \frac{72\zeta}{\phi^4}T_{\mu}^{\nu}T_{\nu}^{\mu}\right]\,.
\end{equation}
Por lo tanto, la acción coincide con el tensor de Weyl al cuadrado, adquiriendo su comportamiento asintótico bien definido para cualquier espacio de AlAdS, cuando se consideran configuraciones correspondientes a $T_{\mu\nu}=0$. Además, como se observó en la última sección, la adición de los contratérminos que completan conformemente a $I_{\phi}$ conduce a una acción finita incluso para configuraciones con $T_{\mu \nu}$ no nulo, siempre y cuando el decaimiento del término $\phi^{-4}T_{\mu}^{\nu}T_{\nu}^{\mu}$ sea lo suficientemente rápido.

Además, existe un punto particular en el espacio de parámetros, es decir, $\alpha=\zeta$, donde la acción se anula idénticamente para todas las soluciones de la teoría para las cuales $T_{\mu \nu}=0$. Por lo tanto, la continuación Euclidiana de estas soluciones tiene una acción on-shell nula, al igual que la solución maximalmente simétrica, y por lo tanto se pueden considerar como parte del mismo estado de vacío de la teoría. Esta condición no implica necesariamente que el campo escalar deba ser trivial, más bien podría representar una configuración stealth. Esto proporciona una noción extendida de criticalidad ya que la acción Euclídea on-shell se anula para soluciones stealth en este punto particular del espacio de parámetros. Este resultado es completamente análogo a Critical Gravity para los campos puramente métricos~\cite{Lu:2011zk}, donde la acción y las cargas conservadas se anulan idénticamente para espacios de Einstein~\cite{Miskovic:2014zja,Anastasiou:2017rjf}. Sin embargo, en ambas teorías el vacío está determinado por espacios tipo Einstein. Aquí, en cambio, la criticalidad surge para configuraciones stealth que pueden o no ser espacios Einstein.

\section{Aplicaciones: Configuraciones Stealth sobre la métrica de Riegert}
Estudiemos las configuraciones stealth para verificar la Renormalización Conforme de manera explícita. Reemplazando la Ec.~\eqref{ansatz} en~\eqref{eom}, encontramos la siguiente solución
\begin{subequations}\label{solution}
\begin{align}
    f(r) &= k + \frac{6mG}{r_0} - \frac{2}{r_0}\left(k+\frac{3mG}{r_0} \right)r - \frac{2mG}{r} - \frac{\lambda r^2}{3}\,,\\
    \phi(r) &= \frac{1}{r-r_0}\sqrt{-\frac{k+\frac{2mG}{r_0}+\frac{\lambda r_0^2}{3}}{2\nu}}\,,
\end{align}
\end{subequations}
donde $m$, $\lambda$ y $r_0$ son constantes de integración. Esta solución existe solo para $\nu\neq0$ y $r_0\neq0$ y, hasta donde sabemos, se encontró por primera vez en~\cite{Brihaye:2009ef}. De hecho, aunque el campo escalar no es trivial, tiene un tensor de energía-momentum nulo. Por lo tanto, concluimos que esta solución representa un campo escalar stealth~\cite{Ayon-Beato:2004nzi,Ayon-Beato:2005yoq,Hassaine:2006gz,Ayon-Beato:2013bsa} sobre la métrica de Riegert~\cite{Riegert:1984zz}; esta última representa el espacio-tiempo más general estático y esféricamente simétrico que resuelve las ecuaciones de campo de CG. Además, esta solución está continuamente conectada al agujero negro de Schwarzschild-AdS topológico cuando $r_0\to\infty$, donde el campo escalar se vuelve constante, es decir, $\phi=\sqrt{-\frac{\lambda}{6\nu}}$.

La solución contiene una singularidad de curvatura en $r=0$, como se puede ver a partir de su escalar de Ricci, que es
\begin{equation}
    R = 4\lambda + \frac{12(3mG+kr_0)}{r_0^2\,r} - \frac{12mG}{r_0\,r^2}\,.
\end{equation}
La singularidad está cubierta por un único horizonte definido como la raíz real del polinomio cúbico $f(r_h)=0$. Para que exista un horizonte de agujero negro, se debe cumplir la condición $r_h>0$. Luego, centrándonos en el caso $k=1$ por simplicidad, identificamos dos casos posibles: (i) $r_0>0$ o (ii) $r_0<0$. En ambos casos, encontramos que $m>0$. En el primero, existe un polo en el campo escalar en $r=r_0$. Luego, exigiendo que el polo escalar se encuentre detrás del horizonte, encontramos $0<r_0<6mG<r_h$ y
\begin{align}
    \frac{(r_0-6mG)(r_0+3mG)^2}{r_0(3 mGr_0)^2 }&<\Lambda <0 &\lor& & \Lambda &=\frac{(r_0-6mG)(r_0+3mG)^2}{r_0(3 mGr_0)^2}\,.
\end{align}
Por otro lado, si $r_0<0$, el campo escalar es regular para $r\in\mathbb{R}_{>0}$. Entonces, de esa forma no es necesario pedir que $r_h>r_0$. Así, encontramos 
\begin{align}
    \bigg(r_0&\leq -2mG& &\land& \Lambda &\leq-\frac{3(r_0+2mG)}{r_0^3}\bigg)& &\lor& \bigg(r_0&>-2mG & &\land&  \Lambda &<0\bigg)\,.
\end{align}
Estas condiciones garantizan la existencia de una solución tipo agujero negro con un campo escalar regular fuera del horizonte. 

La temperatura del agujero negro de la solución~\eqref{solution} viene dada por
\begin{align}\label{beta1}
 T_H = \frac{(r_h-r_0)\left[k(3r_h-r_0) - \lambda r_h^2(r_h-r_0)\right]}{4\pi r_h \left[ 3r_h(r_h-r_0)  +r_0^2\right]} \,,
\end{align} 
Como anticipamos, la invariancia conforme de la acción~\eqref{Lagconformal} hace que la acción Euclídea on-shell sea finita para las configuraciones stealth. Esto se puede ver explícitamente evaluando la ecuación~\eqref{Lagconformal} en la solución~\eqref{solution}, obteniendo
\begin{align}
I_E &= -\frac{16(\alpha-\zeta)\beta\omega_{(k)}m^2G^2}{r_h^3}\left[1-\frac{3r_h}{r_0}\left(1-\frac{r_h}{r_0}\right) \right]\, .
\end{align}
Se puede notar que, aunque el campo escalar no tiene backreaction, su presencia modifica la acción Euclídea on-shell de una manera no trivial. 

Adicionalmente, existe otra solución tipo agujero negro asintóticamente AdS la cual no esta continuamente conectada a la solución de la Ec.~\eqref{solution}. Estas configuraciones fueron presentadas por primera vez en la Ref.~\cite{Herrera:2017ztd}, donde sus cargas conservadas fueron calculadas usando el formalismo de Abbot-Deser-Tekin~\cite{Abbott:1981ff,Deser:2002rt,Deser:2002jk,Deser:2003vh}. En las coordenadas que estamos usando en esta tesis, la solución asintótica de la Ref.~\cite{Herrera:2017ztd} viene dada por
\begin{subequations}\label{solutionsyerko}
\begin{align}
    f(r) &= k + 12\nu\phi_0 -br - \frac{\lambda r^2}{3}\,, \\
    \phi(r) &= \frac{\sqrt{\phi_0}}{r}\,,
\end{align}
\end{subequations}
donde $\phi_0$, $b$ y $\lambda$ son constantes de integración sujetas a la condición $(\zeta-4\alpha)(k+6\nu\phi_0)=0$. Aunque esta solución no está conectada de forma continua a~\eqref{solution}, también representa una configuración de campo escalar stealth. Esta solución posee una singularidad de curvatura en $r=0$, la cual puede ser cubierta por un horizonte ubicado en $r=r_h$ definido mediante el polinomio $f(r_h)=0$. En la sección euclidiana, la ausencia de singularidades cónicas exige que el período del tiempo Euclídeo sea
\begin{align}
    \beta &= \frac{4\pi r_h\ell^2}{r_h^2+k\ell^2}\,,
\end{align}
de donde se puede deducir la temperatura de Hawking. La acción Euclídea on-shell conformemente renormalizada~\eqref{Lagconformal} para la solución en la Ec.~\eqref{solutionsyerko} se verá como 
\begin{align}
I_E = -\frac{16\beta\omega_{(k)}k^2(\alpha-\zeta)}{3r_h},.
\end{align}
Por lo tanto, se puede llegar a la conclusión de que la renormalización conforme proporciona un valor finito para la acción Euclídea on-shell para la solución~\cite{Herrera:2017ztd} también.

\biblio 

\clearpage

\chapter{Conclusión}
    
En esta tesis se estudió cómo la renormalización de las teorías de gravedad escalar-tensor no mínimamente acopladas está dictada por la restauración de la simetría conforme on-shell en el volumen. Consideramos el caso de un campo escalar acoplado conformemente con un potencial cuártico, cuya teoría produce un término de frontera al realizar una transformación de Weyl. Entonces, la simetría conforme se restaura escribiendo el término cinético del escalar de manera no canónica, de tal manera que la acción pueda escribirse en términos del operador Yamabe. La acción resultante es invariante de Weyl, excepto por la configuración escalar constante, donde la transformación conforme se vuelve singular, lo que se evidencia en el hecho de que la acción es divergente para espacio-tiempos AlAdS. La restauración de la simetría de reescalamiento local se logra definiendo el tensor $\Sigma^{\mu \nu}_{\rho \lambda}$, en términos de la métrica y los grados de libertad escalares, de tal manera que sea covariante conforme y tenga el mismo peso conforme que el tensor de Weyl. Entonces, $\Sigma$ al cuadrado es un invariante conformal local de la teoría, que se puede usar para definir la acción de tal manera que la teoría se renormalice y se restaure la invariancia conforme completa, lo que implica que la variación de Weyl se anula exactamente y no hay puntos singulares en la transformación. Se demostró que la acción resultante se renormaliza on-shell, de tal manera que cualquier configuración de campo que satisfaga las EOM de la teoría tenga una acción finita al considerar espacio-tiempos AlAdS con condiciones asintóticas AdS débiles. Además, se observa que el campo escalar no se puede eliminar con una transformación conforme sin modificar el comportamiento de la frontera fuera de la condición AlAdS, lo que cambiaría el estado físico.

También se estudió la teoría que considera la gravedad Einstein-AdS renormalizada escrita en forma de MacDowell-Mansouri en presencia del campo escalar acoplado conformemente renormalizado. En ese caso, la teoría no tiene invariancia conforme debido al sector de Einstein-Hilbert. Sin embargo, para espacios Einstein y campos escalares constantes, se vuelve invariante conforme on-shell. En este caso, la teoría se renormaliza para espaciotiempos Einstein, ya que la acción completa se vuelve proporcional a Weyl al cuadrado. Otras configuraciones que también tienen una acción finita son aquellas con métricas AlAdS tales que sus grados de libertad no Einstein, codificados en el tensor de Ricci sin traza, disminuyen lo suficientemente rápido hacia la frontera conforme. El agujero negro MSTZ~\cite{Martinez:2002ru,Martinez:2005di} pertenece a esta categoría y se calculó el valor de la acción on-shell y la correspondiente temperatura del agujero negro.

Luego, se realizó el mismo análisis en gravedad conforme más campos escalares acoplados conformemente renormalizados, lo cual corresponde a la acción localmente conforme invariante más general construida a partir de contracciones antisimétricas cuadráticas de tensores covariantes de Weyl. De hecho, se demostró explícitamente que esta teoría está renormalizada para espacios de Bach planos, que por virtud de la EOM también son configuraciones stealths. Para este tipo de espacios, la acción está renormalizada para espacios de AlAdS ya que es proporcional a la acción CG. Como ejemplo particular, se consideraron configuraciones stealths sobre la métrica de Riegert y se demostró que su acción Euclidiana on-shell es finita.

En ambas teorías, existen puntos interesantes en el espacio de parámetros donde la acción se puede volver trivial para ciertos tipos de configuraciones métricas y escalares. En particular, para la teoría que incluye gravedad conforme, la acción se anula en el punto crítico de $\alpha = \zeta$ para configuraciones de Bach planas. De manera análoga, la teoría que incluye un sector de Einstein-AdS tiene una acción trivial en el punto crítico de $\alpha = \zeta$ para espacios de Einstein, que corresponden a soluciones a la EOM con un valor constante del campo escalar.

El hecho de que la acción sea trivial implica que todas las cargas asintóticas se anulan de manera idéntica, así como el potencial termodinámico que es proporcional a la acción en la sección Euclidiana. Esto sugiere una novedosa noción de criticidad en teorías escalar-tensoriales, que es diferente de la definición estándar formulada de manera perturbativa en términos de un desacoplamiento de los modos masivos de la métrica del espectro~\cite{Lu:2011zk}. Esta noción de criticidad termodinámica corresponde a puntos en el espacio de la teoría donde el estado fundamental de la teoría se agranda, de tal manera que la configuración máximamente simétrica tiene la misma energía libre que toda una clase de soluciones. De esta manera, forman un espacio de móduli de configuraciones de vacío, que permitirían transiciones espontáneas entre ellas sin costo de energía libre. Esta idea ya fue discutida en el caso de teorías de gravedad pura, para la gravedad de Einstein-AdS en 4D y 6D en las Refs.~\cite{Anastasiou:2016jix,Anastasiou:2017rjf,Anastasiou:2021tlv}. Aunque este punto es muy interesante, su análisis completo requiere un estudio cuidadoso de la termodinámica.

\biblio 

\clearpage

\newpage
\renewcommand\refname{Referencias}          
{\setstretch{1.0}                           
\addcontentsline{toc}{chapter}{Referencias} 
\bibliography{Referencias.bib}              
}

\newpage
\renewcommand{\appendixpagename}{Apéndices}     
\addcontentsline{toc}{chapter}{Apéndices}       






\end{document}